\documentclass[11pt]{article}
\usepackage{amsmath,amssymb}
\usepackage{color}
\newcommand{\ft}[2]{{\textstyle\frac{#1}{#2}}}

\newsavebox{\uuunit}
\sbox{\uuunit}
    {\setlength{\unitlength}{0.825em}
     \begin{picture}(0.6,0.7)
        \thinlines
        \put(0,0){\line(1,0){0.5}}
        \put(0.15,0){\line(0,1){0.7}}
        \put(0.35,0){\line(0,1){0.8}}
       \multiput(0.3,0.8)(-0.04,-0.02){10}{\rule{0.5pt}{0.5pt}}
     \end {picture}}

\allowdisplaybreaks

\numberwithin{equation}{section}

\begin{document}
\begin{titlepage}
\begin{center}
\hfill Nikhef-2014-013 \\
\hfill ITP-UU-14/21 \\
\vskip 6mm

{\LARGE \textbf{Deformations of special geometry:\\[.6ex]
    in search of the topological string }}
\vskip 8mm

\textbf{G.L.~Cardoso$^{a}$, B.~de~Wit$^{b,c}$ and S.~Mahapatra$^{d}$}

\vskip 6mm
$^a${\em  Center for Mathematical Analysis, Geometry and Dynamical Systems,\\
  Department of Mathematics, 
  Instituto Superior T\'ecnico,\\ Universidade de Lisboa,
  Av. Rovisco Pais, 1049-001 Lisboa, Portugal}\\[.2ex]
$^b${\em Nikhef, Science Park 105, 1098 XG Amsterdam, The
  Netherlands}\\[.2ex] 
$^c${\em Institute for Theoretical Physics, Utrecht University,\\
   Leuvenlaan 4, 3584 CE Utrecht, The Netherlands}\\[.2ex]
$^d${\em Physics Department, Utkal University, 
Bhubaneswar 751 004, India}\\[1ex]

{\tt gcardoso@math.ist.utl.pt}\;,\;\,{\tt  B.deWit@uu.nl}\;,\;\,{\tt
  swapna@iopb.res.in}
\end{center}

\vskip .2in
\begin{center} {\bf ABSTRACT } \end{center}
\begin{quotation}\noindent 
  The topological string captures certain superstring amplitudes which
  are also encoded in the underlying string effective action. However,
  unlike the topological string free energy, the effective action that
  comprises higher-order derivative couplings is not defined in terms
  of duality covariant variables. This puzzle is resolved in the
  context of real special geometry by introducing the so-called Hesse
  potential, which is defined in terms of duality covariant variables
  and is related by a Legendre transformation to the function that
  encodes the effective action. It is demonstrated that the Hesse
  potential contains a unique subsector that possesses all the
  characteristic properties of a topological string free energy. Genus
  $g\leq3$ contributions are constructed explicitly for a general
  class of effective actions associated with a special-K\"ahler target
  space and are shown to satisfy the holomorphic anomaly
  equation of perturbative type-II topological string theory.\\[.2ex]
  This identification of a topological string free energy from an
  effective action is primarily based on conceptual arguments and does
  not involve any of its more specific properties. It is fully
  consistent with known results. A general theorem is presented that
  captures some characteristic features of the equivalence, which
  demonstrates at the same time that non-holomorphic deformations of
  special geometry can be dealt with consistently.
  
\end{quotation}
\vfill
\end{titlepage}
\eject
\section{Introduction}
\label{sec:introduction}
\setcounter{equation}{0}
As is well known, Lagrangians for $N=2$ supersymmetric vector
multiplets are encoded in a holomorphic function $F(X)$ whose
arguments correspond to the complex scalar fields $X^I$ of the vector
multiplets. These Lagrangians often play a role as Wilsonian effective
field theories that describe the physics below a certain mass
scale. Homogeneity of the holomorphic function is required whenever
the vector multiplets are coupled to supergravity
\cite{deWit:1984pk}. The physical vector multiplet scalars are then
projectively defined in terms of these variables as a result of the
local scale and $\mathrm{U}(1)$ invariance of the description used in
\cite{deWit:1984pk}. It is possible that the function depends, in
addition, on one or more holomorphic fields, possibly associated with
some other chiral multiplets.  An example of this is the so-called
Weyl multiplet that describes the pure supergravity degrees of freedom
\cite{Bergshoeff:1980is}. When the function $F$ depends on the Weyl
multiplet, then it will also encode a class of higher-derivative
couplings.\footnote{
  These actions are all based on a chiral superspace density, but
  other $N=2$ supersymmetric higher-derivative couplings are known to
  exist (see e.g. \cite{deWit:2010za,Butter:2013lta}). The latter will
  not be considered in this paper.
}  
In case these higher-order derivative couplings are absent, we will
denote the function by $F^{(0)}$, which is always holomorphic and
homogeneous and encodes an action that is at most quadratic in
space-time derivatives. This action will henceforth be referred to as
the `classical action', and its associated non-linear sigma model
parametrizes a special-K\"ahler space.

The abelian vector fields in these actions are subject to
electric/magnetic duality under which the electric field strengths and
their duals transform under symplectic rotations. It is then possible
to convert to a different duality frame, by regarding half of the
rotated field strengths as the new electric field strengths and the
remaining ones as their duals. The latter are then derivable from a
new action. To ensure that the characterization of the new action in
terms of a holomorphic function remains preserved, the scalars of the
vector multiplets are transformed correspondingly. This amounts to
rotating the complex fields $X^I$ and the holomorphic derivatives of
the underlying function $F(X)$ by the same symplectic rotation as the
field strengths and their dual partners
\cite{deWit:1984pk,Cecotti:1988qn}. For reasons that will be described
shortly, we shall refer to the array $(X^I,F_J)$ as the {\it period
  vector}, where $F_J(X)=\partial_J F(X)$. The indices $I, J$ label
the vector multiplets, and cover the range $I,J=0,1,\ldots,n$, so that
the period vector has (complex) dimension $2(n+1)$. Electric/magnetic
duality thus constitutes a group of equivalence transformations that
relate two different Lagrangians (based on two different functions)
giving rise to an equivalent set of equations of motion and Bianchi
identities. A subgroup of these equivalence transformations may
constitute an invariance group, meaning that the Lagrangian and its
underlying function $F(X)$ remain unchanged. We stress that the latter
two quantities do not transform as a {\it function} under these
equivalence transformations.

As it turns out one encounters a similar situation when studying
Calabi-Yau three-folds. The moduli space of these three-folds is a local
product of two submanifolds, describing the metric deformations of the
complex structure and of the K\"ahler class, respectively. The complex
structure moduli determine the shape, and the K\"ahler moduli the size
of the Calabi-Yau three-folds. Usually, when referring to the
Calabi-Yau moduli space, one refers to either one of these two
submanifolds. As it turns out, the corresponding metric deformations
are related to the odd and even harmonic forms, respectively, and the
number of moduli is thus determined by the topology of the Calabi-Yau
three-folds (see e.g.  \cite{Candelas:1990pi}). The latter is specified
by the Hodge numbers $h^{p,\bar p}$ which specify the number of
independent $(p,\bar p)$ harmonic forms. The odd harmonic forms
consist of a $(3,0)$ holomorphic form $\Omega$, and $h^{2,1}$
$(2,1)$-forms, as well as their conjugate $(0,3)$- and
$(1,2)$-forms. Under an infinitesimal change of the complex structure
the $(3,0)$-form $\Omega$ changes into $\Omega$ and the $(2,1)$-forms,
which leads to its {\it periods}, in the following way,
\begin{equation}
  \label{eq:periods}
  X^I= \oint_{A^I} \Omega \,,\qquad F^{(0)}{\!}_I =\oint_{B_I} \Omega \,.
\end{equation}
Here $A^I$ and $B_I$ are an integral basis of homology 3-cycles (in a
symplectic basis) dual to the three-forms. The index $I$ takes the
values $I=0,1,\ldots, h^{2,1}$, corresponding to the $(3,0)$- and the
$(2,1)$-forms and their conjugates. The $X^I$ (or alternatively the
$F^{(0)}{\!}_I$) projectively parametrize the complex-structure
deformations, so that the complex dimension of the corresponding
moduli space equals $h^{2,1}$.~\footnote{
  For completeness we mention that there exists an analogous
  construction for the K\"ahler moduli. While the complex structure
  moduli are associated with the odd cohomology class,
  $\mathbb{H}^{(3,0)}\oplus\mathbb{H}^{(2,1)}\oplus \mathbb{H}^{(1,2)}
  \oplus \mathbb{H}^{(0,3)}$, the K\"ahler moduli are associated with
  the even class,
  $\mathbb{H}^{(0,0)}\oplus\mathbb{H}^{(1,1)}\oplus\mathbb{H}^{(2,2)}
  \oplus \mathbb{H}^{(3,3)}$. The corresponding $2+2\, h^{1,1}$
  coordinates projectively describe the $h^{1,1}$ complex K\"ahler
  moduli \cite{Candelas:1990pi}.  }  

For the periods one can also show (at least in a suitable homology
basis) that there exists a holomorphic, homogeneous, function
$F^{(0)}(X)$, such that $F^{(0)}{\!}_I=\partial_I
F^{(0)}(X)$.~\footnote{ 
  The functions $F^{(0)}$ encoding the moduli space geometry of
  Calabi-Yau three-folds correspond to a restricted class. This paper
  pertains to functions belonging to a more general class. }  %
The duality transformations on the periods simply arise from
symplectic redefinitions of the homology basis of the 3-cycles. It is
worth pointing out that in this case one is dealing with discrete
symplectic transformations, while in the supergravity case the
transformations are continuous (unless one has to account for an
integral lattice of electric and magnetic charges).  This particular
geometry with its associated period vectors and symplectic
transformations is known as {\it special geometry}
\cite{Strominger:1990pd} (for a review, see \cite{Craps:1997gp}). 

The Calabi-Yau moduli space and the supergravity action describing a
Calabi-Yau string compactification are related, because the target
space metric associated with the non-linear sigma model contained in
the corresponding Wilsonian effective action of vector multiplets
coupled to supergravity, must be equal to the metric of the Calabi-Yau
moduli space \cite{Seiberg:1988pf}.  This target space is a so-called
special-K\"ahler space, whose K\"ahler potential is proportional to
\cite{deWit:1984pk},
\begin{equation}
  \label{eq:spec-K-pot}
  K(t,\bar t) \propto \log\left[\frac{\mathrm{i}\big(X^I\,\bar
      F^{(0)}\!{}_I - 
      \bar X^I\, F^{(0)}\!{}_I \big)}{\vert X^0\vert^2} \right]\,,
\end{equation}
and $F^{(0)}(X)$ is the holomorphic function that determines the
supergravity action quadratic in space-time
derivatives. Because $F^{(0)}(X)$ is homogeneous of second degree,
this K\"ahler potential depends only on the `special' coordinates
$t^i=X^i/X^0$ and their complex conjugates, where $i=1,\ldots,n$, so
that we are dealing with a special-K\"ahler space of complex dimension
$n$. In view of the homogeneity, the symplectic rotations acting on
the period vector $(X^I,F^{(0)}{\!}_I)$ induce corresponding
(non-linear) transformations on the special coordinates $t^i$. Up to a
K\"ahler transformations, the K\"ahler potential transforms as a
function under duality.

Yet another quantity that reflects the geometrical features of the
Calabi-Yau moduli space is the topological string. Perturbative string
theory is defined in terms of maps from Riemann surfaces $\Sigma_g$ to
a target space.  When the worldsheet theory has $(2,2)$ supersymmetry
and the target space is Ricci flat, one may construct a topological
version of perturbative string theory by a procedure called twisting
\cite{Witten:1988xj}.  The resulting theory is a cohomological theory,
and correlation functions of observables are independent of the
wordsheet metric on $\Sigma_g$.  When the target-space is a Calabi-Yau
three-fold, the twisted theory is called {\it topological string
  theory} \cite{Bershadsky:1993cx}. There exist two versions of
topological string theory, called the A- and the B-models. In the
A-model, the correlation functions depend only on the K\"ahler moduli,
while in the B-model they only depend on the complex structure moduli.

The topological string is defined in terms of topological free
energies $F^{(g)}$, which are computed from suitable correlators on
orientable Riemann surfaces $\Sigma_g$ of genus $g$.  These free
energies may be formally combined into one single object
${F}(X,\lambda)$, the free energy of topological string theory, which
has the asymptotic expansion,
\begin{equation}
  \label{eq:top-string-F}
  {F}(X,\lambda)  = \sum_{g=0}^{\infty} \lambda^{g-1} \, F^{(g)}(X) \;,
\end{equation}
with $\lambda$ playing the role of a formal (complex) expansion
parameter.  This expression, being a perturbative series in $\lambda$,
is expected to receive non-perturbative corrections in $\lambda$
\cite{Gopakumar:1998ii,Gopakumar:1998jq,Eynard:2008he}. Note that we
have (tentatively) included $F^{(0)}(X)$ in \eqref{eq:top-string-F},
which is the function that encodes the Calabi-Yau moduli space
metric. The functions $F^{(g)}(X)$ are homogeneous functions of the
$X^I$ of degree $-2(g-1)$, so that the $X^0$-dependence can be scaled
out and subsequently be absorbed into the expansion parameter
$\lambda$. In this way ${F}(t,\lambda^\prime)= (X^0)^{-2} \,
{F}(X,\lambda)$ with $\lambda^\prime= \lambda/(X^0)^2$, and
$F^{(g)}(t)= (X^0)^{2g-2} \, F^{(g)}(X)$. This suggests a role of
$\lambda^\prime$ as a loop-counting parameter with
$(\lambda^{\prime})^{-1}\,F^{(0)}(t)$ equal to the classical free energy.

It may seem tempting to identify the expansion \eqref{eq:top-string-F}
in this way with the similar expansion of the effective action in
terms of $W^2$, the square of the lowest component of the Weyl
multiplet, thus generating higher-derivative couplings in the
action. However, this interpretation is inconsistent with the
behaviour one expects for the free energy of the topological
string. The reason is that the genus-$g$ free energies should be
consistent with dualities induced by symplectic rotations of the
periods $(X^I,F^{(0)}{\!}_I)$ of the underlying Calabi-Yau moduli
space. In particular, the $F^{(g)}(X)$ (with $g>0$) should transform
as {\it functions} under these dualities, and the $F^{(g)}(t)$ as
sections. Note that this does not apply to $F^{(0)}$, which does not transform
as a function under electric/magnetic duality. At this point one
concludes that it was premature to include $F^{(0)}$ into the free
energy of the topological string, as the genus-$g$ contributions with
$g>0$ behave as functions under duality, while $F^{(0)}$ does
not. 

On the other hand, the Wilsonian action encoded by the similar
expansion, 
\begin{equation}
  \label{eq:expansion-lea}
  F(X,W^2) = \sum_{g=0}^\infty \;(W^2)^g \,F^{(g)}(X)\,, 
\end{equation}
is subject to different duality transformations, namely those induced
by rotations of the {\it full} period vector $(X^I,F_I)$ rather than
of the `classical' period vector $(X^I,F^{(0)}{\!}_I)$. Consequently
the corresponding coordinates $X^I$ will transform differently under
duality, so that one must conclude that the $X^I$ appearing in the
topological string free energy and the $X^I$ appearing in the
Wilsonian action cannot be identical variables. Hence the coefficient
functions $F^{(g)}$ for the Wilsonian action appearing in
\eqref{eq:expansion-lea} that multiply even powers of the Weyl
multiplet are {\it not} transforming as functions under duality,
unlike those of the topological string. This aspect is most striking
when considering duality symmetries such as S- and T-duality. Under
these dualities the functions $F^{(g)}$ of the topological string are
invariant (possibly up to a scale factor), whereas the analogous
coefficient functions of the Wilsonian action \eqref{eq:expansion-lea}
transform non-linearly and are not invariant. Hence, in spite of the
similarity of the expansions, there is no ground for assuming that the
coefficient functions $F^{(g)}$ of the topological string will
coincide with the corresponding coefficient functions appearing in the
expression \eqref{eq:expansion-lea} that encodes the effective
action. This observation was already made in
e.g. \cite{Cardoso:2008fr,Cardoso:2010gc}, where its consequences were
investigated for dualities that define symmetries of the
model. Nevertheless, we should stress that there must exist a relation
between the effective action and the topological string in view of the
fact that the topological string does capture certain contributions to
string amplitudes, which must in turn be reflected in the effective
action \cite{Bershadsky:1993cx,Antoniadis:1993ze}.

We thus conclude that one seems to be dealing with two different
series expansions of the form \eqref{eq:top-string-F} and
\eqref{eq:expansion-lea}, one pertaining to the topological string
free energy and another one to the effective Wilsonian action with a
class of higher-derivative coupings. In spite of their qualitatively
different behaviour with respect to duality they should somehow
describe the same physics. To make matters more subtle, it is known
that both the topological string and the effective supergravity action
are subject to non-holomorphic modifications. Hence it is reasonable
to expect that these modifications are therefore related as
well. However, so far non-holomorphic deformations have not been
incorporated in the standard treatment of special geometry. The
non-holomorphic modification in the effective action is due to the
integration over massless modes \cite{Dixon:1990pc}, whereas those in
the topological string free energy originate from the pinching of
cycles of the Riemann surfaces \cite{Bershadsky:1993ta}. The need for
non-holomorphic corrections can often be deduced from the lack of
invariance under integer-valued duality symmetries, which requires
modular functions that are not fully holomorphic. This was also
observed when calculating the entropy for BPS black holes with
S-duality invariance \cite{LopesCardoso:1999ur}.

In this paper we will systematically study the connection between the
effective action and the topological string.\footnote{ 
  A preliminary account of our results was published in the
  proceedings of the {\it Frascati School 2011 on Black Objects in
  Supergravity } \cite{Cardoso:2012nh}. }   
Here we should stress that we are just referring to functions that can
{\it potentially} define the topological string free energy in
relation to an underlying effective action. Whether these functions
will actually have a topological string realization is a priori not
known. But the connection that is proposed in this paper seems to be
universal so that it will apply also to those cases where a
topological string realization does exist. We will start from the
holomorphic function that encodes the Wilsonian action, and construct
another quantity that transforms in the same way under duality as the
topological string free energy. Here we are inspired by previous work
on BPS black holes
\cite{LopesCardoso:2006bg,Cardoso:2008fr,Cardoso:2010gc}, where the
so-called Hesse potential emerged as the relevant quantity, defined in
the context of {\it real special geometry}
\cite{Freed:1997dp,Alekseevsky:1999ts,Cortes:2001qd}. The Hesse
potential transforms as a (real) {\it function} under duality. It is
related to the function $F(X)$ that encodes the effective action via a
Legendre transform and it is expressed in terms of duality covariant
variables, so that its behaviour under duality is comparable to what
one observes for the topological string. In hindsight it is not so
difficult to understand this relation by reflecting on the more
familiar case of four-dimensional abelian gauge fields, where the
Lagrangian is a function of the abelian field strengths $F_{\mu\nu}$
and possibly other fields (that we assume to be electrically
neutral). The expressions for the dual field strengths, which are
related to the derivative of the full Lagrangian with respect to the
original field strength, do depend on the specific interaction terms
contained in the Lagrangian.  Therefore the electric/magnetic duality
transformation rules for the original field strengths will depend on
the details of the underlying Lagrangian.  On the other hand the
Hamilonian depends on different quantities, namely the spatial part of
the gauge potential $\boldsymbol{A}$ and the electric displacement
field $\boldsymbol{D}$, where the latter follows from taking the
derivative of the Lagrangian with respect to $\boldsymbol{E}$. The
precise definition of $\boldsymbol{D}$ will thus implicitly depend on
the details of the Lagrangian. Under electric/magnetic duality
$\boldsymbol{B}\propto \boldsymbol{\nabla}\times\boldsymbol{A}$ and
$\boldsymbol{D}$ transform as a dual pair and the Hamiltonian is a
function of these duality covariant variables: they transform into
each other under symplectic rotations in a way that is independent of
the details of the Hamiltonian. In fact, for a theory without higher
derivatives, $\boldsymbol{A}$ and $\boldsymbol{D}$ are the canonical
variables.

Electric/magnetic duality transformations thus act as canonical
transformations and the Hamiltonian will usually decompose into a
number of different {\it functions} that transform consistently under
them. When the canonical transformations constitute
an invariance of the system then these functions will be {\it
invariant}.  As it turns out, the Hesse potential of real special
geometry is the direct analogue of the Hamiltonian. Rather than
depending on the fields $X^I$, it depends on
canonical variables $\phi^I$ and $\chi_I$. As we shall see, these can
again be combined into complex variables in a way that involves the
classical period vector associated with the function
$F^{(0)}(X)$. The duality covariant variables $(\phi^I,\chi_J)$ transform
under the same duality transformations as the classical period vector.

The Hesse potential is related to the function $F(X,W^2)$ via a
Legendre transform, and thus contains the same information as the
effective action. In principle, other relevant quantities that are
related to the underlying Calabi-Yau moduli space, such as the
topological string free energy, can be characterized by functions of
$(\phi^I,\chi_J)$. Precisely as the Hamiltonian discussed above, the
Hesse potential decomposes into different functions that all transform
consistently under duality. The central conjecture of this paper is
that the topological string should coincide with (part of) the Hesse
potential, as this is the only way to explain why it can reproduce
(part of) the effective action. To identify this particular function
we will first consider what happens when the effective action is
purely Wilsonian. As it turns out there is just one function belonging
to the Hesse potential that is `almost harmonic', where the meaning
and implication of the term `almost harmonic' will be explained in due
course. The Hesse potential is nevertheless harmonic in terms of the
(holomorphic) function that encodes the Wilsonian action. Subsequently
it is demonstrated that, upon relaxing the harmonicity constraint on
the function that encodes the effective action, the resulting `almost
harmonic' contribution to the Hesse potential satisfies the same
holomorphic anomaly equation that is known from the topological
string.

The paper is organized as follows. In section
\ref{sec:deformed-special} we characterize possible non-holomorphic
deformations of special geometry in the context of the effective
action based on a theorem that is presented in appendix
\ref{App:theorem}. Subsequently we introduce the formulation of real
special geometry in terms of the Hesse potential, which transforms as
a function under symplectic rotations of its real variables, and
derive a number of results that are important for what follows in
subsequent sections.  Section \ref{sec:eval-hesse-potent} is devoted
to evaluating the Hesse potential in terms of complex duality
covariant variables by carrying out the Legendre transform by
iteration to fourth order, which is sufficient to appreciate its
general structure. At this stage the non-holomorphic contributions can
be understood in the context of a diagrammatic representation of the
Hesse potential as a sum over connected tree graphs.  The Hesse
potential decomposes into an infinite number of terms, which arrange
themselves into an infinite set of functions, all transforming
consistently under electric/magnetic duality.  Some of the expressions
for these terms are collected in appendix \ref{App:funct-H-a-i-geq2}
up to the corresponding order in the iteration.  When the
corresponding effective action is characterized in terms of a
holomorphic function, precisely one of the functions contributing to
the Hesse potential becomes `almost harmonic' in the moduli. This
function is thus the only possible candidate for a topological string
free energy, and in section \ref{sec:non-holom-deform} we demonstrate
that it indeed satisfies a holomorphic anomaly equation which
partially coincides with the holomorphic anomaly equation known for
the topological string \cite{Bershadsky:1993cx}.

Subsequently we relax the harmonicity restriction on the effective
action by allowing a specific non-holomorphic term that transforms as
a symplectic function up to a term that is harmonic. Introducing such
a term induces quite a large variety of additional contributions to
the Hesse potential that leave its characteristic properties intact, 
but the candidate function for the topological string free energy now
satisfies the full holomorphic anomaly equation. Hence this
function has now all the prerequisites for representing the
generating function of the genus-$g$ free energies of the topological
string and we explicitly demonstrate this up to $g\leq3$. The
calculation for $g=3$ is rather involved and it is described in
section \ref{sec:third-order-contr}.  We should stress here that the
logic of our calculations is rather different from the one that is
often followed for the topological string, where the non-holomorphic
corrections are found by integrating the anomaly equation
\cite{Bershadsky:1993cx,Aganagic:2006wq,Grimm:2007tm}, with the
holomorphic contributions playing the role of generalized integration
`constants'. In this paper we {\it construct} the Hesse potential
starting from holomorphic functions, which, in order to ensure that
they transform consistently under duality transformations, will
necessarily contain non-holomorphic contributions.  These
non-holomorphic contributions then turn out to satisfy the holomorphic
anomaly equation. In this way it is obvious that the holomorphic
anomaly arises due to an incompatibility between duality covariance
and holomorphicity. On the other hand we demonstrate that the function
that encodes the effective action, which is holomorphic in the
Wilsonian limit, must also contain corresponding non-holomorphic
corrections of a specific form, which we evaluate order-by-order by
iteration.  All these results are established in the context of a
generic special-K\"ahler space, but we do not wish to imply that in
all these cases an actual topological string realization will exist.

A summary and a discussion of the results is presented in section
\ref{sec:summary-conclusions}. Here we also present a comparison of
our present results with previous work
\cite{Cardoso:2008fr,Cardoso:2010gc} on the FHSV model
\cite{Ferrara:1995yx}. Furthermore we briefly discuss some of the
consequences of the results of this paper for BPS black hole entropy,
especially in connection with its conjectured relation to the
topological string \cite{Ooguri:2004zv}.

There are five appendices. The first appendix \ref{App:theorem}
establishes the consistency of special geometry under non-holomorphic
deformations.  The second appendix \ref{App:funct-H-a-i-geq2} lists a
number of symplectic functions that emerge when evaluating the Hesse
potential by iteration. Appendix \ref{App:transformation-omega} lists
some intermediate results that are relevant for the third-order
calculation described in section \ref{sec:third-order-contr}.
The explicit expressions for the twisted string free energies
$F^{(g)}$ of genus $g\leq3$ are presented in appendix \ref{App:F(g)-s}
based on the construction presented in this paper. Finally in appendix
\ref{App:specific-model} we give further details about the comparison
of the present results to earlier results obtained for the FHSV
model. 

\section{Real and deformed special geometry }
\label{sec:deformed-special}
\setcounter{equation}{0}
In the previous section we introduced holomorphic
functions that encode either the Wilsonian action or the topological
string free energy, as well as a real function known as the Hesse
potential.  While the first two are initially holomorphic, they
eventually acquire non-holomorphic terms caused by the underlying
physics. In the Hesse potential there seems no immediate obstacle to
include such modifications as it is initially defined in terms of real
variables. Let us now reiterate some of the distinctive features of
these three structures and clarify the relevant issues.

The topological string free energy is a function of the Calabi-Yau
moduli which are subject to dualities related to the homology group of
the underlying holomorphic three-form. As explained in section
\ref{sec:introduction}, these moduli are associated with a holomorphic
function $F^{(0)}(X)$, which also encodes a corresponding vector
multiplet Lagrangian with at most two space-time derivatives coupled
to supergravity. The dualities of this Lagrangian are generated by
certain electric/magnetic dualities and they are related to the
(discrete) homology group associated with the Calabi-Yau periods. When
deforming the supergravity, for instance by introducing couplings to
the square of the Weyl multiplet as specified in
\eqref{eq:expansion-lea}, the duality transformations of the moduli
$X^I$ will change their form, whereas the variables $X^I$ in the
topological string will still be associated with
$F^{(0)}(X)$. Therefore, as explained in the previous section, the
supergravity definition and the topological string definition of the
variables $X^I$ will no longer be the same, and correspondingly the
genus-$g$ free energies cannot be identical to the higher-derivative
supergravity couplings.

The topological string free energy contains non-holomorphic
corrections related to the pinchings of cycles in the underlying
Riemann surfaces. These corrections should presumably be related to
the non-holomorphic contributions to the function $F(X,W^2)$ which are
induced by the integration over massless modes, in view of the fact
that the two quantities are known to describe the same (on-shell)
string amplitudes
\cite{Bershadsky:1993cx,Antoniadis:1993ze}. Irrespective of this
relationship the situation regarding the non-holomorphic corrections
to the function $F(X,W^2)$ is subtle. Integrating out the massless
modes leads to interactions that are non-local in generic space-times
and it is not known what the precise dictionary is between
non-holomorphic terms in the function $F$ and the non-local terms in
the Lagrangian. In the supergravity context, non-holomorphic
corrections are most likely related to chiral anomalies associated
with the $\mathrm{U}(1)$ local symmetry that is an essential part of
the superconformal multiplet calculus. These anomalies are cancelled
by the non-holomorphic terms that emerge in the effective action. Such a
phenomenon has been clarified in \cite{deWit:1985bn} for a number of
situations. Another relevant observation is that non-holomorphic
corrections are often required in order to have an exact duality
invariance.

The variables $X^I$ and $W^2$ are only projectively defined, so that
physically relevant results should not depend on uniform rescalings by
a complex number. Hence we can replace the $X^I$ by uniformly rescaled
variables $Y^I$ that differ by a uniform multiplicative complex factor
or field according to a prescription that may depend on the
application that is being considered. Likewise one must also rescale
the expression for $W^2$ by the square of the same factor as for the
$X^I$. The resulting expression is usually denoted by
$\Upsilon$. However, in what follows we will regard $\Upsilon$ as one
of the generalized coupling constants that may play a role. As it
turns out it is not necessary to refer explicitly to such coupling
constants, so that we will suppress them henceforth.

The information encoded in $F(X,W^2)$ can also be encoded in the
context of {\it real} special geometry where the relevant quantity is the
Hesse potential. As was already argued in the introductory section the Hesse
potential represents the Hamiltonian form of the Wilsonian action and
depends on real duality-covariant variables denoted by $\phi^I$ and
$\chi_I$. They can be defined by 
\begin{equation}
  \label{eq:real-special-vars}
  \phi^I = Y^I + \bar Y^{\bar I}\,,\qquad \chi_I = F_I + \bar F_{I}\,.
\end{equation}
Note that the replacement of the original variables $X^I$ by $Y^I$ is
now relevant, as it would not make sense to consider linear
combinations of the original variables $X^I$ and their complex
conjugates in view of the fact that they are projectively
defined.\footnote{ 
  The same strategy was followed previously in the study of BPS black
  holes (see, e.g. \cite{LopesCardoso:2000qm}). The same comment
  applies to the holomorphic derivatives $F_I$. Note that $\bar F_I$
  equals the derivative of $\bar F$ with respect to $\bar Y^I$. At
  this point we refrain from distinguishing holomorphic and
  anti-holomorphic derivatives, $\partial/\partial Y^I$ and
  $\partial/\partial\bar Y^I$, by the use of different types of
  indices. } 
As it turns out, non-holomorphic corrections can be encoded in a real
function $\Omega(Y,\bar Y)$, which is incorporated into the function
$F$ in the following way \cite{Cardoso:2004xf},
\begin{equation}
  \label{eq:F-decomposition}
  F(Y,\bar Y) = F^{(0)}(Y) + 2\mathrm{i}\,\Omega(Y,\bar Y) \,,
\end{equation}
where $F^{(0)}(Y)$ is holomorphic and homogeneous of second degree.
Note that the decomposition \eqref{eq:F-decomposition} is subject to
the equivalence transformation,
\begin{equation}
  \label{eq:holo-eq}
  F^{(0)}(Y) \to F^{(0)}(Y) + g(Y)\,,\qquad \Omega(Y,\bar Y)\to
  \Omega(Y,\bar Y) - \mathrm{Im}\, g(Y)\,,
\end{equation}
which amounts to a shift of $F(Y,\bar Y)$ by an anti-holomorphic
function: $F(Y,\bar Y)\to F(Y,\bar Y)+\bar g(\bar Y)$. This change
does not affect the period vector $(Y^I,F_I)$, which only involves
{\it holomorphic} derivatives, which is the underlying reason for this
equivalence.  When the function $\Omega$ is harmonic, i.e., when it
can be written as the sum of a holomorphic and an anti-holomorphic
function, then one may simply absorb the holomorphic part into the
first term according to \eqref{eq:holo-eq}. We usually refer to
$F^{(0)}(Y)$ as the {\it classical} contribution, because it refers to
the part of the Lagrangian that is quadratic in space-time
derivatives. In that case only the function $\Omega$ will depend on
possible deformation parameters such as $\Upsilon$ and $\bar\Upsilon$
and it may contain harmonic and non-harmonic contributions. The ansatz
\eqref{eq:F-decomposition} may seem somewhat ad hoc, but in fact it
can be derived in a much more general context as proven in the theorem
presented in appendix \ref{App:theorem}, which makes use of the
analogue of the Hesse potential. The first indication for these
results came from the study of BPS black hole entropy
\cite{LopesCardoso:1998wt,Cardoso:2004xf,LopesCardoso:2006bg,
  Cardoso:2008fr,Cardoso:2010gc}.

The new variables \eqref{eq:real-special-vars} have the virtue of
transforming linearly under duality by real symplectic rotations. At
this point it is convenient to define a quantity $\mathcal{H}$ of
$\phi^I$ and $\chi_I$, which contains the same information as the
$F(Y,\bar Y)$ but transforms as a {\it function} under the duality
transformations.  This quantity is the Hesse potential. It is a
generalization of the Hesse potential that was defined in the context
of real special geometry
\cite{Freed:1997dp,Alekseevsky:1999ts,Cortes:2001qd} and follows from
the Legendre transform of $4(\mathrm{Im}\,F^{(0)}+\Omega)$ with
respect to the imaginary part of $Y^I$,
\begin{equation}
  \label{eq:GenHesseP}
  \mathcal{H}(\phi,\chi) = 4 \,\big[ {\rm
    Im}\,F^{(0)} (Y) +\Omega(Y,\bar Y) \big]  + \mathrm{i} \chi_I \,(Y^I-\bar Y^I)  \;. 
\end{equation}
Its generic variation satisfies 
\begin{equation}
  \label{eq:delta-Hesse}
  \delta\mathcal{H} = -\mathrm{i} (F_I -\bar F_I)
  \,\delta\phi^I +\mathrm{i} (Y^I-\bar Y^I) \,\delta\chi_I \,,
\end{equation}
where $F_I$ refers to the holomorphic derivative of
\eqref{eq:F-decomposition}, which confirms that $\mathcal{H}$ is
indeed a function of the duality-covariant variables
$(\phi^I,\chi_I)$. The theorem of appendix \ref{App:theorem}
demonstrates that many of the special geometry properties remain valid
under non-holomorphic deformations. This result had already been
indicated by earlier work on this subject in \cite{Cardoso:2008fr}.

The classical function $F^{(0)}$ is assumed to be holomorphic and
homogeneous of second degree in $Y^I$. In the remainder of the section
we summarize some results for such a function with respect to its
behaviour under electric/magnetic duality that are needed in the next
section.  The electric/magnetic dualities are defined by
$\mathrm{Sp}(2n+2,\mathbb{R})$ rotations of the period vector
$(Y^I,F_I)$, defined in the usual way,
\begin{align}
  \label{eq:em-duality}
  Y^I\to&\, \tilde Y^I = U^I{}_JY^J + Z^{IJ}F_J\,,\nonumber\\
  F_I\to&\, \tilde F_I= V_I{}^JF_J+ W_{IJ}Y^J\,, 
\end{align}
where $U$, $V$, $Z$ and $W$ are the $(n+1)\times(n+1)$ real
submatrices that constitute an element of
$\mathrm{Sp}(2n+2,\mathbb{R})$. Applying these transformations to the
case where $F=F^{(0)}$, so that we are dealing with a homogeneous and
holomorphic function, it follows that $\tilde F^{0)}{}_I$ can be
expressed as the holomorphic derivative of a new holomorphic function,
$\tilde F^{(0)}(\tilde Y)$, with the latter equal to
\begin{align}
  \label{eq:new-F}
  \tilde F^{(0)}(\tilde Y)=&\, F^{(0)}(Y) -\ft12 Y^I F^{(0)}{\!}_I(Y)
  + \ft12 (U^\mathrm{T} W)_{IJ} \,Y^IY^J  \nonumber
  \\
  &\, 
  + \ft12 (U^\mathrm{T} V+ W^\mathrm{T} Z)_I{}^J \, Y^I\,  F^{(0)}{\!}_J(Y)
  + \ft12 (Z^\mathrm{T} V)^{IJ} \,F^{(0)}{\!}_I(Y)\,F^{(0)}{}_J(Y)\,,
\end{align}
which, in general, is difficult to solve explicitly. Note that, when
the function is not homogeneous, there are integration constants
corresponding to either a constant or terms proportional to the
$\tilde Y^I$.  In the presence of non-holomorphic terms the proof of
existence of a new function is much more complicated, but the
arguments, presented in a more generic context in appendix
\ref{App:theorem}, indicate that this is indeed the case, although no
explicit expression has been given in the general case in analogy to
\eqref{eq:new-F}.

Finally we present the transformation rules of the first multiple
derivatives of the function $F^{(0)}$ under the dualities
\eqref{eq:em-duality}, 
\begin{align} 
  \label{eq:F-0-der}
  \tilde F^{(0)}{\!}_{IJ}=&\,(V_I{}^L F^{(0)}{\!}_{LK} + W_{IK})\,
      [\mathcal{S}_0^{-1}]^K{}_J \,,\nonumber\\[.3ex]
   \tilde F^{(0)}{\!}_{IJK}=&\,   [\mathcal{S}_0^{-1}]^L{}_I
   \,[\mathcal{S}_0^{-1}]^M{}_J\,  [\mathcal{S}_0^{-1}]^N{}_K
   \,F^{(0)}{\!}_{LMN} \,,  \nonumber\\[.3ex]
   \tilde F^{(0)}{\!}_{IJKL}=&\,   [\mathcal{S}_0^{-1}]^{(M}{}_{I}
   \,[\mathcal{S}_0^{-1}]^N{}_{J}\,  [\mathcal{S}_0^{-1}]^P{}_K
   \,[\mathcal{S}_0^{-1}]^{Q)}{}_{L} \nonumber\\
   &\,\times \big[F^{(0)}{\!}_{MNPQ} - 3\, F^{(0)}{\!}_{\bullet MN}\,
   \mathcal{Z}_0{}^{\bullet\bullet} \, F^{(0)}{\!}_{PQ\bullet} \big]\,, \nonumber\\[.3ex]
  \tilde F^{(0)}{\!}_{IJKLM}=&\,   [\mathcal{S}_0^{-1}]^{(N}{}_{I}
   \,[\mathcal{S}_0^{-1}]^P{}_{J}\,  [\mathcal{S}_0^{-1}]^Q{}_K
   \,[\mathcal{S}_0^{-1}]^R{}_{L}
   \,[\mathcal{S}_0^{-1}]^{S)}{}_{M}\nonumber\\
   &\,\times
   \big[F^{(0)}{\!}_{NPQRS} - 10\, F^{(0)}{\!}_{\bullet NPQ}\,
   \mathcal{Z}_0{}^{\bullet\bullet} \, F^{(0)}{\!}_{RS\bullet} \nonumber\\
   &\qquad +15\,F^{(0)}{\!}_{\bullet NP}\,
   \mathcal{Z}_0{}^{\bullet\bullet} \, F^{(0)}{\!}_{\bullet\bullet Q}\,\mathcal{Z}_0{}^{\bullet\bullet} \,
   F^{(0)}{\!}_{RS\bullet} \big]\,, \nonumber\\[.3ex] 
   \tilde F^{(0)}{\!}_{IJKLMN}=&\,   [\mathcal{S}_0^{-1}]^{(P}{}_{I}
   \,[\mathcal{S}_0^{-1}]^Q{}_{J}\,  [\mathcal{S}_0^{-1}]^R{}_K
   \,[\mathcal{S}_0^{-1}]^S{}_{L} \,[\mathcal{S}_0^{-1}]^T{}_{M} 
   \, [\mathcal{S}_0^{-1}]^{U)}{}_{N}\nonumber\\
   &\, \times
   \Big[F^{(0)}{\!}_{PQRSTU} - 15\, F^{(0)}{\!}_{\bullet PQRS}\,
   \mathcal{Z}_0{}^{\bullet\bullet} \, F^{(0)}{\!}_{TU\bullet}\nonumber\\
   &\qquad
    -10 \, F^{(0)}{\!}_{\bullet PQR}\,
   \mathcal{Z}_0{}^{\bullet\bullet} \, F^{(0)}{\!}_{STU\bullet}\nonumber\\
   &\qquad +60 \,F^{(0)}{\!}_{\bullet PQR}\,
   \mathcal{Z}_0{}^{\bullet\bullet} \, F^{(0)}{\!}_{\bullet\bullet S}\,\mathcal{Z}_0{}^{\bullet\bullet} \,
   F^{(0)}{\!}_{TU\bullet} \nonumber\\
   &\qquad + 45\,F^{(0)}{\!}_{\bullet PQ}\,
   \mathcal{Z}_0{}^{\bullet\bullet} \, F^{(0)}{\!}_{\bullet\bullet RS}\,\mathcal{Z}_0{}^{\bullet\bullet} \,
   F^{(0)}{\!}_{TU\bullet} \nonumber\\
   &\qquad-90\,F^{(0)}{\!}_{\bullet PQ}\,
   \mathcal{Z}_0{}^{\bullet\bullet} \, F^{(0)}{\!}_{\bullet\bullet R}\,\mathcal{Z}_0{}^{\bullet\bullet} \,
   F^{(0)}{\!}_{\bullet\bullet S} \,\mathcal{Z}_0{}^{\bullet\bullet} \,
   F^{(0)}{\!}_{TU\bullet}  \nonumber\\
   &\qquad -15\,  F^{(0)}{\!}_{XYZ}\nonumber\\
   &\qquad\quad\times \big[\mathcal{Z}_0{}^{X\bullet} \,   F^{(0)}{\!}_{PQ\bullet}\big] 
   \, \big[\mathcal{Z}_0{}^{Y\bullet} \, F^{(0)}{\!}_{RS\bullet} \big]
   \, \big[\mathcal{Z}_0{}^{Z\bullet} \, F^{(0)}{\!}_{TU\bullet}\big] \Big]\,,
\end{align}
which can be obtained by repeated differentiation of the basic
equation \eqref{eq:em-duality} or \eqref{eq:new-F}.  Note that for
clarity we have occasionally replaced indices by bullets in cases
where the index contractions are unambiguous. In the above formulae
the bullets are indices that are simply contracted with the nearest
neighbour bullet as a string. Furthermore we have made use of the
following definitions,
\begin{align}
  \label{eq:def-S}
   \mathcal{S}_0{}^{I}{}_J(Y) =&\, \frac{\partial\tilde Y^I} {\partial
    Y^J}  = U^I{}_J +Z^{IK} F^{(0)}_{KJ}(Y) \,,\nonumber \\
  \mathcal{Z}_0^{IJ}(Y) =&\, [\mathcal{S}_0^{-1}]^I{}_K\, Z^{KJ} \,. 
\end{align}
Because the matrices $U$ and $Z$ are submatrices of a
$(2n+2)$-dimensional symplectic matrix, it follows that
$\mathcal{Z}_0^{IJ}$ is symmetric in $(I,J)$. It also follows that
\begin{align}
  \label{eq:deltaS-Z}
  \delta [\mathcal{S}_0^{-1}]^I{}_J=&\, - \mathcal{Z}_0{}^{IK}\,\delta
  F^{(0)}{\!}_{KL} \,[\mathcal{S}_0^{-1}]^L{}_J\,, \nonumber\\
  \delta \mathcal{Z}_0{}^{IJ} =&\, - \mathcal{Z}_0{}^{IK}\, \delta
  F^{(0)}{\!}_{KL} \, \mathcal{Z}_0{}^{LJ} \,. 
\end{align}
Defining 
\begin{equation}
  \label{eq:def-N}
  N^{(0)}{\!}_{IJ}(Y,\bar Y) =2\,\mathrm{Im}\,[F^{(0)}{\!}_{IJ}(Y)], 
\end{equation} 
we derive the following expression for the behaviour of its inverse
$N^{(0)IJ}$ under duality transformations,
\begin{equation}
  \label{eq:dual-trans-N}
  \tilde N^{(0) IJ} = \mathcal{S}_0{\!}^I{}_K\, 
  \bar{\mathcal{S}}_0{\!}^{J}{}_{L}   \,N^{(0) KL}\,. 
\end{equation}
Using the identity $[\mathcal{S}_0^{-1}]^I{}_K
\,[\bar{\mathcal{S}}_0]^{K}{}_{ J} =\delta^I{\!}_J
-\mathrm{i}\mathcal{Z}_0^{IK}\,N^{(0)}{\!}_{KJ}$, it follows that
\eqref{eq:dual-trans-N} can also be written as, 
\begin{equation}
  \label{eq:dual-trans-N-alt}
  \tilde N^{(0)IJ} =  \mathcal{S}_0{}^I{}_K\, 
  \mathcal{S}_0{}^{J}{}_{ L} \,\big[ N^{(0)KL} -\mathrm{i}
  \mathcal{Z}_0^{KL} \big]= \bar{\mathcal{S}}_0{}^I{}_K\, 
  \bar{\mathcal{S}}_0{}^{J}{}_{ L} \,\big[ N^{(0) KL} +\mathrm{i}
  \bar{\mathcal{Z}}_0^{KL} \big]  \,. 
\end{equation}
These identities will be relevant later on. Incidentally, from the
results presented above one can straightforwardly construct tensors that
transform covariantly under the symplectic transformations, such as 
\begin{equation}
  \label{eq:def-C}
  C_{IJKL} = F^{(0)}{\!}_{IJKL} + 3\mathrm{i}\, N^{(0)MN}\, F^{(0)}{\!}_{M(IJ}\,F^{(0)}{\!}_{KL)N}\,. 
\end{equation}
However, these tensors are not purely holomorphic in view of the
appearance of the matrix $N^{(0)IJ}$. We will encounter such `almost
holomorphic' covariant functions throughout this paper.

Note that in the next section we will introduce different complex
variables denoted by $\mathcal{Y}^I$. Since we will treat $\Omega$ as
a perturbation of the classical function $F^{(0)}$, the full function $F$
will no longer play a role in the various formulae. Therefore, in due
course, we will simply suppress all sub- and superscripts `$0$'
referring to the lowest-order quantities $F^{(0)}$,
$\mathcal{S}_0{}^{I}{}_J$, $\mathcal{Z}_0^{IJ}$, and $N^{(0)IJ}$. Note
also that we are only distinguishing holomorphic and anti-holomorphic
derivatives, $I,J,\ldots$ and $\bar I,\bar J,\ldots$ with a bar when
we are dealing with a real quantity. For instance,
$\bar{\mathcal{S}}_0{}^J{}_K$ is anti-holomorphic, so we will just
keep generic indices $J,K$ (rather than $\bar J,\bar K$), while
$\Omega_I$ and $\Omega_{\bar I}$ need a holomorphic or
anti-holomorphic index to distinguish between the derivative with
respect to $Y^I$ and $\bar Y^I$. The reason for this convention is
that we will often not have the situation where holomorphic and
anti-holomorphic indices are contracted consistently.

\section{The generic structure of the Hesse potential}
\label{sec:eval-hesse-potent}
\setcounter{equation}{0}
In this section we study the generic structure of the Hesse potential,
which requires to carry out a Legendre transform. In practice this can
only be done by iteration. The results will then take the form of an
infinite power series in terms of $\Omega$ and its derivatives. We
will explicitly evaluate the first terms in this expansion up to order
$\Omega^5$, which we expect to suffice for uncovering the general
structure of the full expression. The homogeneous and holomorphic
function $F^{(0)}(Y)$ will not be subject to further restrictions and
will thus encode a generic special-K\"ahler space, while $\Omega$ will
be an arbitrary function of $Y$ and $\bar Y$. The actual calculations
are rather laborious although in principle straightforward; we have
relegated some relevant material to several appendices. Similar
calculations have been performed for more specific cases some time ago
\cite{Cardoso:2008fr,Cardoso:2010gc}.  For instance, the first two
terms to quadratic order in $\Upsilon, \bar\Upsilon$ have been
determined for the FHSV model \cite{Ferrara:1995yx}. As it turned out
some of these terms were consistent with known results obtained from
the topological string by integration of the holomorphic anomaly
equation \cite{Grimm:2007tm}, but the Hesse potential contained
additional terms at this order that were separately S- and T-duality
invariant but did not have an interpretation in the topological string
context. In this paper we will investigate the generic structure of
the Hesse potential and clarify these partial results. We will return
to a discussion of the results for the FHSV model in section
\ref{sec:summary-conclusions}. 

To carry out the Legendre transform by iteration we first choose
convenient variables. Originally the Hesse potential was defined in
terms of the variables $(\phi^I,\chi_I)$ of real special geometry,
whose definition involves the full effective action. It is, however,
more convenient to convert them again to complex variables,
subsequently denoted by $\mathcal{Y}^I$, which coincide precisely with
the fields $Y^I$ that one would obtain from $(\phi^I,\chi_I)$ upon
using just the lowest-order holomorphic function $F^{(0)}$.  The
identification proceeds as follows \cite{Cardoso:2010gc},
\begin{align}
  \label{eq:new-Yfields-Y}
  2\,\mathrm{Re}\, Y^I~ =&\,~\phi^I~=~ 2\,\mathrm{Re}\,\mathcal{Y}^I
  \,,\nonumber\\ 
  2\,\mathrm{Re}\,F_I(Y,\bar Y) ~=&\,~ \chi_I ~=~
  2\,\mathrm{Re}\,F^{(0)}{\!}_I(\mathcal{Y}) \,. 
\end{align}
Since the relation between the variables $\mathcal{Y}^I$ and the real
variables $(\phi^I,\chi_I)$ involves only $F^{(0)}$, their duality
transformations will be directly related. Consequently we will refer to the
variables $\mathcal{Y}^I$ as {\it duality covariant} variables. Under
duality the variables $\mathcal{Y}^I$  transform according to,
\begin{equation}
  \label{eq:duality-new-Y}
  \tilde{\mathcal{Y}}^I= U^I{}_J\,\mathcal{Y}^J  + Z^{IJ}\,F^{(0)}{\!}_J(\mathcal{Y}) =
  \mathcal{S}^I{}_J(\mathcal{Y}) \; \mathcal{Y}^J\,, 
\end{equation}
where we used the homogeneity of $F^{(0)}(\mathcal{Y})$ as well as the
definition of $\mathcal{S}^I{}_J$ given in \eqref{eq:def-S},
except that the expression is now written in terms of the new variables
$\mathcal{Y}^I$. Furthermore we have dropped the subscript by the
replacement $\mathcal{S}_0\to \mathcal{S}$, as we had already
indicated at the end of section \ref{sec:deformed-special}.  

At the classical level, where $\Omega=0$, we obviously have
$\mathcal{Y}^I=Y^I$, but in higher orders the relation between these
moduli is complicated and will involve $\Omega$. Let us therefore
write $\mathcal{Y}^I=Y^I +\Delta Y^I$, where $\Delta Y^I$ is purely
imaginary, and $F= F^{(0)} + 2\mathrm{i}\,\Omega$, so that we can
express \eqref{eq:new-Yfields-Y} in terms of $F^{(0)}$, $\Omega$,
$\mathcal{Y}^I$ and $\Delta Y^I$.  Because the equations will no
longer involve $F$, we will henceforth drop the index `$0$' on
$F^{(0)}$, as mentioned earlier. Consequently all the derivatives of
$F$ will be holomorphic. The equations (\ref{eq:new-Yfields-Y}) can
then be written as,
\begin{align}
  \label{eq:Ycal-fields}
  &
  F_I(\mathcal{Y}-\Delta Y)+\bar F_I(\bar{\mathcal{Y}}+\Delta Y)-
  F_I(\mathcal{Y})-\bar  F_I(\bar{\mathcal{Y}}) = \nonumber \\
  &
  - 2\mathrm{i}\,
  \left[\Omega_I(\mathcal{Y}-\Delta Y,\bar{\mathcal{Y}}+\Delta Y )-
  \Omega_{\bar I}(\mathcal{Y}-\Delta Y,\bar{\mathcal{Y}}+\Delta Y)\right]
   \,,
\end{align}
where we made use of \eqref{eq:F-decomposition}. Upon Taylor
expanding, this equation will lead to an infinite power series in
$\Delta Y^I$, which we can solve by iteration. 
Retaining only the term of first order in $\Delta Y^I$
shows that it is proportional to first derivatives of
$\Omega$. Proceeding to higher orders will then lead to an expression for
$\Delta Y^I$ involving increasing powers of $\Omega$ and $F$ and their
derivatives taken at $Y^I=\mathcal{Y}^I$. Up to fourth order in
$\Omega$ this iteration gives the following expression for $\Delta
Y^I$, 
\begin{align}
  \label{eq:Delta-3-order}
  \Delta Y^I\approx &\, 2\,(\Omega^I-\Omega^{\bar I})\nonumber\\
   &\, -2\mathrm{i} (F+\bar F)^{IJK} 
   (\Omega_J-\Omega_{\bar J}) (\Omega_K-\Omega_{\bar K}) 
   - 8\,\mathrm{Re}(\Omega^{IJ}-\Omega^{I\bar J})\,(\Omega_J-\Omega_{\bar J})
   \nonumber\\
   &\, 
   +\tfrac4{3}\mathrm{i} \left[(F-\bar F)^{IJKL} +3\mathrm{i} (F+\bar
       F)^{IJM} (F+\bar F)_M{}^{KL}\right]\nonumber\\
     &\qquad \qquad
      \times(\Omega_J-\Omega_{\bar  J}) (\Omega_K-\Omega_{\bar K})
     (\Omega_L-\Omega_{\bar L}) \nonumber    \\ 
   &\, 
   + 8\mathrm{i} \left[ 2\,(F+\bar F)^{IJ}{}_K \mathrm{Re}
   (\Omega^{KL}-\Omega^{K\bar L})+ \mathrm{Re}
   (\Omega^{IK}-\Omega^{I\bar K})  (F+\bar F)_{K}{}^{JL} \right]\nonumber\\
   &\qquad \qquad 
    \times (\Omega_J-\Omega_{\bar J})(\Omega_L-\Omega_{\bar L})
    \nonumber  \\
   &\, 
   + 32\,\mathrm{Re}
   (\Omega^{IJ}-\Omega^{I\bar J})\, \mathrm{Re}
   (\Omega_{JK}-\Omega_{J\bar K})\, (\Omega^K-\Omega^{\bar K})  \nonumber\\
   &\, 
   + 8\mathrm{i}\,\mathrm{Im}(\Omega^{IJK} -2\,\Omega^{IJ\bar K} +
   \Omega^{I\bar J\bar K} ) (\Omega_{J}-\Omega_{\bar J})\,
   (\Omega_K-\Omega_{\bar K})  +\mathcal{O}(\Omega^4)\;.
\end{align}
Here indices have been raised by making use of $N^{(0)IJ}$, which was
already defined in \eqref{eq:def-N}, where, for consistency, we will henceforth
change notation and refer to $N^{(0)IJ}$ by $N^{IJ}$. Here we stress
once more that all the derivatives of $F$ and $\Omega$ are taken at
$Y^I=\mathcal{Y}^I$ and $\bar Y^I=\bar{\mathcal{Y}}^I$.

Using the same notation we obtain the following expression for the
Hesse potential \eqref{eq:GenHesseP},
\begin{align}
  \label{eq:Hesse-Y-calY}
  \mathcal{H}(\mathcal{Y},\bar{\mathcal{Y}}) =&\,- \mathrm{i}\big[
  \bar{\mathcal{Y}}^I F_{I} (\mathcal{Y}) -\mathcal{Y}^I\bar F_{I}
  (\bar{\mathcal{Y}}) \big]
  +4\,\Omega(\mathcal{Y},\bar{\mathcal{Y}}) \nonumber \\
  &\, -\mathrm{i}\big[ \mathcal{Y}^I \big( F_I(Y)- F_I(\mathcal{Y})\big) +
  \Delta
  Y^I F_{I} (Y)  -\mbox{h.c.}\big]  \nonumber\\
  &\, + 4\big[\Omega(Y,\bar Y)-\Omega(\mathcal{Y},\bar{\mathcal{Y}}) +
  \Delta Y^I \big(\Omega_I(Y,\bar Y) -\Omega_{\bar I}(Y,\bar
    Y)\big) \big]   \,. 
\end{align}
Here we made use of the homogeneity of $F(Y)$ and of
(\ref{eq:Ycal-fields}). Again this result must be Taylor expanded upon
writing $Y^I=\mathcal{Y}^I-\Delta Y^I$ and $\bar
Y^I=\bar{\mathcal{Y}}^I+\Delta Y^I$. The last two lines of
(\ref{eq:Hesse-Y-calY}) then lead to a power series in $\Delta Y$,
starting at second order in the $\Delta Y$,
\begin{align}
  \label{eq:Hesse-DY-calY}
  \mathcal{H}(\mathcal{Y},\bar{\mathcal{Y}}) \approx &\,- \mathrm{i}[
  \bar{\mathcal{Y}}^I F_{I} (\mathcal{Y}) -\mathcal{Y}^I\bar F_{I}
  (\bar{\mathcal{Y}}) ]
  +4\,\Omega(\mathcal{Y},\bar{\mathcal{Y}}) \nonumber \\
  &\, -N_{IJ}\Delta Y^I\Delta Y^J -\tfrac23\mathrm{i} (F+\bar F)_{IJK}
  \Delta Y^I\Delta Y^J\Delta Y^K \nonumber \\
  &\,  - 4\,\mathrm{Re}(\Omega_{IJ}-\Omega_{I\bar J}) \Delta Y^I\Delta
  Y^J + \tfrac14\mathrm{i} (F-\bar F)_{IJKL} 
  \Delta Y^I\Delta Y^J\Delta Y^K \Delta Y^L  \nonumber\\
  &\, +\tfrac83\mathrm{i}\, \mathrm{Im}(\Omega_{IJK}-3\Omega_{IJ\bar
    K}) \Delta Y^I\Delta Y^J \Delta Y^K  +\cdots\;. 
\end{align}
Inserting the result of the iteration (\ref{eq:Delta-3-order}) into
the expression above leads to the following expression for the Hesse
potential, up to terms of order $\Omega^5$,
\begin{align}
  \label{eq:Hesse-Y-calY-exp}
  \mathcal{H}(\mathcal{Y},\bar{\mathcal{Y}}) \approx &\,- \mathrm{i}[
  \bar{\mathcal{Y}}^I F_{I} (\mathcal{Y}) -\mathcal{Y}^I\bar F_{I}
  (\bar{\mathcal{Y}}) ]
  +4\,\Omega(\mathcal{Y},\bar{\mathcal{Y}}) \nonumber \\
  &\, -4\, \hat N^{IJ}  z_I\,z_J +\tfrac83\mathrm{i}
  (F+\bar F)_{IJK} \hat N^{IL}\hat N^{JM}\hat N^{KN}
  z_L\,z_M\,z_N \nonumber \\
  &\, - \tfrac43\mathrm{i} [(F-\bar F)_{IJKL}+ 3\mathrm{i}(F+\bar
  F)_{IJR}\hat N^{RS}(F+\bar F)_{SKL} ] \nonumber\\
  &\qquad \times \hat N^{IM}\hat
  N^{JN}\hat N^{KP} \hat N^{LQ} 
  z_M\,z_N\,z_P\,z_Q   \nonumber\\
  &\, -\tfrac{32}3\mathrm{i}\, \mathrm{Im}(\Omega_{IJK}-3\,\Omega_{IJ\bar
    K}) \hat N^{IL}\hat N^{JM}\hat N^{KN}\,
  z_L\,z_M\,z_N   +\mathcal{O}(\Omega^5)\;,
\end{align}
where $z_I=\Omega_I-\Omega_{\bar I}$, and where $\hat N^{IJ}$ is
the inverse of the real, symmetric matrix $\hat N_{IJ}$, defined by
\begin{equation}
  \label{eq:hat-N}
  \hat N_{IJ}= 
  N_{IJ} +4\,\mathrm{Re}(\Omega_{IJ}-\Omega_{I\bar J}) \,.
\end{equation}
Upon expanding $\hat N^{IJ}$ we straightforwardly determine the
contributions to the Hesse potential up to fifth order in $\Omega$,
\begin{align}
  \label{eq:Hesse-1-4}
  \mathcal{H}=&\, \mathcal{H}\vert_{\Omega=0} + 4\,\Omega -
  4\,N^{IJ}(\Omega_I\Omega_J +\Omega_{\bar I}\Omega_{\bar J}) +
  8\,N^{IJ} \Omega_I\Omega_{\bar J} \nonumber\\
  &\, 
  +16\,\mathrm{Re}(\Omega_{IJ}-\Omega_{I\bar J}) N^{IK}N^{JL}
    \big(\Omega_K\Omega_L +\Omega_{\bar K}\Omega_{\bar
      L}-2\,\Omega_K\Omega_{\bar L} \big) \nonumber\\
    &\, 
    - \tfrac{16}3
  (F+\bar F)_{IJK}  N^{IL} N^{JM} N^{KN} \,\mathrm{Im}
  (\Omega_L\Omega_M\Omega_N - 3\,\Omega_L\Omega_M\Omega_{\bar N})
  \nonumber \\ 
  &\ - 64 N^{IP} {\rm Re} \left(\Omega_{PQ} - \Omega_{P \bar Q} \right) N^{QR}
  {\rm Re} \left(\Omega_{RK} - \Omega_{R \bar K} \right) N^{KJ}
  \left(\Omega_I \Omega_{J} + \Omega_{\bar I} \Omega_{\bar J} - 2
    \Omega_I \Omega_{\bar J} \right) 
  \nonumber\\ 
  &\ + 64 (F+\bar F)_{IJK}  N^{IL} N^{JM} N^{KP} \,\mathrm{Re}
  \left(\Omega_{PQ} - \Omega_{P \bar Q} 
  \right) N^{QN} \,\mathrm{Im}
  (\Omega_L\Omega_M\Omega_N - 3\,\Omega_L\Omega_M\Omega_{\bar N}) 
   \nonumber\\
  &\,   - \tfrac83\mathrm{i} [(F-\bar F)_{IJKL}+ 3\mathrm{i} (F+\bar
  F)_{R(IJ} N^{RS}(F+\bar F)_{KL)S} ]
   N^{IM}
  N^{JN} N^{KP} N^{LQ} \nonumber\\
 &\, \qquad \qquad \qquad \qquad \qquad 
 \times \mathrm{Re} \left(\Omega_M \Omega_N \Omega_P \Omega_Q - 4
   \Omega_M \Omega_N \Omega_P \Omega_{\bar Q}+ 
  3 \Omega_M \Omega_N \Omega_{\bar P} \Omega_{\bar Q} \right) \nonumber\\
   &\, +\tfrac{64}3 \, \mathrm{Im}(\Omega_{IJK}-3\Omega_{IJ\bar
    K})  N^{IL} N^{JM}  N^{KN}
  \,\mathrm{Im}
  (\Omega_L\Omega_M\Omega_N - 3\,\Omega_L\Omega_M\Omega_{\bar N}) 
     +\mathcal{O}(\Omega^5)\;.
\end{align}
We stress once more that all the quantities in \eqref{eq:Hesse-1-4}
are taken at $Y^I=\mathcal{Y}^I$.

It is clear that the Hesse potential takes a complicated form, but we
note two systematic features. First of all it turns out that the
expression \eqref{eq:Hesse-1-4} can be understood diagrammatically. To
appreciate this, let us return to the definition \eqref{eq:GenHesseP}
of the Hesse potential and rewrite it in a different form,
\begin{align}
  \label{eq:GenHesseP2}
  \mathcal{H}(\phi,\chi) =&\, 4 \,\big[ {\rm Im}\,F(Y) +\Omega(Y,\bar
  Y) \big] + \mathrm{i} \chi_I \,(Y^I-\bar
  Y^I)  \nonumber\\
  =&\, -2\mathrm{i} \, F(\mathcal{Y}-\Delta Y) +2\mathrm{i} \, \bar
  F(\bar{\mathcal{Y}}+\Delta Y) +4\,\Omega(\mathcal{Y}-\Delta
  Y,\bar{\mathcal{Y}}+\Delta Y) \nonumber\\
  &\, +\mathrm{i} \chi_I
  \,\big(\mathcal{Y}^I-\bar{\mathcal{Y}}^I\big) -2\mathrm{i}\, \chi_I
  \, \Delta Y^I\,,
\end{align}
where the purely imaginary quantities $\Delta Y^I$ were introduced in
the beginning of section \ref{sec:eval-hesse-potent} when defining
the iteration procedure. We remind the reader that $\phi^I=2\,
\mathrm{Re}\,Y^I= 2\,\mathrm{Re}\,\mathcal{Y}^I$ and $\chi_I=
2\,\mathrm{Re}\, F_I(\mathcal{Y})$. Substituting these results we
derive
\begin{align}
  \label{eq:GenHesseP3}
  \mathcal{H}(\mathcal{Y},\bar{\mathcal{Y}}) =&\,
  \mathcal{H}(\mathcal{Y},\bar{\mathcal{Y}})\big\vert_{\Omega=0} 
  +N_{IJ}(\mathcal{Y},\bar{\mathcal{Y}}) \,\Delta Y^I\Delta Y^J
 \nonumber\\
  &\, +2\mathrm{i} \,\sum_{n=3}^\infty \,\frac1{n!} \big[ (-)^{n+1}\,F_{I_1\cdots
    I_n}(\mathcal{Y}) +\bar F_{I_1\cdots I_n}(\bar{\mathcal{Y}}) \big]
  \,\Delta Y_1 \cdots  \Delta Y_n 
  \nonumber\\
  &\, +4\,\Omega(\mathcal{Y}-\Delta Y,\bar{\mathcal{Y}}+\Delta Y)   \,. 
\end{align}
It thus follows that $\mathcal{H}(\mathcal{Y},\bar{\mathcal{Y}})-
\mathcal{H}(\mathcal{Y},\bar{\mathcal{Y}})\vert_{\Omega=0} -4\,
\Omega(\mathcal{Y},\bar{\mathcal{Y}})$ can be written as a series
expansion in positive powers of $\Delta Y^I$. Integrating the
exponential of this expression over the (purely imaginary)
fluctuations $\Delta Y^I$, the result can be expressed as an infinite
sum over Feynman diagrams in the standard way with propagators given
by $N^{IJ}$ and vertices by the derivatives of $F$ and $\bar F$ as
well as of the holomorphic and anti-holomorphic derivatives of
$\Omega$. Because the Legendre transformation performed above should
correspond to the tree diagrams, it follows that
$\mathcal{H}(\mathcal{Y},\bar{\mathcal{Y}})-
\mathcal{H}(\mathcal{Y},\bar{\mathcal{Y}})\vert_{\Omega=0} -4\,
\Omega(\mathcal{Y},\bar{\mathcal{Y}})$ will comprise all the {\it
  connected} tree diagrams. In this way one can account for all the
terms in \eqref{eq:Hesse-1-4}, including their combinatorial
factors. Note that the diagrams are not 1PI: by removing the
propagator $N^{IJ}$ each term in the expression will factorize into
two terms! Here we should mention that diagrammatic techniques have
repeatedly played a role in the analysis of the holomorphic anomaly
equation and of the topological string (see e.g.
\cite{Bershadsky:1993cx,Aganagic:2006wq,Grimm:2007tm}).

Another feature is based on the fact that \eqref{eq:Hesse-1-4}
transforms as a function under duality transformations of the fields
$\mathcal{Y}^I$, and so does the first term
$\mathcal{H}\vert_{\Omega=0}$. This observation enables one to
determine how $\Omega$ will transform under dualities. Obviously the
transformation behaviour of $\Omega$ must be non-trivial in view of
the non-linear dependence on $\Omega$ of the Hesse potential. The
evaluation of this transformation proceeds again by iteration.

To demonstrate the procedure, let us review the first few steps. In
lowest order, $\Omega$ must transform as a function, which implies
that
\begin{align}
  \label{eq:transf-Omega1}
  \tilde\Omega(\tilde{\mathcal{Y}}, \bar{\tilde{{\mathcal{Y}}}})
  =&\, \Omega(\mathcal{Y},
  \bar{{\mathcal{Y}}}) +\mathcal{O}(\Omega^2)   \,,\nonumber\\ 
  \tilde\Omega_I(\tilde{\mathcal{Y}}, \bar{\tilde{{\mathcal{Y}}}})
  =&\, [\mathcal{S}^{-1}]^J{}_I(\mathcal{Y}) \,\Omega_J(\mathcal{Y},
  \bar{{\mathcal{Y}}}) +\mathcal{O}(\Omega^2) \,,
\end{align}
where the matrix $\mathcal{S}$ was already defined previously (note that
we have suppressed the subscript `$0$'). Applying this result to the
first few terms of \eqref{eq:Hesse-1-4} and making use of the fact
that $\mathcal{H} -\mathcal{H}\vert_{\Omega=0}$ transforms as a
function, one deduces the next-order result,
\begin{align}
  \label{eq:transf-Omega2}
  &\tilde \Omega -
   \tilde N^{IJ}(\tilde\Omega_I\tilde\Omega_J +\tilde\Omega_{\bar I}\tilde\Omega_{\bar J}) +
  2\,\tilde N^{IJ} \tilde\Omega_I\tilde \Omega_{\bar
    J}+\mathcal{O}(\tilde\Omega^3)  \nonumber\\
  &\quad =\Omega -
  N^{IJ}(\Omega_I\Omega_J +\Omega_{\bar I}\Omega_{\bar J}) +
  2\,N^{IJ} \Omega_I\Omega_{\bar J} +\mathcal{O}(\Omega^3)\,, 
\end{align}
where on the left-hand side the functions depend on the transformed
fields $\tilde{\mathcal{Y}}^I$, while on the right-hand side they
depend on the original fields $\mathcal{Y}^I$.  

Using the exact relations
\eqref{eq:def-S}-\eqref{eq:dual-trans-N-alt}, suppressing also the
subscript `$0$' in the symmetric matrix $\mathcal{Z}_0^{IJ}$, one
discovers that the first equation in \eqref{eq:transf-Omega1} receives
the following correction in second order in $\Omega$,
\begin{equation}
  \label{eq:second-order variation-Omega}
  \tilde \Omega(\tilde{\mathcal{Y}},\bar{\tilde{\mathcal{Y}}})  =
  \Omega - \mathrm{i}\big( \mathcal{Z}^{IJ} 
  \,\Omega_I \Omega_J - \bar{\mathcal{Z}}^{IJ}
  \,\Omega_{\bar I} \Omega_{\bar J} \big) +\mathcal{O}(\Omega^3)
  \,,  
\end{equation}
which in turn gives rise to the following result for derivatives
of $\Omega$,
\begin{align}
  \label{eq:transf-der-Omega2}
  \tilde\Omega_I(\tilde{\mathcal{Y}},\bar{\tilde{\mathcal{Y}}})=&\,
  [\mathcal{S}^{-1}]^J{}_I \Big[ \Omega_J 
  +\mathrm{i} F_{JKL} \,\mathcal{Z}^{KM}\Omega_M\,\mathcal{Z}^{LN}
  \Omega_N -2\mathrm{i} \Omega_{JK}\mathcal{Z}^{KL} \Omega_L
  +2\mathrm{i} \Omega_{J\bar K}\bar{\mathcal{Z}}^{KL}
  \Omega_{\bar L} \Big]\nonumber\\
  &\,
  +\mathcal{O}(\Omega^3)\,,\nonumber \\
  \tilde\Omega_{IJ}(\tilde{\mathcal{Y}},\bar{\tilde{\mathcal{Y}}})
  =&\, [\mathcal{S}^{-1}]^K{}_I 
  [\mathcal{S}^{-1}]^L{}_J  \Big[ \Omega_{KL} -F_{KLM}
  \,\mathcal{Z}^{MN} \Omega_N  \Big] +\mathcal{O}(\Omega^2)
  \,,\nonumber\\ 
  \tilde\Omega_{I\bar
    J}(\tilde{\mathcal{Y}},\bar{\tilde{\mathcal{Y}}}) =&\, [\mathcal{S}^{-1}]^K{}_I
  [\bar{\mathcal{S}}^{-1}]^{\bar L}{}_{\bar J} \,\Omega_{K\bar
    L}  +\mathcal{O}(\Omega^2) \,.
\end{align}
Here and henceforth we make frequent use of \eqref{eq:deltaS-Z}. 

The iteration can be continued by including the terms of order
$\Omega^3$, making use of \eqref{eq:transf-der-Omega2} for derivatives
of $\Omega$, to obtain the expression for $\tilde\Omega$ up to terms
of order $\Omega^4$. In the next iterative step one then derives the
effect of a duality transformation on $\Omega$ up to terms of order
$\Omega^5$. Before presenting this result, we wish to observe that
terms transforming as a proper function under duality, will not
contribute to this result. This is precisely what happens to the term
proportional to $N^{IJ} \Omega_I\Omega_{\bar J}$ that appears in
\eqref{eq:Hesse-1-4}, which transforms as a function under symplectic
transformations in this order of the iteration. Consequently this term
does not contribute to \eqref{eq:second-order variation-Omega}. As it
turns out an infinite set of contributions to the Hesse potential will
be generated that transform separately as functions under duality. By
separating those from \eqref{eq:Hesse-1-4}, we do not change the
transformation behaviour of $\Omega$ but we can extract certain
functions from the Hesse potential in order to simplify its
structure. However, these functions must be constructed also by
iteration, order by order in $\Omega$.

We have evaluated this decomposition in terms of separate functions in
detail, which leads to 
\begin{align}
  \label{eq:Hesse-decomp}
  \mathcal{H} =&\, \mathcal{H}^{(0)} + \mathcal{H}^{(1)} +
  \mathcal{H}^{(2)} + \big(\mathcal{H}^{(3)}_1 + {\mathcal{H}}^{(3)}_2
  + \mathrm{h.c.}\big) + \mathcal{H}^{(3)}_3 +
  \mathcal{H}^{(4)}_1  +\mathcal{H}^{(4)}_2 +\mathcal{H}^{(4)}_3 \nonumber\\
  &\, +\big(\mathcal{H}^{(4)}_4+ \mathcal{H}^{(4)}_5 +
  \mathcal{H}^{(4)}_6 + \mathcal{H}^{(4)}_7 + \mathcal{H}^{(4)}_8 +
  \mathcal{H}^{(4)}_9 +\mathrm{h.c.}\big) \ldots\,,
\end{align}
where the $\mathcal{H}^{(a)}_i$ are certain expressions to be defined
below, whose leading term is of order $\Omega^a$. For higher values of
$a$ it turns out that there exists more than one function with the
same value of $a$, and those will be labeled by $i=1,2,\ldots$. Of all
the combinations $\mathcal{H}^{(a)}_i$ appearing in
\eqref{eq:Hesse-decomp}, $\mathcal{H}^{(1)}$ is the only one that
contains $\Omega$, while all the other combinations contain
derivatives of $\Omega$. Obviously, $\mathcal{H}^{(0)}$ equals,
\begin{equation}
  \label{eq:Hesse-0}
  \mathcal{H}^{(0)} = - \mathrm{i}[
  \bar{\mathcal{Y}}^I F_{I} (\mathcal{Y}) -\mathcal{Y}^I\bar F_{I}
  (\bar{\mathcal{Y}}) ]\,,
\end{equation}
whereas $\mathcal{H}^{(1)}$ at this level of iteration is given by,
\begin{align}
  \label{eq:Hesse-1}
  \mathcal{H}^{(1)}=&\, 4\,\Omega - 4\,N^{IJ}(\Omega_I\Omega_J
    +\Omega_{\bar I}\Omega_{\bar J})
    \nonumber\\
    &\, +16\,\mathrm{Re} \Big[ \Omega_{IJ} (N \Omega)^I (N \Omega)^J
    \big]  + 16\, \Omega_{I \bar J} \, (N \Omega)^I ( N \bar \Omega)^J
    \nonumber\\ 
    &\, - \tfrac{16}3 \,\mathrm{Im} \left[ F_{IJK} (N \Omega)^I (N
      \Omega)^J (N \Omega)^K
    \right] \nonumber\\
    &\, - \tfrac43 \mathrm{i} \Big[ \left(F_{IJKL} + 3 \mathrm{i}
      F_{R(IJ} N^{RS} F_{KL)S} \right) (N \Omega)^I (N \Omega)^J (N
    \Omega)^K (N \Omega)^L
     - {\rm h.c.} \Big] \nonumber\\
     &\, 
    - \tfrac{16}3 \left[ \Omega_{IJK} (N \Omega)^I (N \Omega)^J (N
      \Omega)^K  + {\rm h.c.}  \right] \nonumber\\
      &\, 
    - 16\left[ \Omega_{IJ\bar K} (N \Omega)^I (N \Omega)^J (N \bar \Omega)^K
      + {\rm h.c.}     \right] \nonumber\\
      &\ - 16 \mathrm{i} \,\Big[ F_{IJK} N^{KP} \,\Omega_{PQ}  (N
      \Omega)^I (N \Omega)^J (N \Omega)^Q 
      - {\rm h.c.} \Big]      \nonumber\\
     & \, - 16 \Big[ (N \Omega)^P  \, \Omega_{PQ} \, N^{QR}
     \Omega_{RK} \, (N \Omega)^K  
           + {\rm h.c.} \Big] \nonumber\\
       & \, - 16 \Big[ (N \Omega)^P \, \Omega_{PQ} \, N^{QR} 
         \Omega_{R \bar K} \, (N \bar \Omega)^K  \nonumber\\
      & \, \qquad \qquad +  N^{IP} \, \Omega_{P \bar Q} \, N^{QR} \left( 
      \Omega_{\bar R K} \, (N \Omega)^K 
      +  \Omega_{\bar R \bar K} \, (N \bar \Omega)^K \right) \Omega_I
       + {\rm h.c.} \Big]\nonumber\\
        &\ - 16 \mathrm{i} \,\Big[ F_{IJK} N^{KP} \,\Omega_{P\bar Q}  
             (N \Omega)^I ( N \Omega)^J (N \bar \Omega)^Q
            - {\rm h.c.} \Big]  + \mathcal{O}(\Omega^5)\,. 
\end{align}
Here we have used the notation $(N \Omega)^I = N^{IJ} \Omega_J$, $(N
\bar\Omega)^I = N^{IJ} \Omega_{\bar J}$.  Index symmetrizations, such
as in $F_{R(IJ} N^{RS} F_{KL)S}$, are always of strength one. For
instance, in this example, where there are three independent
combinations, one includes a factor $1/3$. The expressions for the
higher-order functions $\mathcal{H}^{(a)}_i$ with $a=2,3,4$ are given
in appendix \ref{App:funct-H-a-i-geq2}. These expressions have been
obtained by requiring that they constitute functions under symplectic
transformations, order by order in $\Omega$. There exist other
expressions that do transform as functions under duality in this
approximation, but which do not appear in $\mathcal{H}$. We have
included two examples of such functions in appendix
\ref{App:funct-H-a-i-geq2}, one of which will be relevant later on.

Because $\mathcal{H}^{(1)}$ transforms as a function under symplectic
transformations, we can deduce the transformation behavior of $\Omega$
up to order $\Omega^5$ by generalizing \eqref{eq:transf-Omega1} to
higher orders. In this way one derives,
 \begin{align}
  \label{eq:Omega-tilde4}
  \tilde \Omega(\tilde{\mathcal{Y}}, \bar{\tilde{{\mathcal{Y}}}}) =&\,
  \Omega - \mathrm{i}\big( \mathcal{Z}^{IJ} 
  \,\Omega_I \Omega_J - \bar{\mathcal{Z}}^{IJ}
  \,\Omega_{\bar I} \Omega_{\bar J} \big)\nonumber \\
  &\, +\tfrac23\big( F_{IJK} \,\mathcal{Z}^{IL}\Omega_L
  \,\mathcal{Z}^{JM}\Omega_M \,\mathcal{Z}^{KN}\Omega_N +
  \mathrm{h.c.}\big) 
  \nonumber\\
  &\, -2 \big(\Omega_{IJ} \, \mathcal{Z}^{IK}\Omega_K
  \mathcal{Z}^{JL}\Omega_L  +\mathrm{h.c.}\big)
  +4\,\Omega_{I\bar J} \,
  \mathcal{Z}^{IK}\Omega_K\,  \bar{\mathcal{Z}}^{JL}\Omega_{\bar L}  \nonumber\\
  &\,+ \Big[ - \tfrac{1}{3}\mathrm{i} F_{IJKL} (\mathcal{Z} \Omega)^I
  (\mathcal{Z} \Omega)^J (\mathcal{Z} \Omega)^K
  (\mathcal{Z} \Omega)^L \nonumber\\
  &\, \qquad + \tfrac{4}{3}\mathrm{i} \Omega_{IJK} (\mathcal{Z}
  \Omega)^I (\mathcal{Z} \Omega)^J (\mathcal{Z}
  \Omega)^K \nonumber\\
  & \, \qquad + \mathrm{i}\, F_{IJR} \, {\cal Z}^{RS} \, F_{SKL} \,
  ({\cal Z} \Omega)^I ({\cal Z} \Omega)^J
  ({\cal Z} \Omega)^K ({\cal Z} \Omega)^L  \nonumber\\
  & \, \qquad - 4 \mathrm{i} \,\Omega_{IJ \bar K} \, (\mathcal{Z}
  \Omega)^I \, (\mathcal{Z} \Omega)^J
  (\bar{\mathcal{Z}} \bar \Omega)^{\bar K} \nonumber\\
  & \qquad - 4 \mathrm{i} \, F_{IJK} {\cal Z}^{KP} \,\Omega_{PQ}
  \,({\cal Z} \Omega)^I ({\cal Z} \Omega)^J \
  ({\cal Z} \Omega)^Q \nonumber\\
  &\qquad + 4 \mathrm{i} \, F_{IJK} \mathcal{Z}^{KP} \,\Omega_{P\bar
    Q} (\mathcal{Z} \Omega)^I ( \mathcal{Z} \Omega)^J
  (\bar{\mathcal{Z}}\bar \Omega)^{\bar Q}
  \nonumber\\
  & \, \qquad + 4 \mathrm{i} \, (\mathcal{Z} \Omega)^P \, \Omega_{PQ}
  \, \mathcal{Z}^{QR} \left(\Omega_{RK} \, (\mathcal{Z}\Omega)^K
    -   2 \Omega_{R \bar K} \, (\bar{\mathcal{Z}} \bar \Omega)^{\bar
      K} \right)  \nonumber\\ 
  & \, \qquad - 4 \mathrm{i} \, (\mathcal{Z} \Omega)^P \, \Omega_{P
    \bar Q} \, \bar{\mathcal{Z}}^{\bar Q \bar R}
      \Omega_{\bar R K} \, (\mathcal{Z} \Omega)^K 
       + {\rm h.c.} \Big]  + \mathcal{O}(\Omega^5)\,,
\end{align}
where the functions on the right-hand side depend on the fields
$\mathcal{Y}^I$ and $\bar{\mathcal{Y}}^I$. We used the obvious
notation $\big(\mathcal{Z}\Omega\big)^I= \mathcal{Z}^{IJ}\Omega_J$ and
$\big(\bar{\mathcal{Z}}\bar\Omega\big)^I= \bar{\mathcal{Z}}^{\bar
  I\bar J}\bar\Omega_{\bar J}$. It is remarkable that the matrix
$N^{IJ}$ no longer appears in this relation. This is due to a subtle
interplay of the various contributions to this result, which involves
the ones coming from \eqref{eq:transf-der-Omega2}. On closer
inspection, the result \eqref{eq:Omega-tilde4} turns out to be
identical (modulo an overall factor 4) to $\mathcal{H}^{(1)}$ given in
\eqref{eq:Hesse-1} upon making the replacement $N^{IJ}\to \mathrm{i}
\mathcal{Z}^{IJ}$ and/or $N^{IJ}\to -\mathrm{i}
\bar{\mathcal{Z}}^{\bar I\bar J}$, where the precise form depends
on the type of index contractions (i.e holomorphic or
anti-holomorphic) to $N^{IJ}$. The fact that none of the other
functions $\mathcal{H}^{(a)}$ contribute is perhaps not surprising
because those are manifestly symplectic functions where no
$\mathcal{Z}$ dependent variations are generated, whereas in
$\mathcal{H}^{(1)}$ such terms are generated and have to be absorbed
into the transformation rule of $\Omega$. Clearly this is an
intriguing result for which we have not found a general proof 
within our approach, although we are aware
of the fact that similar properties have been encountered in
\cite{Aganagic:2006wq}.

With the above result \eqref{eq:Omega-tilde4} one can continue with
the iterations by extending \eqref{eq:transf-der-Omega2} to the next
order, finding
\begin{align}
  \label{eq:transf-Omega3}
  \tilde\Omega_I(\tilde{\mathcal{Y}},\bar{\tilde{\mathcal{Y}}}) =&\,
  [\mathcal{S}^{-1}]^J{}_I \Big[ \Omega_J +\mathrm{i} F_{JKL}
  \,\left( \mathcal{Z} \Omega\right)^K\,(\mathcal{Z} \Omega)^L
  -2\mathrm{i}\, \Omega_{JK} (\mathcal{Z} \Omega)^K +2\mathrm{i}
  \Omega_{J\bar K} (\bar{\mathcal{Z}}
  \bar{\Omega})^{\bar K} \nonumber\\
  &\,+ \tfrac23 F_{JKLP} (\mathcal{Z} \Omega)^K (\mathcal{Z}
  \Omega)^L (\mathcal{Z} \Omega)^P + 2\, F_{KLP} (\mathcal{Z}
  \Omega)^K{}_J (\mathcal{Z} \Omega)^L
  (\mathcal{Z} \Omega)^P \nonumber\\
  &\, + 4\, F_{JKL} (\mathcal{Z} \Omega)^K (\mathcal{Z}
  \Omega)^L{}_{P} (\mathcal{Z} {\Omega})^{P} - 4\, F_{JKL}
  (\mathcal{Z} \Omega)^K (\mathcal{Z} \Omega)^L{}_{\bar P}
  (\mathcal{\bar Z}
  {\bar \Omega})^{\bar P} \nonumber\\
  & \, - 2 \,F_{JKL} \mathcal{Z}^{LP} F_{PQS} (\mathcal{Z} \Omega)^K
  (\mathcal{Z} \Omega)^Q (\mathcal{Z} \Omega)^S + 2\, {\bar F}_{\bar
    K \bar L \bar P} (\bar{\mathcal{Z}} \bar \Omega)^{\bar K}{}_J
  (\bar{\mathcal{Z}}  \bar \Omega)^{\bar L} (\bar{\mathcal{Z}} \bar
  \Omega)^{\bar P} \nonumber\\ 
  &\, - 2 \,\Omega_{JKL} (\mathcal{Z} \Omega)^K (\mathcal{Z}
  \Omega)^L- 4 \,\Omega_{KL} (\mathcal{Z} \Omega)^K{}_J (\mathcal{Z}
  \Omega)^L - 2\, \Omega_{J \bar K \bar L}
  (\bar{\mathcal{Z}} \bar \Omega)^{\bar K} (\bar{\mathcal{Z}} \bar
  \Omega)^{\bar L} \nonumber\\ 
  &\, - 4 \,\Omega_{\bar K \bar L} (\bar{\mathcal{Z}} \bar
  \Omega)^{\bar K}{}_J (\bar{\mathcal{Z}} \bar \Omega)^{\bar L} + 4\,
  \Omega_{JK \bar L} (\mathcal{Z} \Omega)^K (\bar{\mathcal{Z} }
  \bar{\Omega})^{\bar L} + 4\, \Omega_{K \bar L} (\mathcal{Z}
  \Omega)^K{}_J
  (\bar{\mathcal{Z} } \bar{\Omega})^{\bar L} \nonumber\\
  & \, + 4\, \Omega_{K \bar L} (\mathcal{Z} \Omega)^K
  (\bar{\mathcal{Z} } \bar{\Omega})^{\bar L}{}_J \Big]
  +\mathcal{O}(\Omega^4)\,,\nonumber \\
  \tilde\Omega_{IJ}=&\, [\mathcal{S}^{-1}]^K{}_I
  [\mathcal{S}^{-1}]^L{}_J  \Big[ \Omega_{KL} -F_{KLM}
  \,\mathcal{Z}^{MN} \Omega_N   + \mathrm{i} F_{KLMN} (\mathcal{Z}
  \Omega)^M (\mathcal{Z} \Omega)^N \nonumber\\ 
  & \, + 2 \mathrm{i}\, F_{KMN}  (\mathcal{Z} \Omega)^M{}_L
  (\mathcal{Z} \Omega)^N 
   - 2 \mathrm{i}\, F_{KMN} \mathcal{Z}^{MP} F_{PQL}  (\mathcal{Z}
   \Omega)^Q  (\mathcal{Z} \Omega)^N \nonumber\\  
   & - 2 \mathrm{i}\, \Omega_{KLP} (\mathcal{Z} \Omega)^P -  2
   \mathrm{i}\, \Omega_{KP} (\mathcal{Z} \Omega)^P{}_L 
   + 2 \mathrm{i}\, \Omega_{KP} \mathcal{Z}^{PQ} F_{QLS}
   (\mathcal{Z} \Omega)^S \nonumber\\ 
   & \,+  2 \mathrm{i}\, \Omega_{KL \bar P}  (\bar{\mathcal{Z}}
   \bar{\Omega})^{\bar P} 
   +  2 \mathrm{i} \,\Omega_{K  \bar P} (\bar{\mathcal{Z}} \bar{\Omega})^{\bar P}{}_L
  \Big] +\mathcal{O}(\Omega^3)
  \,,\nonumber\\ 
  \tilde\Omega_{I\bar J}=&\, [\mathcal{S}^{-1}]^K{}_I
  [\bar{\mathcal{S}}^{-1}]^{\bar L}{}_{\bar J} \,\Big[ \Omega_{K\bar
    L}  + 2\, \mathrm{i} F_{KMN} (\mathcal{Z} \Omega)^M{}_{\bar L}
  (\mathcal{Z} \Omega)^N 
    - 2 \mathrm{i}\, {\bar F}_{\bar L \bar P \bar N}
    (\bar{\mathcal{Z}} \bar{\Omega})^{\bar N}{}_{K}  
    (\bar{ \mathcal{Z}} \bar \Omega)^{\bar P} \nonumber\\
     & \, - 2 \mathrm{i}\, \Omega_{KM{\bar L}} (\mathcal{Z} \Omega)^M 
    - 2 \mathrm{i} \,\Omega_{KM} (\mathcal{Z} \Omega)^M{}_{\bar L} + 
    2 \mathrm{i} \,\Omega_{K \bar L \bar M}  (\bar{\mathcal{Z}} \bar
    \Omega)^{\bar M}     + 
    2 \mathrm{i} \,\Omega_{K \bar M }  (\bar{\mathcal{Z}} \bar
    \Omega)^{\bar M}{}_{\bar L}      \Big]    \nonumber\\
    & \, +\mathcal{O}(\Omega^3) \,,
\end{align}
where $ (\mathcal{Z} \Omega)^M{}_{L} = \mathcal{Z}^{MN} \Omega_{NL}$,
$(\bar{\mathcal{Z}} \bar{\Omega})^{\bar P}{}_L =
\bar{\mathcal{Z}}^{\bar P \bar N} {\Omega}_{\bar N L}$,
$({\mathcal{Z}} \Omega)^L{}_{\bar P} = {\mathcal{Z}}^{LK} \Omega_{K
  {\bar P}}$, etc. These results will then contribute to the
determination of the next order contribution to
\eqref{eq:Omega-tilde4}. The fact that $\Omega$ transforms
non-linearly under dualities, while we are at the same time
considering an expansion in terms of $\Omega$ and its derivatives,
suggests to introduce a formal expansion parameter $\alpha$ and expand
$\Omega= \sum_{n=1}^\infty \, \alpha^{n-1} \, \Omega^{(n)}$, so that
we are obtaining relations between products of different coefficient
functions $\Omega^{(n)}$ order-in-order in $\alpha$. At this stage
there is no direct need for this, but we will follow this strategy in
the next section where matters become somewhat more involved.

Rather than proceeding with this iteration procedure, we will simply
assume that all characteristic features noted in the results above,
will continue to hold in higher orders as well. An obvious conclusion
is then that the quantity $\Omega$ does {\it not} transform as a
function under symplectic transformations in view of the result
\eqref{eq:Omega-tilde4}. This is in agreement with our earlier claims,
for instance in \cite{Cardoso:2008fr,Cardoso:2010gc}. On the other hand
we expect that $\Omega$ must belong to a restricted class and, in
particular, it should have a well-defined harmonic limit that will
define the Wilsonian action.  To understand how this may come about,
let us first start from an $\Omega$ that is harmonic in the variables
$\mathcal{Y}^I$ and their complex conjugates 
$\bar{\mathcal{Y}}^I$. It is then reasonable to expect that also
$\tilde \Omega$ will be harmonic in the new variables, so that both
$\Omega$ and $\tilde\Omega$ can be written as a sum of a holomorphic
and an anti-holomorphic function in their respective variables. Indeed
this property is confirmed by \eqref{eq:Omega-tilde4}, or
alternatively by the last equation in \eqref{eq:transf-Omega3}. Hence
we conclude that the condition
\begin{equation}
  \label{eq:harmonic-restr}
  \Omega_{I\bar J}=0\,,
\end{equation}
is preserved under symplectic transformations. In the next section we
first discuss the consequences of this harmonicity constraint, before
considering modifications thereof.

\section{Non-holomorphic deformations and the anomaly equation}
\label{sec:non-holom-deform}
\setcounter{equation}{0}
At the end of the previous section we mentioned the possibility that
$\Omega$ is an harmonic function of the variables $\mathcal{Y}^I$ and
$\bar{\mathcal{Y}}^I$, a property that is consistent with respect to
symplectic transformations. However, the function of interest is
$\mathcal{H}^{(1)}(\mathcal{Y},\bar{\mathcal{Y}})$, which is not
harmonic in this case but which can still be decomposed in a way that
is rather similar to the decomposition of a harmonic function. Namely
one can write,
\begin{align}
  \label{eq:harmonic}
  \Omega(\mathcal{Y},\bar{\mathcal{Y}}) =&\,  \omega(\mathcal{Y})+
  \bar\omega(\bar{\mathcal{Y}})\,, \nonumber \\
  \mathcal{H}^{(1)}(\mathcal{Y},\bar{\mathcal{Y}}) =&\,
  {h}(\mathcal{Y},\bar{\mathcal{Y}}) +
  \bar{h}(\bar{\mathcal{Y}},\mathcal{Y})  \,,
\end{align}
where the function ${h}(\mathcal{Y},\bar{\mathcal{Y}})$ depends
holomorphically on $\omega$ and  equals, 
\begin{align}
  \label{eq:def-h}
    {h}(\mathcal{Y},\bar{\mathcal{Y}})=&\, 4\,\omega - 4\,N^{IJ}\,\omega_I\omega_J    
    \nonumber\\
    &\, +8\, \omega_{IJ} (N \omega)^I (N \omega)^J
    + \tfrac{8}3\mathrm{i} \,F_{IJK} (N \omega)^I (N \omega)^J (N \omega)^K
     \nonumber\\
    &\, - \tfrac43 \mathrm{i} \left(F_{IJKL} + 3 \mathrm{i}
      F_{R(IJ} N^{RS} F_{KL)S} \right) (N \omega)^I (N \omega)^J (N
    \omega)^K (N \omega)^L  \nonumber\\
     &\, 
    - \tfrac{16}3 \omega_{IJK} (N \omega)^I (N \omega)^J (N
      \omega)^K   
       - 16 \mathrm{i} \,F_{IJK} N^{KP} \,\omega_{PQ}  (N
      \omega)^I (N \omega)^J (N \omega)^Q   \nonumber\\
     & \, - 16 \, (N \omega)^P  \, \omega_{PQ} \, N^{QR}
     \omega_{RK} \, (N \omega)^K    + \mathcal{O}(\omega^5)\,. 
\end{align}
Because both $\mathcal{H}^{(1)}$ and $\tilde\Omega$, given in
\eqref{eq:Hesse-1} and \eqref{eq:Omega-tilde4}, are now harmonic in
the $\omega$, it follows that $\omega$ must transform under symplectic
transformations in direct correspondence with \eqref{eq:Omega-tilde4}, 
 \begin{align}
  \label{eq:omega-tilde4}
  \tilde \omega(\tilde{\mathcal{Y}}) =&\, \omega - \mathrm{i}\,
  \mathcal{Z}^{IJ} \omega_I \,\omega_J +\tfrac23 F_{IJK}
  \,(\mathcal{Z}\omega)^I \,(\mathcal{Z}\omega)^J
  \,(\mathcal{Z}\omega)^K
  \nonumber\\
  &\, -2 \,\omega_{IJ} \, (\mathcal{Z}\omega)^I\,
  (\mathcal{Z}\omega)^J   \nonumber\\
  &\, - \tfrac{1}{3}\mathrm{i} F_{IJKL} (\mathcal{Z} \omega)^I
  (\mathcal{Z} \omega)^J (\mathcal{Z} \omega)^K
  (\mathcal{Z} \omega)^L \nonumber\\
  &\, + \tfrac{4}{3}\mathrm{i} \,\omega_{IJK} (\mathcal{Z} \omega)^I
  (\mathcal{Z} \omega)^J (\mathcal{Z}
  \omega)^K \nonumber\\
  & \, + \mathrm{i}\, F_{IJK} \, {\cal Z}^{KL} \, F_{LMN} \, ({\cal Z}
  \omega)^I ({\cal Z} \omega)^J
  ({\cal Z} \omega)^M ({\cal Z} \omega)^N  \nonumber\\
  & \,- 4 \mathrm{i} \, F_{IJK} {\cal Z}^{KP} \,\omega_{PQ} \,({\cal
    Z} \omega)^I ({\cal Z} \omega)^J \
  ({\cal Z} \omega)^Q   \nonumber\\
  &\, +4\mathrm{i} ({\cal Z} \omega)^I \omega_{IJ} {\cal Z}^{JK}
  \omega_{KL} ({\cal Z} \omega)^L + \mathcal{O}(\omega^5)\,,
\end{align}
so that $\omega$ transforms holomorphically.  Obviously
$h(\mathcal{Y},\bar{\mathcal{Y}})$ must be a symplectic
function. However, its dependence on $\bar{\mathcal{Y}}$ resides
exclusively in the complex matrix $N^{IJ}$. Therefore we first study
the dependence of ${h}(\mathcal{Y},\bar{\mathcal{Y}})$ on $N^{IJ}$,
and derive the following equation
\begin{equation}
  \label{eq:der-N-of-h}
  \frac{\partial h}{\partial N^{IJ}} =  -\tfrac14 \partial_I h\,\partial_J
  h\,, 
\end{equation}
which we have verified up to terms of order $\omega^5$. This result
can easily be understood on the basis of the diagrammatic
interpretation that we have presented in the previous section. From it
one straightforwardly determines the non-holomorphic derivative of
${h}(\mathcal{Y},\bar{\mathcal{Y}})$,
\begin{equation}
  \label{eq:non-holo-eq}
  \partial_{\bar I} {h}(\mathcal{Y},\bar{\mathcal{Y}}) = \tfrac14
  \mathrm{i} \bar  F_{IKL} \, N^{KM} N^{LN}\, 
  \partial_M{h}(\mathcal{Y},\bar{\mathcal{Y}})\, 
  \partial_N{h}(\mathcal{Y},\bar{\mathcal{Y}})\, .
\end{equation}
The equation \eqref{eq:non-holo-eq} partially coincides with what is
known as the holomorphic anomaly equation for the topological string
and represents the terms that are induced by the pinching of a cycle
of the underlying Riemann surface resulting in two disconnected
Riemann surfaces \cite{Bershadsky:1993cx}. This restricted anomaly
equation is also obtained when considering the $N=2$ chiral superspace
action for abelian vector multiplets in the presence of a chiral
background field \cite{deWit:1996ix}. When expanding this
action in terms of the background, the holomorphic expansion
coefficient functions do not transform as functions under
duality. This can be resolved by covariantizing the Taylor expansion
with a suitable connection, but in that case the expansion coefficient
functions are no longer holomorphic. As it turns out these modified
coefficient functions then satisfy the same anomaly equation
\eqref{eq:non-holo-eq}. Note that this equation is integrable, so that
no additional constraints are implied. Clearly the holomorphic anomaly
is due to the fact that there is a conflict between symplectic
covariance and holomorphicity.

The above result would justify the identification of
$\mathcal{H}^{(1)}$ with the topological string, except that the
holomorphic anomaly equation is still incomplete. This implies that we
have to somehow relax the assumption that $\Omega$ is harmonic in
$\mathcal{Y}$, while still expressing it in terms of a holomorphic
function $\omega(\mathcal{Y})$ in such a way that $\mathcal{H}^{(1)}$
will remain harmonic (possibly up to a separate non-harmonic function,
as we shall see in section \ref{sec:third-order-contr}) in terms of
$\omega$. However, modifying the ansatz \eqref{eq:harmonic} must be
consistent with duality, in the sense that the modification will hold
for {\it the whole class of functions} $\Omega$ that are related by
duality. Our task is therefore to demonstrate that the present
framework can be extended so as to induce the remaining term in the
anomaly equation that is related to pinchings of cycles of the Riemann
surface that reduce the genus by one unit.

As it turns out, a consistent extension can be constructed by
introducing non-harmonic terms whose variation under symplectic
transformations is still harmonic. Such a modification does
preserve the present framework in a way that is consistent with
duality. For instance we could choose the following ansatz for $\Omega$,
\begin{equation}
  \label{eq:ansatz-Omega1}
   \Omega(\mathcal{Y},\bar{\mathcal{Y}}) =\omega(\mathcal{Y})+
  \bar\omega(\bar{\mathcal{Y}}) + \alpha\,\ln\det[N_{IJ}]
  +\beta\,\Psi(\mathcal{Y},\bar{\mathcal{Y}}) \,,
\end{equation}
where $\alpha$ and $\beta$ are arbitrary real parameters and $\Psi$ a
non-holomorphic function of $\mathcal{Y}$ and
$\bar{\mathcal{Y}}$. Note that we assume that the two deformations do
not depend on the holomorphic function $\omega$ or its complex
conjugate, because we insist on harmonicity with respect to 
$\omega$.

The deformation proportional to $\beta$ is the easiest to deal with,
so let us consider this one first. Since $\Psi$ is a given function
one cannot simply substitute the ansatz \eqref{eq:ansatz-Omega1} with
$\alpha=0$ into the expression for the function $\mathcal{H}^{(1)}$,
because this would require $\Psi$ to change non-trivially in order to
satisfy \eqref{eq:Omega-tilde4}. Hence we must introduce additional
terms into $\Omega$ to ensure that the modified $\Omega$ will still
transform according to \eqref{eq:Omega-tilde4} without modifying the
holomorphicity of $\omega$ and leaving $\Psi$ as a function. As its
turns out, this can be done in the next order by including the
following additional terms,
\begin{equation}
  \label{eq:mod-Omega-beta}
  \Omega(\mathcal{Y},\bar{\mathcal{Y}}) =\omega+
  \bar\omega  
  +\beta\,\Psi(\mathcal{Y},\bar{\mathcal{Y}}) +
  N^{IJ} \big[2\beta\,\partial_I\omega \,\partial_J\Psi +
  \beta^2\,\partial_I   \Psi\,\partial_J\Psi + \mathrm{h.c.}\big] \,.
\end{equation}
So far we have been performing an interation in powers of $\Omega$. It
is formally consistent to treat $\beta$ and $\omega$ as being of
the same order, but then we must assume that $\omega$ can obtain terms
of even higher order in $\beta$ in view of the non-linear
transformation rules, Hence we assume $\omega$ can be expanded in
terms of the parameter $\beta$, 
\begin{equation}
  \label{eq:series-exp-omega-beta}
  \omega(\mathcal{Y})\longrightarrow \omega(\mathcal{Y},\beta) =
  \sum_{n=1}^\infty \;\omega_n(\mathcal{Y})\,\beta^{n-1}\,,
\end{equation}
although we will keep this expansion implicit in what follows. It is
now straightforward to see that $\omega$ transforms as follows under
symplectic transformations
\begin{equation}
  \label{eq:transf-omega-beta}
  \tilde\omega = \omega - \mathrm{i}\,
  \mathcal{Z}^{IJ} \omega_I \,\omega_J + \mathcal{O}(\omega^3)\,,
\end{equation}
which agrees with \eqref{eq:omega-tilde4}. Subsequently we consider
the function $\mathcal{H}^{(1)}$ in this order of iteration,
\begin{equation}
  \label{eq:H-1-beta}
   \mathcal{H}^{(1)} = h+\bar h  +4\beta\,\Psi + \mathcal{O}(\omega^3)
   \,,
\end{equation}
where $h$ coincides with \eqref{eq:def-h} in this order of iteration. 
Consequently the addition of such a deformation leaves the holomorphic
anomaly equation \eqref{eq:non-holo-eq} unaltered. 

In view of this result we continue with the first modification in
\eqref{eq:ansatz-Omega1} proportional to $\alpha$. In principle the analysis proceeds in a
similar way as in the previous case, but here one has to also 
investigate the consistency in first order. However, it is easy to see
how consistency can be achieved because we have
\begin{equation}
  \label{eq:var-ln-det-N}
  \ln\det[\tilde N_{IJ}]=
  \ln\det[N_{IJ}] - \ln\det\big[\mathcal{S}\big] -
  \ln\det\big[\bar{\mathcal{S}}\big]\,, 
\end{equation}
where $\mathcal{S}$ was defined in \eqref{eq:def-S}. Because
$\mathcal{S}$ is holomorphic, the effect of the non-harmonic
modification $\alpha\,\ln\det[N_{IJ}]$ under duality can simply be absorbed
by assigning the following transformation to $\omega$,
\begin{equation}
  \label{eq:transf-omega}
  \tilde\omega(\tilde{\mathcal{Y}}) =\omega(\mathcal{Y}) + \alpha\,
  \ln\det\big[\mathcal{S}\big] \,,
\end{equation}
up to terms of higher order in $\Omega$ (or $\omega$) and
$\alpha$. Hence in lowest order
$\Omega(\mathcal{Y},\bar{\mathcal{Y}})$ transforms as a function, so
that our previous analysis remains unaffected as $\tilde\Omega=\Omega
+\mathcal{O}(\Omega^2)$. This is confirmed by the following. First of
all, derivatives of the holomorphic function $\omega(\mathcal{Y})$
remain holomorphic but they do acquire extra terms in their
transformation rules, as is shown in
\begin{align}
  \label{eq:trans-omega-der-1}
  \tilde\omega_I  =&\,  [\mathcal{S}^{-1}]^J{}_I \,\big(\omega_J
  +\alpha \,F_{JKL} \mathcal{Z}^{KL} \big) +\mathcal{O}(\alpha^2) \,,
  \nonumber\\ 
   \tilde\omega_{IJ}  =&\, [\mathcal{S}^{-1}]^K{}_I
    [\mathcal{S}^{-1}]^L{}_J\, \big[ \omega_{KL} - F_{KLM}
    \mathcal{Z}^{MN} (\omega_N +\alpha F_{NPQ} \,\mathcal{Z}^{PQ} )
    \nonumber\\ 
    &\,\qquad\qquad\qquad\qquad + \alpha\big(F_{KLMN}
    - F_{KMP}\, F_{LNQ}\, \mathcal{Z}^{PQ}\big) \mathcal{Z}^{MN} 
     \big] +\mathcal{O}(\alpha^2)  \,.
\end{align}
Furthermore the first few derivatives of $\Omega$ are now equal to 
\begin{align}
  \label{eq:ddd-Omega}
  \Omega_I=&\, \omega_I -\mathrm{i}\alpha\, F_{IJK}\,N^{JK}
  +\mathcal{O}(\alpha^2)  \,,\nonumber\\
  \Omega_{IJ}=&\, \omega_{IJ} +\alpha\, F_{IKL}\,F_{JMN} \,N^{KM}
  N^{LN}  -\mathrm{i}\alpha F_{IJKL}\,N^{KL}  +\mathcal{O}(\alpha^2)  \,,\nonumber\\
  \Omega_{I\bar J}=&\, -\alpha\, F_{IKL}\,\bar F_{JMN} \,N^{KM} N^{LN}
  +\mathcal{O}(\alpha^2) \,,
\end{align} 
which, in leading order, transform consistently under duality
transformations (i.e. according to \eqref{eq:transf-Omega3}). To show
this one makes use of \eqref{eq:trans-omega-der-1} and
\eqref{eq:F-0-der}. Hence we conclude that to lowest order in
$\Omega$, the non-holomorphic deformation \eqref{eq:ansatz-Omega1} is
consistent. Observe that the last equation \eqref{eq:ddd-Omega}
constitutes a deviation from the harmonicity condition
\eqref{eq:harmonic-restr}.

In due course we will also need the result for the transformation of
the third- and fourth-derivatives of $\omega$, implied by
\eqref{eq:transf-omega}, 
\begin{align}
  \label{eq:trans-omega-ders}
     \tilde\omega_{IJK}  =&\, [\mathcal{S}^{-1}]^M{}_{\!(I}
     [\mathcal{S}^{-1}]^N{}_J [\mathcal{S}^{-1}]^P{}_{\!K)}\,
     \nonumber\\
     &\,\times\big[ \omega_{MNP}  
     -3\,  F_{MN\bullet} \,\mathcal{Z}^{\bullet\bullet}\,
     \big(\omega_{P\bullet}  +\alpha\, F_{PQR\bullet}\,
     \mathcal{Z}^{QR} \big) \nonumber\\
     &\,\quad
     -\big(F_{MNP\bullet}    -3\, F_{MN\bullet}\,
     \mathcal{Z}^{\bullet\bullet} F_{P\bullet\bullet} \big)
     \mathcal{Z}^{\bullet\bullet} \,\big(\omega_\bullet 
    +\alpha
    \,F_{QR\bullet}\,\mathcal{Z}^{QR} \big)
    \nonumber\\ 
    &\, \quad
    + \alpha\big( F_{MNPQR} -3\, F_{MNQ\bullet} \,
    \mathcal{Z}^{\bullet\bullet}\, F_{PR\bullet}   
    +3\,F_{MN\bullet} \, \mathcal{Z}^{\bullet\bullet}\,
    F_{Q\bullet\bullet}\, \mathcal{Z}^{\bullet\bullet}\,
    F_{\bullet PR}  \big)  \mathcal{Z}^{QR} \nonumber\\  
    &\,\quad
    + 2\, \alpha\, F_{MQ \bullet}\, \mathcal{Z}^{\bullet
      \bullet}\, F_{\bullet \bullet N}\, \mathcal{Z}^{\bullet \bullet} \,F_{PR \bullet}
    \mathcal{Z}^{QR} \big] +\mathcal{O}(\alpha^2) \,,  \nonumber\\[.2ex]
    \tilde\omega_{IJKL} =&\, [\mathcal{S}^{-1}]^M{}_{\!(I}
    [\mathcal{S}^{-1}]^N{}_J [\mathcal{S}^{-1}]^P{}_{\!K}
    [\mathcal{S}^{-1}]^Q{}_{\!L)}\,
    \nonumber\\
    &\,\times\big[ \omega_{MNPQ} -6\, F_{MN\bullet}
    \,\mathcal{Z}^{\bullet\bullet}\, \big(\omega_{PQ\bullet} +\alpha\,
    F_{PQUV\bullet}\,
    \mathcal{Z}^{UV} \big)\nonumber\\
    &\,\quad -4\,\big( F_{MNP\bullet} -3\,F_{MN\bullet}\, \mathcal{Z}^{\bullet\bullet}
    F_{P\bullet\bullet} \big)  \,\mathcal{Z}^{\bullet\bullet}\,
    \big(\omega_{Q\bullet} +\alpha\, F_{Q\bullet UV}\,
    \mathcal{Z}^{UV}\big)   \nonumber\\
     &\,\quad
    +3\,F_{MN\bullet}\,\,\mathcal{Z}^{\bullet\bullet}\,\big(\omega_{\bullet\bullet}
    +\alpha\,F_{\bullet\bullet UV}\,\mathcal{Z}^{UV} \big)
    \,\mathcal{Z}^{\bullet\bullet} \, F_{\bullet PQ}
    \nonumber\\
    &\,\quad -\big(F_{MNPQ\bullet} -4\, F_{MNP\bullet}\,
    \mathcal{Z}^{\bullet\bullet} F_{Q\bullet\bullet} -6\,
    F_{MN\bullet}\, \mathcal{Z}^{\bullet\bullet} F_{PQ\bullet\bullet}
    \big) \mathcal{Z}^{\bullet\bullet} \,\big(\omega_\bullet +\alpha
    \,F_{UV\bullet}\,\mathcal{Z}^{UV} \big) \nonumber\\
    &\,\quad 
    - 12\, F_{MN\bullet} \,\mathcal{Z}^{\bullet\bullet}\,
    F_{P\bullet\bullet} \mathcal{Z}^{\bullet\bullet}\,
    F_{Q\bullet\bullet}
    \,\mathcal{Z}^{\bullet\bullet}\,\big(\omega_\bullet +\alpha
    \,F_{UV\bullet}\,\mathcal{Z}^{UV} \big) \nonumber\\
    &\,\quad
    -3\, \big(\mathcal{Z}^{R\bullet}\,F_{MN\bullet}\big) \,
    \big(\mathcal{Z}^{S\bullet}\,F_{PQ\bullet}\big) \,  F_{RST} \,
    \mathcal{Z}^{T\bullet}\, \big(\omega_\bullet +\alpha
    \,F_{UV\bullet}\,\mathcal{Z}^{UV} \big) 
    \nonumber\\
 &\, \quad
    + \alpha\big[F_{MNPQUV} \mathcal{Z}^{UV} 
    -4\, F_{MNPUV} \,\mathcal{Z}^{U\bullet} F_{Q\bullet\bullet} \mathcal{Z}^{\bullet V}   
    -3\, F_{MNXU} \, F_{PQYV} \, \mathcal{Z}^{XY}\mathcal{Z}^{UV}
    \big] \nonumber\\
    &\,\quad
    +4\alpha \big[ 3\big( F_{MN\bullet\bullet} \mathcal{Z}^{\bullet\bullet}
    F_{P\bullet\bullet} \mathcal{Z}^{\bullet \bullet}
    F_{Q\bullet\bullet}  \mathcal{Z}^{\bullet\bullet}\big) 
    +F_{MNP\bullet}
    \mathcal{Z}^{\bullet X}
    \,\big(F_{X\bullet\bullet}\mathcal{Z}^{\bullet\bullet} 
    F_{Q\bullet\bullet}\mathcal{Z}^{\bullet\bullet} \big)  \big]
    \nonumber\\
    &\,\quad 
    +6\alpha\,\big[2\, F_{MN\bullet} \mathcal{Z}^{\bullet X} \,
    \big(F_{XP\bullet\bullet}\mathcal{Z}^{\bullet \bullet}
    F_{Q\bullet\bullet}\mathcal{Z}^{\bullet\bullet}  \big) 
      +  F_{MN\bullet} \mathcal{Z}^{\bullet X} \,
      \big(F_{X\bullet\bullet}\mathcal{Z}^{\bullet \bullet} 
      F_{PQ\bullet\bullet}\mathcal{Z}^{\bullet\bullet} \big) \big]\nonumber\\
      &\,\quad -6\alpha \,\big(F_{M\bullet\bullet}\,\mathcal{Z}^{\bullet\bullet}\,
      F_{N\bullet\bullet}\, \mathcal{Z}^{\bullet \bullet}\,
      F_{P\bullet\bullet} \,\mathcal{Z}^{\bullet \bullet}\,
      F_{Q\bullet\bullet} \, \mathcal{Z}^{\bullet\bullet}\big)  \nonumber\\
      &\, \quad
     -12 \alpha\, F_{MN\bullet} \,\mathcal{Z}^{\bullet\bullet}\,
      F_{P\bullet\bullet} \,\mathcal{Z}^{\bullet X} \, 
      \big(F_{X\bullet\bullet} \,\mathcal{Z}^{\bullet \bullet}
      F_{Q\bullet\bullet} \,\mathcal{Z}^{\bullet \bullet}\big)
    \nonumber\\
   &\, \quad
     -12 \alpha\, F_{MN\bullet} \,\mathcal{Z}^{\bullet X}\,
      \big(F_{X\bullet\bullet} \,\mathcal{Z}^{\bullet \bullet}
      F_{P\bullet\bullet} \,\mathcal{Z}^{\bullet \bullet}
      F_{Q\bullet\bullet} \,\mathcal{Z}^{\bullet \bullet}\big)
    \nonumber\\
    &\,\quad
    -3\alpha\,F_{MN\bullet}\,\mathcal{Z}^{\bullet X}\,
    (F_{X\bullet\bullet} \mathcal{Z}^{\bullet \bullet}
    F_{Y\bullet\bullet} \mathcal{Z}^{\bullet \bullet})\,
    \mathcal{Z}^{Y\bullet}\, F_{\bullet PQ} +\mathcal{O}(\alpha^2) \,. 
\end{align}

To continue this scheme to higher orders in $\Omega$ is not an easy
task. So far we have been working order-by-order in powers of
$\Omega$, but now we are dealing also with additional terms that are
proportional to the parameter $\alpha$. Within the iterative procedure
that we have been following it is consistent to formally treat
$\omega$ and $\alpha$ as being of the same order as $\Omega$. Counting
in this way shows that the corrections in \eqref{eq:trans-omega-ders}
are of first order in $\alpha$. Because the equations that we are
dealing with are non-linear it is therefore imperative that the
$\omega$ itself can in principle contain contributions of arbitrary
order in $\alpha$, a possibility that we have already been alluding to
below equation \eqref{eq:transf-Omega3}. Therefore we will in addition
assume that $\omega$ can be expanded in terms of $\alpha$,
\begin{equation}
  \label{eq:series-exp-omega}
  \omega(\mathcal{Y})\longrightarrow \omega(\mathcal{Y},\alpha) =
  \sum_{n=1}^\infty \;\omega_n(\mathcal{Y})\,\alpha^{n-1}\,,
\end{equation}
although we will keep this expansion implicit in what follows.
Assuming that $\omega$ incorporates the higher-order terms in $\alpha$
we can proceed to higher orders by iteration to obtain the extension
of the original almost-harmonic ansatz \eqref{eq:ansatz-Omega1},
possibly up to terms that separately constitute proper functions under
symplectic transformations.

We thus continue to derive the terms of order $\alpha^2$, which can be
found from \eqref{eq:Omega-tilde4},
\begin{align}
  \label{eq:holom-Omega-diff-2}
  \tilde\Omega- \Omega= &\, -\mathrm{i} \big( \mathcal{Z}^{IJ}
  \Omega_I\Omega_J - \bar{\mathcal{Z}}^{IJ}
  \Omega_{\bar I}\Omega_{\bar J} \big) + \mathcal{O}(\Omega^3)\nonumber\\
  =&\, -\mathrm{i} \mathcal{Z}^{IJ}\big[\omega_I \,\omega_J
  -2\mathrm{i} \alpha\,\omega_I \, F_{JKL} N^{KL} -\alpha^2\,
  F_{IKL}\,F_{JMN}\,N^{KL}\,N^{MN} \big] + \mathrm{h.c.} \nonumber\\
  &\, +  \mathcal{O}(\alpha^3)\,, 
\end{align}
where we note that the right-hand side is of order $\alpha^2$, as both
$\alpha$ and $\omega$ are counted as being of the same order as
$\Omega$.  Our task is now to include further modifications into
$\Omega$, just as we did earlier in \eqref{eq:ansatz-Omega1}, so that
the change of $\Omega$ under a symplectic transformation becomes
consistent with the right-hand side of \eqref{eq:holom-Omega-diff-2},
up to a term that is harmonic, which can then be absorbed into the
variation of $\omega$. The problem here is, however, that the
expression \eqref{eq:holom-Omega-diff-2} is linear in
$\mathcal{Z}^{IJ}$ but involves also terms that are linear or
quadratic in $N^{IJ}$. It is clear that one cannot construct a
suitable addition to $\Omega$ depending exclusively on $\omega_I$,
$N^{IJ}$ and $F_{IJK}$.  There exists, however, an alternative, namely
to include higher derivatives of $\omega$. According to
\eqref{eq:trans-omega-der-1} the second derivative $\omega_{IJ}$ leads
to similar variations, suggesting another possible
modification. Indeed, it follows that $N^{IJ}\omega_{IJ}$,
$F_{IJKL}N^{IJ} N^{KL}$ and $F_{IJK}\,F_{LMN}\,N^{IL}N^{JM}N^{KN}$
(and complex conjugates where appropriate) are the terms that one may
add to $\Omega$ so that only a holomorphic variation will remain in
\eqref{eq:holom-Omega-diff-2}. And indeed, one can verify by explicit
calculation that $\Omega$ should be written as (up to a symplectic
function of $\mathcal{Y}$ and $\bar{\mathcal{Y}}$),
\begin{align}
  \label{eq:ansatz-nonholo-Omega-2}
  \Omega(\mathcal{Y},\bar{\mathcal{Y}}) =&\, \alpha\,\ln N
  \nonumber\\ 
  &\, + \Big[\omega +2\,\alpha\, N^{IJ} \omega_{IJ} -\alpha^2 \big[
    \mathrm{i}F_{IJKL} -\tfrac23 
     F_{IKM} F_{JLN}   N^{MN}  \big]   N^{IJ}N^{KL} +\mathrm{h.c.}\Big] \nonumber\\
   &\,+ \mathcal{O}(\alpha^3) \,,
\end{align}
where here and henceforth we use the definition
$N\equiv\det[N_{IJ}]$. Hence the holomorphic function $\omega$ is now
accompanied by a variety of specific non-holomorphic modifications
which will contribute to the effective action. Indeed, with this
result for $\Omega(\mathcal{Y},\bar{\mathcal{Y}})$, explicit
evaluation shows that \eqref{eq:holom-Omega-diff-2} is satisfied up to
order $\alpha^3$ provided that $\omega$ transforms (holomorphically)
according to
\begin{align}
  \label{eq:trans-omega-2}
  \tilde\omega=&\, \omega + \alpha\,  \ln\det\big[\mathcal{S}\big]
   +\mathrm{i} \mathcal{Z}^{IJ} \big[  2\,\alpha
   \,\omega_{IJ} -\big(\omega_I
   +\alpha\,F_{IKL}\mathcal{Z}^{KL}\big) \big(\omega_J
   +\alpha\,F_{JMN}\mathcal{Z}^{MN}\big) \big]  \nonumber\\ 
   &\, + \mathrm{i}\alpha^2  \big[ F_{IJKL}     -\tfrac23 F_{IKM} F_{JLN}
   \mathcal{Z}^{MN}  \big]\mathcal{Z}^{IJ}  \mathcal{Z}^{KL} +
   \mathcal{O}(\alpha^3)  \,.
\end{align}
To see this one makes use of the transformations of multiple
derivatives of the holomorphic function $F(\mathcal{Y})$, listed in
\eqref{eq:F-0-der}, the transformation of $N^{IJ}$ as given in
\eqref{eq:dual-trans-N-alt}, and the transformation rule for
$\omega_{IJ}$ specified in the second equation of
\eqref{eq:trans-omega-der-1}. Incidentally, the result
\eqref{eq:ansatz-nonholo-Omega-2} is in line with the ansatz
\eqref{eq:series-exp-omega} according to which the function $\omega$
is expanded in powers of $\alpha$. Furthermore it turns out that the
result \eqref{eq:ansatz-nonholo-Omega-2} takes the form of a sum over
connected 1PI diagrams, unlike the corresponding result for
$\mathcal{H}^{(1)}$.

Let us now return to the almost harmonic function
$\mathcal{H}^{(1)}(\mathcal{Y},\bar{\mathcal{Y}};N)$, which now
decomposes according to 
\begin{equation}
  \label{eq:almost-harm}
    \mathcal{H}^{(1)}(\mathcal{Y},\bar{\mathcal{Y}};N) =
    4 \alpha\,\ln N   + 
   {h}(\mathcal{Y},\bar{\mathcal{Y}}) +
  \bar{h}(\bar{\mathcal{Y}},\mathcal{Y})  \,,
\end{equation}
which turns out to be a harmonic function of $\omega$. Indeed, 
making use of \eqref{eq:ansatz-nonholo-Omega-2} one finds that
${h}(\mathcal{Y},\bar{\mathcal{Y}})$ now takes the form, 
\begin{align}
  \label{eq:def-h-mod}
    {h}(\mathcal{Y},\bar{\mathcal{Y}})=&\, 4\,\omega -
    4\,\big(\omega_I -\mathrm{i}\alpha\,F_{IKL} N^{KL}\big)\, N^{IJ}\, 
    \big( \omega_J   -\mathrm{i}\alpha\,F_{JMN} N^{MN}\big)  
    \nonumber\\
    &\, +8\,\alpha\, \omega_{IJ} N^{IJ}  - 4\,\alpha^2 \big[\mathrm{i}F_{IJKL}
    -\tfrac23 F_{IKM} F_{JLN}   N^{MN}  \big]   N^{IJ}N^{KL}   +
     \mathcal{O}(\alpha^3)  \,. 
\end{align}
We stress that, by construction, $\mathcal{H}^{(1)}$ remains a
symplectic function in the presence of the term $4\alpha\,\ln N$ in
\eqref{eq:almost-harm}. Furthermore $h$ will separately transform as a
function {\it beyond linear order in} $\alpha$ and derivatives of $h+
4\alpha\,\ln N$ will still transform as proper tensors. This must be
the case because the transformations of $h$ beyond the linear order
depend only on $\mathcal{S}$ through the tensor $\mathcal{Z}^{IJ}$,
and likewise $\bar h$ depends only on $\bar{\mathcal{S}}$ through the
tensor $\bar{\mathcal{Z}}^{IJ}$. As we shall see, the non-holomorphic
derivative of $h$ will transform as a vector as it is not of first
order in $\alpha$.  We also observe that the transformation of
$\omega$ as specified in \eqref{eq:trans-omega-2} follows precisely
from the expression for $\tfrac14 h (\mathcal{Y},\bar{\mathcal{Y}})$
upon replacing $N^{IJ}$ by $\mathrm{i}\mathcal{Z}^{IJ}$, with the
exception of the term $\alpha\, \ln\det\big[\mathcal{S}\big]$ that is
related to the explicitly non-harmonic term in
\eqref{eq:ansatz-nonholo-Omega-2}. This is in line with the phenomenon
noted below equation \eqref{eq:Omega-tilde4}.

Now we return to the holomorphic anomaly equation. Following the
discussion at the beginning of this section we first determine,
\begin{align}
  \label{eq:N-derivative-h}
  \frac{\partial{h}}{\partial N^{IJ}} =&\, -\tfrac14 \partial_I\big(h
  +4\alpha\,\ln  N \big) \, 
  \partial_J\big(h +4\alpha\,\ln  N \big) \nonumber\\  
  &\,+ 2\,\alpha\, D_I\partial_J\big(h +4\alpha\,\ln  N \big)  
  +\mathcal{O}(\alpha^3) \,. 
\end{align}
Here we have introduced a covariant derivative which ensures
covariance under the symplectic transformations. On a holomorphic
vector, $V_I$, this covariant derivative takes the form,
\begin{equation}
  \label{eq:K-connection}
  D_IV_J =\partial_I V_J - \Gamma_{IJ}{}^K\, V_K\,,  
\end{equation}
where $\Gamma_{IJ}{}^K$ is Christoffel connection associated with the
K\"ahler metric $g_{I\bar J}= \partial_I \partial_{\bar J}
K(\mathcal{Y},\bar{\mathcal{Y}})= N_{IJ}$, with $K$ the K\"ahler
potential \footnote{ 
  Note that there is no uniformity in the literature regarding the
  overall sign of $K$. See, e.g. \cite{Cardoso:2008fr}. } 
\begin{equation}
  \label{eq:kahler-pot}
  K(\mathcal{Y},\bar{\mathcal{Y}}) = -\mathrm{i}\big[ \bar{\mathcal{Y}}^{\bar I}
    F_{I}(\mathcal{Y}) - \mathcal{Y}^I\bar
  F_{\bar I}(\bar{\mathcal{Y}}) \big] \,. 
\end{equation}
Observe that, for a K\"ahler space the non-vanishing connection
components are $\Gamma_{IJ}{}^K$ and its complex conjugate
$\Gamma_{\bar I\bar J}{}^{\bar K}$.  The non-vanishing (up to complex
conjugation) connection and curvature components are then equal to
\begin{align}
  \label{eq:metric-conn-curv}
  \Gamma_{IJ}{}^K =&\, g^{K\bar L} \partial_{I} g_{J\bar L} = -\mathrm{i}
  F_{IJL}\,N^{LK} \,, \nonumber \\
  R_{\bar I JK}{}^L =&\,\partial_{\bar I}\, \Gamma_{JK}{}^L
  = - N^{LM} \,\bar F_{\bar I\bar M\bar N}\, N^{NP} \,F_{PJK} \,. 
\end{align}
We remind the reader that combinations of higher derivatives of the
holomorphic function $F(\mathcal{Y})$ that involve also the matrix
$N^{IJ}$ can transform covariantly under symplectic transformations,
as was pointed out at the end of section \ref{sec:deformed-special}
(see e.g. \eqref{eq:def-C}). 

We should also stress that the above discussion pertains to the
underlying K\"ahler geometry. Obviously the special diffeomorphisms related to
the symplectic transformations form a subgroup of the group of
holomorphic diffeomorphisms. This is confirmed by evaluating the
transformation of the connection under a symplectic transformation,
using the results presented in section \ref{sec:deformed-special},
\begin{equation}
  \label{eq:sympl-var-connection}
  \Gamma_{IJ}{}^K\longrightarrow  [\mathcal{S}^{-1}]^L{\!}_I\,
  [\mathcal{S}^{-1}]^M{\!}_J \,\big[ \Gamma_{LM}{}^N  \mathcal{S}^K{\!}_N
  - \partial_L \mathcal{S}^K{\!}_M\big] \,.  
\end{equation}

After these comments and clarifications we return to equation
\eqref{eq:N-derivative-h}. Noting the lack of holomorphicity in
$h$ resides in $N^{IJ}$, we can now determine the anti-holomorphic
derivative of $h$,
\begin{align}
  \label{eq:nonholo-derivative-h}
  \partial_{\bar I} h =&\,  \mathrm{i} \,\bar F_{IKL} N^{KM}
  N^{LN} \, \big[ \tfrac14 \partial_M\big(h  +4\alpha\,\ln  N \big) \, 
  \partial_N\big(h +4\alpha\,\ln  N \big) - 2\,\alpha\,
  D_M\partial_N\big(h +4\alpha\,\ln  N \big) \big]  \nonumber\\  
  &\,  +\mathcal{O}(\alpha^3) \,, 
\end{align}
where the right-hand side contains no terms linear in $\alpha$. However,
one could consider the mixed derivative of $h  +4\alpha\,\ln  N$,
which does contain terms {\it linear} in $\alpha$ given by
\begin{equation}
  \label{eq:mixed-deriv-alpha}
  \partial_{\bar I}\partial_J \big(h +4\alpha\,\ln  N\big) =  -
  4\alpha\, N^{KL} N^{MN} F_{JKM}  \bar F_{ILN} \,. 
\end{equation}
The expression on the right-hand side is precisely equal to $4\alpha\,
R_{\bar IJ}$, where $R_{\bar IJ}= R_{\bar IKJ}{}^K$ equals the Ricci
tensor of the special K\"ahler manifold whose value follows from the
second equation in \eqref{eq:metric-conn-curv}. The equations
\eqref{eq:nonholo-derivative-h} and \eqref{eq:mixed-deriv-alpha} are
the familiar holomorphic anomaly equations of the topological string.

In this section we introduced a deformation of $\Omega$ proportional
to the parameter $\alpha$ which induced further corrections to
$\Omega$ of higher orders in $\alpha$. This deformation was not itself
a proper function, but its variation under a symplectic transformation
was harmonic. Obviously this deformation was not unique because the
effect of the symplectic transformation would remain the same upon
adding a proper symplectic function to the deformation. Therefore we
also considered adding a separate non-harmonic function to $\Omega$
(c.f. \eqref{eq:ansatz-Omega1}). We concluded that this modification
must lead to new terms in $\Omega$, as shown in
\eqref{eq:mod-Omega-beta}, but they contribute only to the Hesse
potential by an additive contribution of the original non-harmonic
function. However, when defining $\mathcal{H}^{(1)}$ we agreed that
such additions should be included as a separate symplectic functions
in the expansion of the Hesse potential in terms of independent
functions. Hence this additive term will not affect 
$\mathcal{H}^{(1)}$. This aspect is essential for deriving the
holomorphic anomaly equation. However, when the non-harmonic function
is of first order in the deformation parameter, it will not contribute
to the holomorphic anomaly equation, because it does not generate
higher-order terms under the iteration, while
\eqref{eq:nonholo-derivative-h} only receives contributions beyond the
first order. The first-order contributions are instead governed by the
separate equation \eqref{eq:mixed-deriv-alpha}.

We will not analyze this issue in further detail here, but there is
one type of deformation of the anomaly equation that is worth
recalling. Suppose that on the right-hand side of
\eqref{eq:nonholo-derivative-h} we change $h+4\alpha\,\ln
N$ by adding a function of the K\"ahler potential defined in
\eqref{eq:kahler-pot} (which equals the Hesse potential with
$\Omega=0$). This K\"ahler potential satisfies the following
identities,
\begin{equation}
  \label{eq:def-K}
  \partial_I K=
  N_{IJ}\,\bar{\mathcal{Y}}^J\,,\qquad \partial_I\partial_{\bar J} K=
  N_{IJ}\,, \qquad D_{I}\partial_JK=0\,,
\end{equation}
which all have a geometrical meaning. Adding a function of $K$ to
$h+4\alpha\,\ln N$ will introduce terms proportional to $N_{MP}
\bar{\mathcal{Y}}^P$ or $N_{NP} \bar{\mathcal{Y}}^P$ on the right-hand
side of \eqref{eq:nonholo-derivative-h} that cancel when contracted
with the overall factor $\bar F_{IKL} N^{KM} N^{LN}$. 

Finally, as is shown in \eqref{eq:trans-omega-2}, the transformation
rule of $\omega$ has now acquired new terms of order $\alpha^2$. The
derivatives of $\omega$ will thus receive corresponding
contributions. In particular \eqref{eq:trans-omega-der-1} and
\eqref{eq:trans-omega-ders} will change. For the calculations in
section \ref{sec:third-order-contr} it is relevant to present the full
expressions for the variations of $\omega_I$ and $\omega_{IJ}$ up to
order $\alpha^3$. In view of their length we have listed these
equations in appendix \ref{App:transformation-omega}.

\section{Evaluating the third-order contributions}
\label{sec:third-order-contr}
\setcounter{equation}{0}
There are good reasons for evaluating also the contributions of order
$\alpha^3$. One of them is that $\Omega$ and $\mathcal{H}^{(1)}$ are no
longer obviously `partially harmonic' in higher orders. Another one is
that the contributions of third order have never been fully  worked out
explicitly for the topological string.

Let us again start with \eqref{eq:Omega-tilde4}, but now approximated to
terms of order $\Omega^3$, 
\begin{align}
  \label{eq:Omega-tilde4-approx-3}
  \tilde \Omega(\tilde{\mathcal{Y}}, \bar{\tilde{{\mathcal{Y}}}}) =&\,
  \Omega - \mathrm{i}\big( \mathcal{Z}^{IJ} 
  \,\Omega_I \Omega_J - \bar{\mathcal{Z}}^{IJ}
  \,\Omega_{\bar I} \Omega_{\bar J} \big)\nonumber \\
  &\, +\tfrac23\big( F_{IJK} \,\mathcal{Z}^{IL}\Omega_L
  \,\mathcal{Z}^{JM}\Omega_M \,\mathcal{Z}^{KN}\Omega_N +
  \mathrm{h.c.}\big) 
  \nonumber\\
  &\, -2 \big(\Omega_{IJ} \, \mathcal{Z}^{IK}\Omega_K
  \mathcal{Z}^{JL}\Omega_L  +\mathrm{h.c.}\big)
  +4\,\Omega_{I\bar J} \,
  \mathcal{Z}^{IK}\Omega_K\,  \bar{\mathcal{Z}}^{JL}\Omega_{\bar L}
  +\mathcal{O}(\Omega^4)\,, 
\end{align}
which must hold irrespective of the precise form of $\Omega$. To
evaluate the right-hand side we must first determine $\Omega_I$ to 
order $\alpha^2$, which follows from \eqref{eq:ansatz-nonholo-Omega-2},
\begin{align}
  \label{eq:Omeg-derivative}
    \Omega_I =&\, \omega_I -\mathrm{i} \alpha\,F_{IJK}\,N^{JK}
    \nonumber\\
     &\, +2\,\alpha \big[ \omega_{IJK} \,N^{JK}
    +\mathrm{i}\big(\omega_{JK}+\bar\omega_{JK}\big) N^{JL}
      F_{I LM} N^{MK}  \big] \nonumber\\
      &\, -\alpha^2 \big[\mathrm{i}F_{IJKLM} -\tfrac43 F_{IJLN} N^{NP} 
      F_{KMP}\big] N^{JK} N^{LM} \nonumber\\
      &\, +2\,\alpha^2 \big(F_{JKLM} -\bar F_{JKLM}\big) N^{JK} N^{LN}
      N^{MP}\, F_{INP} 
      \nonumber\\
      &\, +2\mathrm{i} \alpha^2 \big( F_{JKL} F_{MNP} +\bar F_{JKL}
      \bar F_{MNP} \big) N^{JM} N^{KN}  N^{LQ} N^{PR} F_{IQR}    
      +\mathcal{O}(\alpha^3) \nonumber\\
      = &\, \omega_I -\mathrm{i} \alpha\,F_{IJK}\,N^{JK}  +2\,\alpha
      \, \omega_{IJK} \,N^{JK}   \nonumber\\ 
     &\, +2 \mathrm{i}\alpha\, F_{I JK}\,N^{JM}N^{KN} \big[ \omega_{MN}
     + \alpha\, F_{MPR}\, F_{NQS} N^{PQ} N^{RS} 
    -\mathrm{i}\alpha\,F_{MNPQ} N^{PQ} \big]  \nonumber\\
      &\, -\alpha^2 \big[\mathrm{i}F_{IJKLM} -\tfrac43 F_{IJLN} N^{NP} 
      F_{KMP}\big] N^{JK} N^{LM} \nonumber\\
       &\, +2 \mathrm{i}\alpha\, F_{I JK}\,N^{JM}N^{KN} \big[
       \bar\omega_{ MN}
     + \alpha\,\bar F_{MPR}\, \bar F_{NQS} N^{PQ} N^{RS} 
    +\mathrm{i}\alpha\,\bar F_{MNPQ} N^{PQ} \big]  \nonumber\\
&\,  +\mathcal{O}(\alpha^3)\,.
\end{align}
We now observe that the above expression is no longer 'almost
holomorphic' and thus deviates from the results obtained before. The
troublesome terms are contained in the last line of
\eqref{eq:Omeg-derivative}, which turns out to be equal to
\begin{align}
  \label{eq:nonhol-mixed-Omega}
   2 \mathrm{i}\alpha\, F_{I JK}\,N^{JM}N^{KN} \,\Omega_{\bar M\bar
     N}  +\mathcal{O}(\alpha^3)\,, 
\end{align}
where we made use of the second equation in \eqref{eq:ddd-Omega}. It
will be convenient to keep writing these non-holomorphic contributions
in terms of non-holomorphic derivatives of $\Omega$. The crucial point
to note is, however, that we have extracted an explicit factor of
$\alpha$, whereas so far $\alpha$ appeared only implicitly in
$\Omega$. As we will demonstrate shortly, one consequence of our
analysis is that the Hesse potential will involve additional
symplectic functions, but now multiplied by explicit powers of
$\alpha$.

Substituting the above result \eqref{eq:Omeg-derivative} and the last two equations of
\eqref{eq:ddd-Omega} into \eqref{eq:Omega-tilde4-approx-3}, we obtain,
\begin{align}
  \label{eq:Omega-tilde4-explicit-3}
  &\tilde \Omega(\tilde{\mathcal{Y}}, \bar{\tilde{{\mathcal{Y}}}}) -
  \Omega(\mathcal{Y},\bar{\mathcal{Y}}) \nonumber\\
  &=\Big\{ - \mathrm{i} \big(\omega_I -\mathrm{i}
  \alpha\,F_{I\bullet\bullet}\,N^{\bullet\bullet}\big) \mathcal{Z}^{IJ} \big(\omega_J
  -\mathrm{i}
  \alpha\,F_{J\bullet\bullet}\,N^{\bullet\bullet}\big)  \nonumber\\
  &\, +\tfrac23 F_{IJK} \,\mathcal{Z}^{IL} \big(\omega_L -\mathrm{i}
  \alpha\,F_{L\bullet\bullet}\,N^{\bullet\bullet}\big)
  \,\mathcal{Z}^{JM} \big(\omega_M -\mathrm{i}
  \alpha\,F_{M\bullet\bullet}\,N^{\bullet\bullet}\big)
  \mathcal{Z}^{KN}\big(\omega_N
  -\mathrm{i} \alpha\,F_{N\bullet\bullet}\,N^{\bullet\bullet}\big) 
  \nonumber\\
  & -4 \mathrm{i} \alpha \big(\omega_I -\mathrm{i}
  \alpha\,F_{I\bullet\bullet}\,N^{\bullet\bullet}\big) \mathcal{Z}^{IJ} \big[ \omega_{J\bullet\bullet}
  \,N^{\bullet\bullet}
  +\mathrm{i}\omega_{K\bullet}\,  N^{\bullet\bullet} 
  F_{J\bullet\bullet} N^{\bullet K} \big]
  \nonumber\\ 
  & +2\mathrm{i}\alpha^2 \big(\omega_I -\mathrm{i}
  \alpha\,F_{I\bullet\bullet}\,N^{\bullet\bullet}\big)
  \mathcal{Z}^{IJ} \big[\mathrm{i}F_{JKLMN} -\tfrac43 F_{JKMP} N^{PQ} 
      F_{LNQ}\big] N^{KL} N^{MN}    \nonumber \\
 & -4\mathrm{i}\alpha^2 \big(\omega_I -\mathrm{i}
  \alpha\,F_{I\bullet\bullet}\,N^{\bullet\bullet}\big)
  \mathcal{Z}^{IJ}  F_{JPQ} N^{MP}  N^{NQ} N^{KL} 
  \,F_{KLMN}       \nonumber \\
 & +4\,\alpha^2 \big(\omega_I -\mathrm{i}
  \alpha\,F_{I\bullet\bullet}\,N^{\bullet\bullet}\big)
  \mathcal{Z}^{IJ}  F_{JRS}    N^{MR} N^{QS} N^{KN} N^{LP}  \,
   F_{KLM} F_{NPQ}     \nonumber \\
  &\, -2 \big(\omega_I -\mathrm{i}
  \alpha\,F_{I\bullet\bullet}\,N^{\bullet\bullet}\big)
  \mathcal{Z}^{IJ} \big(\omega_{JK} +\alpha F_{JMN}F_{KPQ} N^{MP}
  N^{NQ} 
  -\mathrm{i}\alpha
  F_{JK\bullet\bullet}N^{\bullet\bullet}\big)\nonumber\\
  &\qquad\qquad \times 
  \mathcal{Z}^{KL}  \big(\omega_L -\mathrm{i}
  \alpha\,F_{L\bullet\bullet}\,N^{\bullet\bullet}\big)  
   +\mathrm{h.c.} \Big\}  \nonumber \\
  & +4\,\alpha\big\{\Omega_I \,\mathcal{Z}^{IJ}\,F_{JKL} \,N^{KM}N^{LN}
  \,\Omega_{\bar M\bar N}  +\mathrm{h.c.}\big\} \nonumber\\
  & -4\,\alpha\,\Omega_I \,\mathcal{Z}^{IJ}\,F_{JKL} \,N^{KM}N^{LN}\,
  \bar F_{PMN} \,\bar{\mathcal{Z}}^{PQ} \,\Omega_{\bar Q}
  +\mathcal{O}(\alpha^4)\,. 
\end{align} 
The last two lines are not `almost harmonic'. The first of these two 
lines arises as a result of the non-holomorphic terms noted in
\eqref{eq:nonhol-mixed-Omega}, and the last line originates from the
manifestly non-harmonic term present at the end of the expression
\eqref{eq:Omega-tilde4-approx-3} (which has been included above upon
replacing $\Omega_{I\bar J}$ by the corresponding expression given in
\eqref{eq:ddd-Omega}). 

It is now straightforward to verify with the help of
\eqref{eq:transf-Omega3} that these two lines are precisely generated
upon assuming that $\Omega$ will contain a term
$-4\,\alpha\,\Omega_{IJ} \,\,N^{IK}N^{JL}\, \Omega_{\bar K\bar L}$ at
this order of iteration. This is quite a non-trivial result, because
we are not just rewriting the expression
\eqref{eq:Omega-tilde4-approx-3} that was originally expressed in
terms of $\Omega$ and its derivatives, into a similar expression!
Rather, as already mentioned, we have now extracted an explicit power
of $\alpha$, whereas so far the parameter $\alpha$ only appeared
implicitly in $\Omega$. This signals a new pattern that will become
more manifest shortly.

Using the previous results of the transformation rules of the function
$\omega$ and its derivatives, exhibited in \eqref{eq:trans-omega-ders}
and \eqref{eq:trans-omega-ders-second-order}, as well as the
transformation rules \eqref{eq:F-0-der} for multiple derivatives of the
holomorphic function $F$, one can, after a fair amount of non-trivial
manipulations, determine the expression for
$\Omega(\mathcal{Y},\bar{\mathcal{Y}})$, up to a symplectic function
of $\mathcal{Y}$ and $\bar{\mathcal{Y}}$, 
\begin{align}
  \label{eq:ansatz-nonholo-Omega-3}
  \Omega(\mathcal{Y},\bar{\mathcal{Y}}) =&\, \alpha\,\ln N
  \nonumber\\
  &+ \Big\{\omega +2\,\alpha\,N^{IJ}  \omega_{IJ} 
  -2\,\alpha \,N^{IJ} N^{KL} \,\big[
  \omega_{IK} \,\omega_{JL} -\alpha \,\omega_{IJKL} \big]\nonumber\\
  &\qquad + \tfrac83
  \mathrm{i} \alpha^2\, F_{IJK} N^{IL} N^{JM} N^{KN} \, \omega_{LMN}  \nonumber\\
  &\qquad-\alpha^2 \big[ \mathrm{i}F_{IJKL} -\tfrac23
  F_{IKM} F_{JLN}   N^{MN}  \big]   N^{IJ}N^{KL} \nonumber\\
  & \qquad +4\mathrm{i}\,\alpha^2 \,N^{PQ} \big(F_{PQIJ} +\mathrm{i}
  F_{PIM} N^{MN} F_{Q JN}\big) N^{IK}N^{JL} \,\omega_{KL}
  \nonumber\\
  & \qquad -\tfrac23 \alpha^3 N^{IJ} \big[ \mathrm{i} F_{IJKLMN} N^{KL}
  N^{MN}  - 4\, F_{IJKLM} \, N^{KN} N^{LP} N^{MQ} \, F_{NPO}  \big]
  \nonumber\\
  & \qquad  +2\,\alpha^3 \big[N^{IJ} F_{IJKL}  N^{KP} N^{LQ} F_{PQRS} \,
  N^{RS} \nonumber\\
  & \qquad\qquad\qquad\qquad\
  +\tfrac13 F_{IJKL} \, N^{IM} N^{JN} N^{KP} N^{LQ} \, F_{MNPQ} \big]
  \nonumber\\     
  &\qquad  +4\mathrm{i} \alpha^3 F_{IJKL} \big[ N^{IJ} N^{KM}N^{LN}F_{MPQ} N^{PR} N^{QS}
  F_{RSN}\nonumber\\
  &\qquad\qquad\qquad \qquad + N^{IM}N^{JN}N^{KP}N^{LQ}
  F_{MNR}N^{RS}F_{PQS} \big] \nonumber\\ 
  &\qquad  -2\,\alpha^3   F_{IJK} F_{LPQ} F_{RST} F_{UVW} N^{IL}
  N^{JP} N^{KR} N^{QU} N^{SV} N^{TW} 
  \nonumber\\
  &\qquad -\tfrac43\alpha^3  F_{IJK} F_{LPQ} F_{RST} F_{UVW} N^{IL}  N^{JR}
  N^{KU} N^{PS} N^{QV} N^{TW}   + \mathrm{h.c.}\Big\} \nonumber\\
 & - 4\,\alpha \, \Omega_{IJ} N^{IK} N^{JL} \Omega_{\bar K \bar L} 
 + \mathcal{O}(\alpha^4) \,.
\end{align}
It is clear that the terms that are independent of the holomorphic
function $\omega(\mathcal{Y})$ are becoming more and more numerous in
higher orders. Note that the above result is `almost harmonic', with
the exception of the last term. Furthermore the `almost harmonic'
terms take again the form of a sum over 1PI diagrams. 

In the limit $\alpha\to 0$ the expression
\eqref{eq:ansatz-nonholo-Omega-3} for $\Omega$ reduces to the original
harmonic expression that we started from initially in
\eqref{eq:harmonic}. With the exception of the last term in
\eqref{eq:ansatz-nonholo-Omega-3}, which will recombine with other
terms in due course, the almost harmonic terms have to be included
into the expression for the function that encodes the effective
action. Hence they imply that the original non-harmonic modification
$\ln N$ in \eqref{eq:ansatz-Omega1} is incomplete and must
be modified order-by-order by additional non-harmonic term. These
terms will thus contribute to the effective action, where they are
expected to encode non-local interactions associated with the massless
modes.

We remind the reader that $\Omega$ is not a symplectic function and
the next step is to determine the symplectic function
$\mathcal{H}^{(1)}$ in third order of $\alpha$, which follows upon
substitution of the above result for $\Omega$ into
\eqref{eq:Hesse-1}. Let us first concentrate on the terms that are not
`almost harmonic'. They originate from three different sources. First
there is the last term in \eqref{eq:ansatz-nonholo-Omega-3} (which
appears with an additional factor 4 in $\mathcal{H}^{(1)}$), then
there are explicit non-harmonic terms in the expression
\eqref{eq:Hesse-1} for $\mathcal{H}^{(1)}$, and finally there are the
non-holomorphic contributions in \eqref{eq:Omeg-derivative} that were
summarized in \eqref{eq:nonhol-mixed-Omega}, which induce
corresponding modifications in $\mathcal{H}^{(1)}$. These three
contibutions are
\begin{align}
  \label{eq:der-Omega-contr}
  & - 16\,\alpha \, \Omega_{IJ} N^{IK} N^{JL} \Omega_{\bar K \bar L} \nonumber\\
  &-16\, \alpha\,\Omega_I N^{IJ} F_{JMN} N^{MP} N^{NQ} \bar F_{PQK}
  N^{KL}\,\Omega_{\bar L} \nonumber\\
  & -16\,\mathrm{i}\alpha\,\Omega_I N^{IJ}F_{JKL} \,N^{KM}N^{LN}
  \,\Omega_{\bar M\bar N} +\mathrm{h.c.}  
\end{align}
and they combine into
\begin{align} 
  \label{eq:comb} 
   -16\,\alpha\,\big(\Omega_{IJ}
   +\mathrm{i} F_{IJM} N^{MP} \Omega_P 
  \big) N^{IK}\,N^{JL} 
  \big(\Omega_{\bar K\bar L} -\mathrm{i} \bar F_{KLN} N^{NQ}
  \Omega_{\bar{\mathrm{Q}}}  \big) \,,
\end{align}
which equals precisely $-16\,\alpha$ times the non-harmonic symplectic
function $\mathcal{G}_1 (\mathcal{Y},\bar{\mathcal{Y}})$ that has been
listed in appendix \ref{App:funct-H-a-i-geq2}, up to terms of order
$\Omega^3$.  The function $\mathcal{H}^{(1)}$ thus acquires the form
\begin{equation}
  \label{eq:almost-harm-G}
    \mathcal{H}^{(1)}(\mathcal{Y},\bar{\mathcal{Y}};N) =
    4\, \alpha\ln N   + 
   {h}(\mathcal{Y},\bar{\mathcal{Y}}) +
  \bar{h}(\bar{\mathcal{Y}},\mathcal{Y})  
  -16\,\alpha \, \mathcal{G}_1 (\mathcal{Y},\bar{\mathcal{Y}})
+ \mathcal{O}(\alpha^4) \,, 
\end{equation} 
where 
\begin{align}
  \label{eq:def-h-mod3}
    {h}(\mathcal{Y},\bar{\mathcal{Y}})=&\, 4\,\omega -
    4\,\big(\omega_I -\mathrm{i}\alpha\,F_{IKL} N^{KL}\big)\, N^{IJ}\, 
    \big( \omega_J   -\mathrm{i}\alpha\,F_{JMN} N^{MN}\big)  
    \nonumber\\
    &\, +8\,\alpha\, \omega_{IJ} N^{IJ} -8\,\alpha \,N^{IJ} N^{KL} \,\big[
  \omega_{IK} \,\omega_{JL} -\alpha \,\omega_{IJKL} \big]\nonumber\\
  &\,+ \tfrac{32}3
  \mathrm{i} \alpha^2\, F_{IJK} N^{IL} N^{JM} N^{KN} \, \omega_{LMN}  \nonumber\\
  &\, +16\mathrm{i}\,\alpha^2 \,N^{\bullet\bullet}
  \big(F_{\bullet\bullet IJ} +\mathrm{i} F_{\bullet IM} N^{MN}
  F_{\bullet JN}\big) N^{IK}N^{JL} \,\omega_{KL}
  \nonumber\\
  &\, -16\,\alpha \,\omega_{I\bullet\bullet} N^{\bullet\bullet}
  \,N^{IJ}\, \big(\omega_J-\mathrm{i}\alpha\, F_{J\bullet\bullet}
  N^{\bullet\bullet}\big)  \nonumber\\
  &\, +8\, \big(\omega_I-\mathrm{i}\alpha\,
  F_{I\bullet\bullet}N^{\bullet\bullet} \big) N^{IJ}
  \,\omega_{JK}\,N^{KL} \big(\omega_L-\mathrm{i}\alpha\,
  F_{L\bullet\bullet}N^{\bullet\bullet}\big)\nonumber\\
  &\, + \tfrac83 \mathrm{i} \, F_{IJK} N^{IL} N^{JM} N^{KN}
  \big(\omega_L-\mathrm{i}\alpha\,
  F_{L\bullet\bullet}N^{\bullet\bullet}\big) \nonumber\\
  &\,\qquad\qquad \times \big(\omega_M-\mathrm{i}\alpha\,
  F_{M\bullet\bullet}N^{\bullet\bullet}\big)
  \big(\omega_N-\mathrm{i}\alpha\,
  F_{N\bullet\bullet}N^{\bullet\bullet}\big) \nonumber\\
  &\, -16\mathrm{i}\,\alpha\, \big(\omega_I-\mathrm{i}\alpha\,
  F_{I\bullet\bullet}N^{\bullet\bullet}
  \big) N^{IJ} F_{JKL} \,N^{KM}N^{LN} \,\omega_{MN}  \, \nonumber\\
  &\, +8\mathrm{i}\,\alpha^2\, \big(\omega_I-\mathrm{i}\alpha\,
  F_{I\bullet\bullet}N^{\bullet\bullet}
  \big)\,N^{IJ}  \nonumber\\
  &\,\qquad\qquad \times \big(F_{JKL\bullet\bullet}
  +\tfrac43\mathrm{i}F_{JKM\bullet}\, N^{MN}
  F_{LN\bullet} \big)  \,N^{KL} N^{\bullet\bullet}    \, \nonumber\\
  &\,-8\mathrm{i}\,\alpha\, \big(\omega_I-\mathrm{i}\alpha\,
  F_{I\bullet\bullet}N^{\bullet\bullet} \big) N^{IJ} \big(
  F_{JK\bullet\bullet}+\mathrm{i} F_{JP\bullet}N^{PQ}
  F_{KQ\bullet}\big) N^{\bullet\bullet}\nonumber\\
  &\,\qquad\qquad \times N^{KL} \big(\omega_L-\mathrm{i}\alpha\,
  F_{L\bullet\bullet}N^{\bullet\bullet}\big)\nonumber\\
  &\, -16\,\alpha^2\, \big(\omega_I-\mathrm{i}\alpha\,
  F_{I\bullet\bullet}N^{\bullet\bullet}\big) \,
  N^{IJ} F_{JKL} \,N^{KM}N^{LN} \nonumber\\
  &\,\qquad\qquad\times \big(F_{MN\bullet\bullet}
  +\mathrm{i} F_{MP\bullet}N^{PQ} F_{NQ\bullet} \big) N^{\bullet\bullet}  \, \nonumber\\
  &\, - 4\,\alpha^2 \big[\mathrm{i}F_{IJKL}
    -\tfrac23 F_{IKM} F_{JLN}   N^{MN}  \big]   N^{IJ}N^{KL}\nonumber\\
  &\, -\tfrac83 \alpha^3 N^{IJ} \big[ \mathrm{i} F_{IJKLMN} N^{KL}
  N^{MN}  - 4\, F_{IJKLM} \, N^{KN} N^{LP} N^{MQ} \, F_{NPQ}  \big]
  \nonumber\\
  &\, +8\,\alpha^3 \big[N^{IJ} F_{IJKL}  N^{KP} N^{LQ} F_{PQRS} \,
  N^{RS} \nonumber\\
  &\qquad\qquad\qquad\qquad
  +\tfrac13 F_{IJKL} \, N^{IM} N^{JN} N^{KP} N^{LQ} \, F_{MNPQ} \big]
  \nonumber\\     
  &\, +16\mathrm{i}\,\alpha^3 F_{IJKL} \big[ N^{IJ} N^{KM}N^{LN}F_{MPQ} N^{PR} N^{QS}
  F_{RSN}\nonumber\\
  &\qquad\qquad\qquad\qquad  + N^{IM}N^{JN}N^{KP}N^{LQ}
  F_{MNR}N^{RS}F_{PQS} \big] \nonumber\\ 
  &\, -8\,\alpha^3   F_{IJK} F_{LPQ} F_{RST} F_{UVW} N^{IL}  N^{JP} N^{KR} N^{QU} N^{SV} N^{TW}
  \nonumber\\
  &\,-\tfrac{16}3\alpha^3  F_{IJK} F_{LPQ} F_{RST} F_{UVW} N^{IL}  N^{JR}
  N^{KU} N^{PS} N^{QV} N^{TW}   \nonumber\\
  &\, + \mathcal{O}(\alpha^4) \,.
\end{align}
As before the $\omega(\mathcal{Y})$ will transform holomorphically such
that its explicit transformation rule follows from
\eqref{eq:def-h-mod3} upon making the substitution $N_{IJ}\to
\mathrm{i} \mathcal{Z}^{IJ}$. We have verified by explicit calculation
that this is indeed the case, which provides an explicit check on the
calculation.

At this point one can again determine the holomorphic anomaly equation
following the same steps as before. As it turns out the result
coincides with \eqref{eq:nonholo-derivative-h}, but now valid up to order
$\alpha^4$. This can be seen as an indication that the holomorphic
anomaly equation will not acquire further corrections in higher
orders.

\section{Summary and conclusions}
\label{sec:summary-conclusions}
\setcounter{equation}{0}
Based on the observation that the duality transformations act
differently on the function that encodes the effective action than on
the topological free energy, we have proposed a conceptual framework
based on the Hesse potential of real special geometry to understand
the relation between the two. Subsequently we have studied the Hesse
potential by iteration for a generic effective action, first starting
from a Wilsonian effective action and subsequently by considering the
effect of non-harmonic deformations. The Hesse potential decomposes
into an infinite series of symplectic functions and we established
that the topological string free energy could reside in precisely one
of them. This function is then subject to the holomorphic anomaly
equation, irrespective of its dynamical content.

The results of an explicit iteration of the genus $g\leq3$ topological
string free energy fully confirms the correctness of the proposal. We
should again stress that we concentrate on the generic features of this
relationship, rather than on specific models. The relations that we
find are thus universal, but it is not assumed that the
resulting expression for the topological string free energy will have
an actual realization as a topological string model. This is the
reason that we do not make contact with specific aspects of the
topological string, such as the wave function approach and the issue
of background dependence \cite{Witten:1993ed}. 

One implication of our result is that we are also able to relate
the non-holomorphic terms associated to the effective action to the
ones that appear in the topological string free energy. This is
perhaps not so surprising in view of the fact that there is a 
qualitative relation between the pinching of a cycle that decreases the
genus of the Riemann surface in the topological string and the
integration over massless modes in the effective action! But it is
important to realize that, while our construction demonstrates how to
construct the topological string free energy from a given effective
action, the inverse is clearly not possible because the effective
action is equivalent to the full Hesse potential, while the topological
string free energy constitutes only part of the Hesse potential.

At several occasions we already mentioned that the results of this
paper are consistent with our previous work
\cite{Cardoso:2008fr,Cardoso:2010gc}, where we analyzed the same
issues by using a variety of different strategies. It is therefore of
interest to compare the present results with the results of the
past. To highlight some interesting issues we therefore reconsider the
earlier results on the FHSV model, which were based on imposing the
exact S- and T-dualities of this model on the effective action. There
we used a slightly different perturbative procedure and we worked in a
parametrization based on special coordinates.  Subsequently we
determined the Hesse potential by iteration, in a way that is similar
to what was done in the present paper. We then discovered that the
Hesse potential did indeed contain terms that cannot belong to the
topological string free energy at genus-2, because they do not
depend (anti-)holomorphically on the topological string coupling. As
we now know, those are the contributions that do not belong to the
function $\mathcal{H}^{(1)}$, but at that stage such a systematic
classification was not available. Nevertheless, the terms that did
depend (anti-)holomorphically on the topological string coupling were
consistent with the results obtained from the FHSV topological string
\cite{Grimm:2007tm}, except that the proportionality constant remained
ambiguous in view of the fact that the corresponding expression was
duality invariant, so that its leading contribution could be changed
by a corresponding change in the effective action where we had only
imposed the requirement of invariance.  Interestingly, the present
approach which emphasizes covariance rather than invariance, clarifies
this result. To appreciate all this we have summarized some of details of
the derivation of the genus-2 FHSV topological string free energy in
appendix \ref{App:specific-model}.

Finally we wish to return to the issue of BPS black hole entropy in
supersymmetric theories with eight supercharges, which formed a major
motivation for the present work. In
\cite{LopesCardoso:1998wt,LopesCardoso:2000qm} a general formula for
BPS black hole entropy was given based on Wald's definition of black
hole entropy \cite{Wald:1993nt}, which was covariant under dualities
and incorporated the higher-derivative corrections to the Weyl
multiplet that we already referred to in section
\ref{sec:introduction}. (Incidentally, there is now increasing
evidence that other higher-derivative couplings will not contribute to
BPS black hole entropy by virtue of certain non-renormalization
theorems \cite{deWit:2010za,Butter:2013lta}.) The formula of
\cite{LopesCardoso:1998wt} was reinterpreted in \cite{Ooguri:2004zv}
in terms of a mixed partition function which was subsequently related
to the topological string. However, this relationship depended
crucially on the assumption that the topological free energy and the
function that encodes the supergravity action are directly related, or
perhaps even identical! As we have been trying to emphasize in this
paper, the topological string does capture certain string amplitudes
that should also follow from the effective action. But this does not
imply that the topological string and the action are given by the same
function.

We should perhaps add here that it is possible to present the
supergravity input in the form of the Hesse potential (analogous to
converting a Lagrangian into a Hamiltonian description), for which one
can define a modified black hole partition function associated with
the canonical ensemble \cite{LopesCardoso:2006bg}. This would offer an
effective way to make contact with the topological string, were it not
for the fact that the black hole solutions from which one starts in
supergravity are, by definition, solutions of the full effective
action. Therefore they should involve the full Hesse potential, which,
as we have shown in this paper, consists of an infinite series of
symplectic functions of which just one will correspond to the
topological string free energy. Finally we note that the work of this
paper pertains specifically to theories with eight supercharges, while
a substantial part of the literature on BPS black holes is based on
theories with sixteen supercharges (although often treated in a
reduction to eight supercharges). An extension of the work of this
paper for theories to sixteen supersymmetries should therefore be of
interest.

\subsection*{Acknowledgements}
We acknowledge helpful discussions with Murad Alim, Michele Cirafici,
Edi Gava, Thomas Grimm, Babak Haghighat, Albrecht Klemm, Thomas
Mohaupt, Kumar Narain, Hirosi Ooguri, \'Alvaro Osorio, Ashoke Sen,
Samson Shatashvili, Marcel Vonk, Edward Witten and Maxim Zabzine. The
work of G.L.C. is partially funded by Funda\c{c}\~{a}o para a
Ci\^{e}ncia e a Tecnologia (FCT/Portugal) through project
PEst-OE/EEI/LA0009/2013 and through the grants PTDC/MAT/119689/2010
and EXCL/MAT-GEO/0222/2012. The work of B.d.W. is supported by the ERC
Advanced Grant no. 246974, {\it ``Supersymmetry: a window to
  non-perturbative physics''}. This work is also supported by the COST
action MP1210 {\it ``The String Theory Universe''}. \\
We thank our respective institutes for hospitality during the course
of this work. S.M. thanks the Alexander von Humboldt Stiftung for a
reinvitation grant which enabled her to visit the Max Planck Institut,
M\"unchen, and Dieter L\"ust for offering hospitality. We also thank
the Max Planck Institut f\"ur Gravitationsphysik
(Albert-Einstein-Institute) for hospitality extended to us during the
completion of this work.
 

\begin{appendix}
\section{Non-holomorphic deformation of special geometry}
\label{App:theorem}
\setcounter{equation}{0}
In this appendix we prove the following theorem.\\[.5ex]
{\sf A.1. Theorem}:\\[.1ex]
Given a Lagrangian $\mathcal{L}(\phi,\dot \phi)$ depending on $n$
coordinates $\phi^i$ and $n$ velocities $\dot \phi^i$, with
corresponding Hamiltonian $\mathcal{H}(\phi,\pi)= \dot \phi^i\, \pi_i
-\mathcal{L}(\phi,\dot \phi)$, there exists a description in terms of
complex coordinates $x^i=\tfrac12(\phi^i+\mathrm{i}\dot \phi^i)$ and a
complex function $F(x,\bar x)$, such that,
\begin{align}
  \label{eq:theorem-prop}
  2\, \mathrm{Re}\,x^i =&\, \phi^i\,,\nonumber\\
    2\, \mathrm{Re}\,F_i(x,\bar x) =&\, \pi_i\,, \quad\mbox{where}\;
    F_i = \frac{\partial F(x,\bar x)}{\partial x^i}\,.
\end{align}
The function $F(x,\bar x)$ is defined up to an anti-holomorphic function
and can be decomposed into a holomorphic and a purely imaginary
non-harmonic function,
\begin{equation}
  \label{eq:F(x)}
  F(x,\bar x) = F^{(0)}(x) + 2\mathrm{i} \Omega(x,\bar x)\,.
\end{equation}
The equivalence transformations take the form, 
\begin{equation}
  \label{eq:ambiguity}
  F^{(0)}\to F^{(0)} + g(x)\,, \qquad \Omega\to \Omega- \mathrm{Im}\,
  g(x)\,, 
\end{equation}
which results in $F(x,\bar x)\to F(x,\bar x)+\bar g(\bar x)$.\\
The Lagrangian can then be expressed in terms of $F$ and $\Omega$,
\begin{equation}
  \label{eq:L-sympl}
  \mathcal{L}= 4 [\mathrm{Im}\, F -\Omega] \,,
\end{equation}
so that the Hamiltonian takes the form 
\begin{equation}
  \label{eq:H-sympl}
  \mathcal{H} =-4\, \big[\mathrm{Im}\, F -\Omega\big] + 2\, \pi_i \,
  \mathrm{Im}\,x^i  \,.
\end{equation}
This expression is identical to the expression for the Hesse
potential given in \eqref{eq:GenHesseP}, up to an overall minus
sign. Alternatively the Hamiltonian can be written as
\begin{equation}
  \label{eq:alt-expression-hesse}
  \mathcal{H} =
  -\mathrm{i}(x^i \,\bar F_{\bar\imath} -\bar x^{\bar\imath} \,F_i)
  -4\,\mathrm{Im} 
  [F^{(0)}-\tfrac12 x^i\,F^{(0)}_i] -2(2\,\Omega
  -x^i\Omega_i -\bar x^{\bar\imath} \Omega_{\bar\imath}) \,,
\end{equation}
where $F_i=\partial F/\partial x^i$, $\bar F_{\bar\imath}=\partial\bar
F/\partial \bar x_{\bar\imath}$, and similarly for the functions $F^{(0)}$ and
$\Omega$. When the function $F^{(0)}(x)$ is homogeneous of second degree,
the second term will vanish. The third term is a measure of the
deviation from homogeneity of $\Omega$. This decomposition is known
from the entropy function for BPS black holes
\cite{LopesCardoso:2006bg}. 

Furthermore, the $2n$-vector $(x^i,F_i)$ turns out to define a
complexification of the phase-space coordinates $(\phi^i,\pi_i)$ that
transforms precisely as $(\phi^i,\pi_i)$ under canonical (symplectic)
reparametrizations, 
\begin{equation}
  \label{eq:symplectic}
  \begin{pmatrix} 
    x^i\\[2mm] F_i(x,\bar x)
  \end{pmatrix}
  \longrightarrow
  \begin{pmatrix} 
    \tilde x^i\\[2mm] \tilde F_i(\tilde x,\bar{\tilde x})
  \end{pmatrix}
  =
  \begin{pmatrix} 
    U^i{}_j& Z^{ij}\\[2mm]  W_{ij} & V_i{}^j 
  \end{pmatrix}
  \begin{pmatrix} 
    x^j\\[2mm] F_j(x,\bar x)
  \end{pmatrix}\;,
\end{equation}
where the real matrix is an element of
$\mathrm{Sp}(2n,\mathbb{R})$. Observe that for the real part of the
vector $(x^i,F_i)$, the above transformation is the standard canonical
transformation on coordinates and momenta. The equation
(\ref{eq:symplectic}) is integrable so that the symplectic transformation
leads to new functions $\tilde F^{(0)}$ and $\tilde\Omega$. \\[1ex]
{\sf A.2. Proof:}\\[.1ex]
The proof of this theorem proceeds as follows. First note the following
complex vectors, 
\begin{equation}
  \label{eq:def-x-y}
  x^i = \tfrac12\Big(\phi^i +
  \mathrm{i}\frac{\partial\mathcal{H}}{\partial \pi_i} \Big)\,,\qquad
  y_i = \tfrac12\Big(\pi_i -
  \mathrm{i}\frac{\partial\mathcal{H}}{\partial \phi^i} \Big)\,,
\end{equation}
constructed out of two canonical pairs, one comprising the
variables $\phi^i,\pi_i$ and the other one the derivatives of the
Hamiltonian, which transform in the same way under canonical
transformations (here we use that the Hamiltonian transforms as a function under
canonical transformations). 

In view of the inverse Legendre relation, $\dot \phi^i
= \partial\mathcal{H}/\partial \pi_i$, the complex $x^i$
in~\eqref{eq:def-x-y} coincide with the $x^i$ defined
previously. Furthermore, when writing the Lagrangian as a function of
the $x^i$ and $\bar x^{\bar\imath}$, it follows that
\begin{equation}
  \label{eq:x-derivative-L}
   \frac{\partial\mathcal{L}(x,\bar  x)}{\partial x^i} = -2\mathrm{i} y_i\,.  
\end{equation}
Here we used that the Legendre transformation leading to the
Hamiltonian yields $\partial\mathcal{L}/\partial\phi^i =
-\partial\mathcal{H}/\partial \phi^i$ (where on the right-hand side
$\pi_i$ is kept constant and on the left-hand side $\dot\phi^i$ is
kept constant). Observe that we did not make use of the equations of
motion.

Subsequently we write $\mathcal{L}$ as the sum of a harmonic and a
non-harmonic function,
\begin{equation}
  \label{eq:n-n-harmonic-L}
  \mathcal{L}= -2\mathrm{i}\big[F^{(0)}(x)-\bar F^{(0)}(\bar x)\big]
  +4\, \Omega(x,\bar x) = 4 [\mathrm{Im}\, F -\Omega] \,,
\end{equation}
so that \eqref{eq:x-derivative-L} reads
\begin{equation}
y_i=\frac{\partial}{\partial x^i} \big[ F^{(0)}(x) + 2\mathrm{i} \Omega(x,\bar x) \big] \;.
\end{equation}
Thus $y_i=\partial_i F(x,\bar
x)$ with $F(x,\bar x) = F^{(0)}(x) + 2\mathrm{i} \Omega(x,\bar x)$, up
to an arbitrary anti-holomorphic function, and ${\rm Re}\, y_i = \pi_i$. The Hamiltonian then
follows from \eqref{eq:H-sympl}, which leads to the expression
\eqref{eq:alt-expression-hesse}. Hence we now have shown that
$(x^i,F_i)$ equals the vector $(x^i,y_i)$ which transforms under
canonical transformations according to \eqref{eq:symplectic}.

What remains to be proven is that the result of the transformation
\eqref{eq:symplectic} is integrable. The vector $(x^i,y_i)$ transforms
according to \eqref{eq:symplectic} into a vector $(\tilde x^i,\tilde
y_i)$ while the Hamiltonian, which depends on $(x^i+\bar x^i, y_i
+\bar y_i)$, transforms as a function under canonical transformations,
so that $\tilde{\mathcal{H}}(\tilde x^i+\bar{\tilde x}^i, \tilde
y_i+\bar{\tilde y}_i) = \mathcal{H}(x^i+\bar x^i, y_i +\bar y_i)$. The
dual quantities $(\tilde x^i,\tilde y_i)$ and the new Hamiltonian
$\tilde{\mathcal{H}}$ will satisfy the same relation as the original
quantities. The new Lagrangian $\tilde{\mathcal{L}}$ which follows
from an inverse Legendre transformation of the new Hamiltonian, will depend on
$\tilde x^i$ and $\bar{\tilde x}^i$. Applying the same steps as before
we then find the new function $\tilde F(\tilde x,\bar{\tilde x})$. 

There is one subtlety here and that is that the decomposition of the
function $F$ into $F^{(0)}$ and $\Omega$ is ambiguous. The ambiguity
is resolved by noting that the symplectic transformation
\eqref{eq:symplectic} can also be applied to the the vector $(x^i,
F_i{}^{(0)}(x))$.  In that case the new function $F^{(0)}$ can be
determined separately, as the holomorphic case is known to be
integrable, and it is given in \eqref{eq:new-F}, up to a constant and
terms linear in $\tilde x^i$.  The latter terms can be determined
explicitly, for instance, by using that $F^{(0)} -\tfrac12 x^i
F_i{}^{\!(0)}$ transforms as a function under duality. Having
determined the functions $\tilde F^{(0)}$ and $\tilde F$, the
non-harmonic function $\tilde\Omega$ follows. This completes the proof
of the theorem. \\[1ex] 
{\sf A.3. Corollary:}\\[.1ex]
Let us derive the well-known result (see, for instance,
\cite{Gaillard:1981rj}) that the first-order derivative of the Lagrangian
with respect to some parameter (such as a coupling constant) transforms
as a function under symplectic transformations
\eqref{eq:symplectic}. We denote this parameter by $g$ and note
that $\partial_g\mathcal{H}(\phi,\pi;g)$ transforms as a function
under canonical transformations (which do not depend on $g$, but they
act on $g$-dependent quantities as shown in \eqref{eq:symplectic}) for
any value of $g$. Subsequently, take the derivative of the Hamiltonian
with respect to $g$ keeping $\phi^i$ and $\pi_i$ fixed. Consequently
one derives
\begin{align}
  \label{eq:2ham-lambda}
  \frac{\partial\mathcal{H}(\phi,\pi;g)}{\partial g} =&\,
  \Big(\pi_i 
  -  \frac{\partial\mathcal{L}(\phi,\dot\phi;g)}{\partial\dot\phi^i}
  \Big)  \,\frac{\partial\dot\phi^i}{\partial g}
  - \frac{\partial\mathcal{L}(\phi,\dot\phi;g)}{\partial g}
  = - \frac{\partial\mathcal{L}(\phi,\dot\phi;g)}{\partial g}
  \,,
\end{align}
which proves the assertion. 

\section{The symplectic functions $\boldsymbol{\mathcal{H}^{(a)}_i}$
  for $\boldsymbol{a\geq2} $ and some other functions that do not
  initially appear in $\boldsymbol{\mathcal{H}}$ }
\label{App:funct-H-a-i-geq2}
\setcounter{equation}{0}
Here we collect the explicit results for various functions
$\mathcal{H}^{(a)}_i$ that appear in \eqref{eq:Hesse-decomp}. These
functions have been determined by iteration in orders of $\Omega$ and
its derivatives. We present the terms of the iterative expansion up to
$\mathcal{O}(\Omega^5)$. At the end of this appendix we will be
presenting two more functions, $\mathcal{G}_1$ and $\mathcal{G}_2$,
that did not initially show up in the iterative procedure for the
Hesse potential carried out in this paper. These will only be given up
to order $\Omega^3$. Note that here and elsewhere we only use indices
$\bar I,\bar J,\ldots$ when they are necessary. For instance, we will
write $F_{IJK}$ and $\bar F_{IJK}$ because $F$ is holomorphic and
$\bar F$ is anti-holomorphic, so that there is no need for using
holomorphic or anti-holomorphic indices, whereas for the derivatives
of the real quantity $\Omega$ we write $\Omega_I$ and
$\Omega_{\bar{I}}$ to distinguish holomorphic and anti-holomorphic
derivatives. The reason is that $N^{IJ}$ has no unique assignment of
(anti)holomorphic indices, so that there will never be a consistent
pattern of contractions based on holomorphic and anti-holomorphic
indices. Denoting $(N\Omega)^I \equiv N^{IJ} \, \Omega_J$ and $(N
\bar\Omega)^I \equiv N^{IJ} \, \Omega_{\bar J}$, we have obtained the
following expressions, 
\begin{align}
  \label{eq:H-from-2}
  \mathcal{H}^{(2)}=&\,8\, N^{IJ}\Omega_I \Omega_{\bar J}
  -16\,\big[\Omega_{IJ} 
  (N\bar\Omega)^I (N\Omega)^J +\,\Omega_{I\bar J} (N\bar\Omega)^I
  (N\bar\Omega)^J + {\rm h.c.} \big] \nonumber\\
  &\, - 8 \mathrm{i} \,\big[F_{IJK} (N\bar\Omega)^I (N\Omega)^J
  (N\Omega)^K - {\rm h.c.} \big] \nonumber\\
  &\, + \tfrac{16}3 \mathrm{i} \big[ \left( F_{IJKL} + 3 \mathrm{i}
    F_{IJM} N^{MN} F_{NKL} \right) \, (N\Omega)^I (N\Omega)^J
  (N\Omega)^K (N \bar \Omega)^L - {\rm h.c.} \big]
  \nonumber\\
  &\, + 16\, \big[  \Omega_{IJK} \,  (N\Omega)^I (N\Omega)^J (N \bar
  \Omega)^K + {\rm h.c.} \big] \nonumber\\ 
  &\,  + 16\, \big[ \left( \Omega_{IJ\bar K} + \mathrm{i} F_{IJM} N^{MN}
    \Omega_{N \bar K} \right) \, \left( (N\Omega)^I (N\Omega)^J (N
    \Omega)^K + 2 (N\Omega)^I (N \bar \Omega)^J (N \bar \Omega)^K
  \right)    \big] \nonumber\\
 &\,  + 16\, \big[ \left( \Omega_{\bar I\bar J K} - \mathrm{i} \bar
   F_{IJM} N^{MN}
    \Omega_{\bar N K} \right) \, \left( (N\bar\Omega)^I
    (N\bar\Omega)^J (N \bar\Omega)^K + 2 (N\bar\Omega)^I (N \Omega)^J
    (N \Omega)^K  \right) \big] \nonumber\\
  &\, + 32\, \Big[\Omega_{IK} \, N^{KL} \, \Omega_{LJ} \, (N \Omega)^I (N
  \bar \Omega)^J + {\rm h.c.} \Big]
  \nonumber\\
  &\, + 32\, \Omega_{IK} \, N^{KL} \, {\Omega}_{\bar L \bar J} \, (N
  \Omega)^I (N \bar \Omega)^J
  \nonumber\\
  & \, + 16 \mathrm{i} \Big[ F_{IJK} \, N^{KL} \, \Omega_{LM} \, \left(
    (N \Omega)^I (N \Omega)^J (N \bar \Omega)^M + 2 (N \Omega)^M (N
    \Omega)^I (N \bar \Omega)^J \right) - {\rm h.c.} \Big]
  \nonumber\\
  & \, + 16 \mathrm{i} \Big[ F_{IJK} \, N^{KL} \, \Omega_{\bar L \bar
    M} \, (N \Omega)^I (N \Omega)^J (N \bar \Omega)^M - {\rm h.c.}
  \Big]
  \nonumber\\
  & \, +8\, (N \Omega) ^I \, (N \Omega)^J \, F_{IJM} N^{MN} {\bar
    F}_{NKL} (N \bar \Omega)^K (N \bar \Omega)^L 
  \nonumber\\
   & \, + 32 \Big[(N \Omega)^I \, \Omega_{IJ} \, N^{JK} 
   \Omega_{K \bar L} (N \Omega)^L + {\rm h.c.} \Big] \nonumber\\
  & + 32 \Big[(N \bar \Omega)^I \, \Omega_{IJ} \, N^{JK} 
  \Omega_{K \bar L} (N \bar \Omega)^L + {\rm h.c.} \Big] \nonumber\\
   & + 32 \Big[(N \Omega)^I \, \Omega_{IJ}\,  N^{JK} \Omega_{\bar K L} 
   (N \Omega)^L + {\rm h.c.} \Big] \nonumber\\
   & + 16 \mathrm{i} \Big[ (N\Omega)^I (N\Omega)^J F_{IJK} N^{KL} 
   \Omega_{\bar LM}  (N\Omega)^M - {\rm h.c.} \Big]   \nonumber\\
   &\, + 32 \Big[(N \Omega)^I  \Omega_{I\bar J} \,  N^{JK} \, 
   \Omega_{\bar K L}\, (N \bar\Omega)^L + {\rm h.c.} \Big] \nonumber\\
 &\, + 32 \Big[(N \Omega)^I  \Omega_{I\bar J} \,  N^{JK} \, 
   \Omega_{K \bar L} 
   \, (N \bar \Omega)^L  \Big]   \;,      \\[.4ex]
   \mathcal{H}^{(3)}_1=&\,- \tfrac83\mathrm{i} F_{IJK}
   (N\bar\Omega)^I(N\bar\Omega)^J(N\bar\Omega)^K \nonumber\\
   &\, +8 \mathrm{i}\, F_{IJK} (N\bar\Omega)^I (N\bar\Omega)^J \,
   N^{KL}  \nonumber\\
   &\,\quad \times\left[2 \Omega_{\bar L \bar M} (N \bar{\Omega})^M +
     2\, \Omega_{\bar LM} (N \Omega)^M - \mathrm{i}  \bar{F}_{LMN} (N
     \bar{\Omega})^M (N \bar{\Omega})^N \right]  \,,   \\[.4ex] 
  \mathcal{H}^{(3)}_2=&\, 
  8\, \left(\Omega_{IJ} + \mathrm{i} F_{IJK} (N
    \Omega)^K \right) (N \bar{\Omega})^I (N \bar{\Omega})^J 
      \nonumber\\
     &\, - \tfrac83 \mathrm{i}  \left(F_{IJKL} + 3 \mathrm{i}
       F_{M(IJ} N^{MN} F_{KL)N} \right)    \nonumber\\
     &\,\qquad \times 
          \big[ 3 (N \Omega)^I (N \Omega)^J (N \bar \Omega)^K (N \bar
            \Omega)^L
            - 2 (N \bar \Omega)^I (N \bar \Omega)^J (N \bar \Omega)^K
            (N \Omega)^L  \big] \nonumber\\
          &\, - \tfrac{16}3 \, \Omega_{IJK} \left( 3\,(N
              \bar{\Omega})^I (N \bar{\Omega})^J (N \Omega)^K - (N
              \bar{\Omega})^I (N \bar{\Omega})^J (N \bar \Omega)^K
            \right)  \nonumber\\
          &\, - 16\,\Omega_{IJ\bar K} (N \bar{\Omega})^I (N
            \bar{\Omega})^J (N \bar \Omega)^K \nonumber\\
          &\ - 16 \mathrm{i} \,F_{IJK} N^{KL} \,\Omega_{LM}
          \nonumber\\
          &\,\qquad \times  \big[
          - (N \bar \Omega)^I ( N \bar \Omega)^J (N \bar \Omega)^M
          + (N \bar \Omega)^I ( N \bar \Omega)^J (N \Omega)^M + 2 (N
          \bar \Omega)^I ( N
          \Omega)^J (N \bar \Omega)^M \big]  \nonumber\\
          & \, - 16 \, (N \bar \Omega)^I \, \Omega_{IJ} \, N^{JK}
          \Omega_{KL} \, (N \bar \Omega)^L \nonumber\\
          & \, - 32 \, (N \Omega)^I \, \left( \Omega_{IJ} + \mathrm{i}
            F_{IJK} (N \Omega)^K \right) \, N^{JL} \left( \Omega_{\bar
              L \bar M} - \mathrm{i} {\bar F}_{LMN} (N \bar \Omega)^N
          \right) \, (N \Omega)^M
          \nonumber\\
          & \, + 16 \mathrm{i} (N \Omega)^I (N \Omega)^J F_{IJK}N^{KL}
          \left({\Omega}_{\bar L \bar M} - \mathrm{i} {\bar F}_{LMN}
            (N \bar \Omega)^{\bar N} \right) (N \Omega)^M
          \nonumber\\
          & \, - 16\, (N \Omega)^I \Omega_{\bar I J} N^{JK}
          \Omega_{K \bar M} (N \Omega)^M  \nonumber\\
          & \, - 32 \, (N \bar \Omega)^I \left( \Omega_{IJ} +
            \mathrm{i} F_{IJK} (N \Omega)^K \right) N^{JL}
          \Omega_{\bar L M} (N \Omega)^M \nonumber\\
          & - 16 \mathrm{i} \,( N \bar \Omega)^I \, (N \bar \Omega)^J
          F_{IJK} N^{KL} \Omega_{L \bar M} (N
          \bar\Omega)^M  \;,  \\[.4ex]
  \mathcal{H}^{(3)}_3=&\,- 16\,\Omega_{I\bar J} (N\bar\Omega)^I (N
  \Omega)^J  \nonumber\\ 
   &\,+ 16\,\Big[ 2 (N \bar{\Omega})^I (N \Omega)^J \left(\Omega_{IK} 
     N^{KL} \Omega_{L \bar J}  + \Omega_{I \bar J M} (N \Omega)^M
   \right) \nonumber\\ 
  &\, \qquad + (N \bar{\Omega})^I \Omega_{I \bar J} N^{JK}
  \left( \mathrm{i} F_{KLM} (N \Omega)^L (N \Omega)^M + 2 \Omega_{KL}
    (N \Omega)^L + 2 \Omega_{K \bar L} (N \bar{\Omega})^L \right)
  \nonumber\\
   & \qquad + 2 \mathrm{i} (N \bar \Omega)^I (N \Omega)^J F_{IJK}
   N^{KL} \Omega_{L \bar M} (N \Omega)^M  + {\rm h.c.} \Big]\,,
   \\[.4ex] 
  \mathcal{H}^{(4)}_1=&\, 32 \, (N \bar \Omega)^I \left(
    \Omega_{IJ} + \mathrm{i} {F}_{IJK} (N \Omega)^K \right)
  N^{JL} \left( \Omega_{\bar L \bar M} - \mathrm{i} {\bar
      F}_{\bar L \bar M \bar N} (N \bar \Omega)^N
  \right) (N \Omega)^M \,,  \\[.4ex]
  \mathcal{H}^{(4)}_2=&\, 32 \, (N \Omega)^I \,
  \Omega_{\bar{I} J} \, N^{JK} \, \Omega_{\bar{K} L}
  \, (N \bar \Omega)^L  \\[.4ex]
  \mathcal{H}^{(4)}_3=&\, 8 \, F_{IJM} N^{MN} {\bar F}_{NKL} \, (N
  \bar \Omega)^I (N \bar \Omega)^J (N 
  \Omega)^K (N \Omega)^L \,, \\[.4ex]
  \mathcal{H}^{(4)}_4=&\, - \tfrac43 \mathrm{i} \left(
    F_{IJKL} + 3 \mathrm{i} \, F_{MIJ} N^{MN} F_{KLN} \right)
  (N \bar \Omega)^I (N \bar \Omega)^J (N \bar \Omega)^K (N
  \bar \Omega)^L \,,  \\[.4ex]
  \mathcal{H}^{(4)}_5=& \, - 16 \mathrm{i} \, F_{IJK} N^{KQ} \Omega_{\bar QL} (N 
  \bar \Omega)^L \,(N \bar \Omega)^I (N \bar \Omega)^J \,,
  \\[.4ex] 
  \mathcal{H}^{(4)}_6=&\, - 16 \mathrm{i}\, F_{IJK} N^{KL}
  \left( \Omega_{\bar L \bar M} - \mathrm{i} {\bar
      F}_{LMN} (N \bar \Omega)^N \right)
  (N \bar \Omega)^I (N \bar \Omega)^J (N  \Omega)^M \,,
  \\[.4ex] 
  \mathcal{H}^{(4)}_7=&\, 16 \,\left( \Omega_{IJ\bar K} +
    \mathrm{i} F_{IJL} \, N^{LM} \, \Omega_{M \bar K}\right)
  \, (N \bar \Omega)^I (N \bar \Omega)^J (N \Omega)^K \,,
  \\[.4ex] 
    \mathcal{H}^{(4)}_8 =&\,
    32 \, (N \bar \Omega)^I 
    \left( \Omega_{IJ} + \mathrm{i} {F}_{IJK}  (N \Omega)^K \right) N^{JL} 
    \Omega_{\bar LM} \,  (N \bar \Omega)^M \,,
    \\[.4ex] 
   \mathcal{H}^{(4)}_9  =&\,
   - 16 \mathrm{i} \, (N  \bar \Omega)^I  (N  \bar \Omega)^J F_{IJK}
   N^{KL} \Omega_{\bar LM} (N  \bar \Omega)^M     \;.
\end{align}

As indicated above there are also other functions that do not
initially appear in $\mathcal{H}$. We give two examples below up to
terms of order $\Omega^3$.
\begin{align}
  \label{eq:G-fcts}
  \mathcal{G}_1 =&\, \big(\Omega_{IJ} +\mathrm{i} F_{IJM} N^{MP} \Omega_P
  \big) N^{IK}\,N^{JL} 
  \big(\Omega_{\bar K\bar L} -\mathrm{i} \bar F_{KLN} N^{NQ}
  \Omega_{\bar{\mathrm{Q}}}  \big) \,,\\ 
  \mathcal{G}_2 =&\, \Omega_{I\bar J} N^{IL}\,N^{JK} \Omega_{K \bar L}
   \,. 
\end{align}
Note that the functions $\mathcal{G}_{1,2}$ take the form of 1PI
connected diagrams, whereas the functions $\mathcal{H}^{(0)}_i$ do
not. 

\section{Transformation rules of $\boldsymbol{\omega_I}$ and
  $\boldsymbol{\omega_{IJ}}$ to order $\boldsymbol{\alpha^2}$ }
\label{App:transformation-omega}
\setcounter{equation}{0}
In this appendix we list the transformation rules of some of the
derivatives of the function $\omega$. For the first four multiple
derivatives those were already given in \eqref{eq:trans-omega-der-1}
and \eqref{eq:trans-omega-ders} to order $\alpha$. However, the
transformation rule of $\omega$ itself is known to order
$\alpha^2$ (cf. \eqref{eq:trans-omega-2}) so that also the derivatives
can be determined in that order. In section
\ref{sec:third-order-contr} we in fact need the transformation rules
for $\omega_I$ and $\omega_{IJ}$ to order $\alpha^3$. In view of their
length we display these transformations in this appendix. The results
read as follows,
\begin{align}
  \label{eq:trans-omega-ders-second-order}
  \tilde\omega_I =&\, [\mathcal{S}^{-1}]^J{}_I \,\Big[ \omega_J
  +\alpha \,F_{JKL} \mathcal{Z}^{KL} +2\mathrm{i} \alpha
  \,\omega_{JKL} \, \mathcal{Z}^{KL}      \nonumber\\
  &\, \qquad\qquad - 2\mathrm{i} \big(\omega_{J\bullet} + \alpha
  \,F_{J\bullet KL} \mathcal{Z}^{KL} \big)
  \,\mathcal{Z}^{\bullet\bullet} \big(\omega_\bullet 
  +\alpha \,F_{\bullet MN} \mathcal{Z}^{MN}\big)  \nonumber\\
  &\,\qquad\qquad +\mathrm{i} \big(\omega_\bullet +\alpha \,F_{\bullet KL}
  \mathcal{Z}^{KL}\big) \mathcal{Z}^{\bullet\bullet}\,
  F_{J\bullet\bullet} \,\mathcal{Z}^{\bullet\bullet}
  \big(\omega_\bullet
  +\alpha \,F_{\bullet MN} \mathcal{Z}^{MN}\big) \nonumber\\
    &\,\qquad\qquad -2\mathrm{i} \alpha\mathcal{Z}^{K\bullet}
  \,F_{J\bullet\bullet}\, \mathcal{Z}^{\bullet L}\, \big[\omega_{KL}
  - F_{KL\bullet} \,\mathcal{Z}^{\bullet\bullet} \big(\omega_\bullet
  +\alpha \,F_{\bullet MN} \mathcal{Z}^{MN}\big) \big] \nonumber\\
  &\,\qquad\qquad
  -2\mathrm{i} \alpha^2\mathcal{Z}^{K\bullet}
  \,F_{J\bullet\bullet}\, \mathcal{Z}^{\bullet L}\,\big(F_{KLMN} -
  F_{KM\bullet} \,\mathcal{Z}^{\bullet\bullet} \,F_{LN\bullet}\big)
  \,\mathcal{Z}^{MN} 
  \nonumber\\
  &\, \qquad\qquad + \mathrm{i}\alpha^2 \big[ F_{JKLMN} -\tfrac43
  F_{JKM\bullet} \,\mathcal{Z}^{\bullet\bullet}
  \,F_{LN\bullet}\big] \,\mathcal{Z}^{KL}\,\mathcal{Z}^{MN} \Big]  +
  \mathcal{O}(\alpha^3) \,,   \nonumber\\[.2ex] 
 \tilde\omega_{IJ} =&\, [\mathcal{S}^{-1}]^K{}_{(I}\,
        [\mathcal{S}^{-1}]^L{\!}_{J)}\,\nonumber\\
        &\,\times \Big[ \omega_{KL} - F_{KLM} \mathcal{Z}^{MN} (\omega_N
        +\alpha F_{NPQ} \,\mathcal{Z}^{PQ} )
        \nonumber\\
        &\,\quad + \alpha\big(F_{KLMN}
        - F_{KMP}\, F_{LNQ}\, \mathcal{Z}^{PQ}\big) \mathcal{Z}^{MN} 
        \nonumber\\
        &\,\quad + 2 \mathrm{i}\alpha\,\big[\big(\omega_{KLPQ} -F_{KL\bullet}\,
        \mathcal{Z}^{\bullet\bullet}\, \omega_{\bullet PQ}\big)
        \mathcal{Z}^{PQ} 
                 - 2  \,(\omega_{K\bullet\bullet}\, \mathcal{Z}^{\bullet\bullet}
        F_{L\bullet\bullet} \mathcal{Z}^{\bullet\bullet})
         \big] \nonumber\\
        &\,\quad -2\mathrm{i} \big(\omega_{K\bullet} +\alpha 
        F_{K \bullet MN} 
        \mathcal{Z}^{MN}
        \big)
        \mathcal{Z}^{\bullet\bullet}  \big(\omega_{L\bullet} +\alpha
        F_{L\bullet PQ} \, \mathcal{Z}^{PQ}\big)  \nonumber\\
        &\,\quad +2\mathrm{i} \, F_{KL\bullet} \mathcal{Z}^{\bullet\bullet}
        \big(
        \omega_{\bullet \bullet} +\alpha F_{\bullet \bullet MN}
         \mathcal{Z}^{MN}\big)\,
        \mathcal{Z}^{\bullet\bullet}\,  \big(\omega_{\bullet} +\alpha
        F_{\bullet PQ} \,\mathcal{Z}^{PQ}\big) \nonumber\\
        &\,\quad -2\mathrm{i} \big(\omega_{KL\bullet} +\alpha
        F_{KL\bullet MN} \mathcal{Z}^{MN} \big)\,
        \mathcal{Z}^{\bullet\bullet}  \,\big(\omega_\bullet +\alpha
        F_{\bullet PQ}\mathcal{Z}^{PQ} \big) \nonumber\\
        &\,\quad +4\mathrm{i} \big(\omega_{K\bullet} +\alpha
        F_{K\bullet MN}\mathcal{Z}^{MN} \big)\, \mathcal{Z}^{\bullet\bullet}\,
        F_{L\bullet\bullet} \,\mathcal{Z}^{\bullet\bullet}\,
        \big(\omega_\bullet +\alpha F_{\bullet PQ} \,\mathcal{Z}^{PQ}\big) \nonumber\\
        &\,\quad +\mathrm{i} \big(\omega_{\bullet} +\alpha F_{\bullet
          MN} \,\mathcal{Z}^{MN} \big) \,\mathcal{Z}^{\bullet\bullet}
        \, \big[F_{\bullet KL\bullet} - 2\,
        F_{\bullet K\bullet} \mathcal{Z}^{\bullet\bullet}F_{\bullet L\bullet}
        \big]\, \mathcal{Z}^{\bullet\bullet}\,
        \big(\omega_\bullet +\alpha F_{\bullet PQ}\,\mathcal{Z}^{PQ}
        \big) \nonumber\\ 
        &\,\quad  
        -\mathrm{i} F_{RST} \,(\mathcal{Z}^{R\bullet} \,F_{\bullet KL} \big)\,
        \big[\mathcal{Z}^{S\bullet} \big(\omega_{\bullet} +\alpha
        F_{\bullet MN}\, \mathcal{Z}^{MN}\big)\big]\,
        \big[\mathcal{Z}^{T\bullet} \big(\omega_{\bullet} +\alpha
        F_{\bullet PQ}\,\mathcal{Z}^{PQ}\big)\big] \nonumber\\
        &\,\quad
        + 4\mathrm{i}\alpha \big(F_{X\bullet\bullet}
        \mathcal{Z}^{\bullet\bullet} 
        F_{K\bullet\bullet}\mathcal{Z}^{\bullet\bullet}\big)
        \mathcal{Z}^{X \bullet}\, 
        \big(\omega_{L\bullet} +\alpha F_{L\bullet MN} \,\mathcal{Z}^{MN}\big)
        \nonumber\\
        &\,\quad 
        + 4\mathrm{i}\alpha \big(F_{XK\bullet\bullet}
        \mathcal{Z}^{\bullet\bullet}
        F_{L\bullet\bullet} \mathcal{Z}^{\bullet\bullet} \big)
        \mathcal{Z}^{X\bullet} \,\big(\omega_{\bullet} +\alpha
        F_{\bullet MN} \,\mathcal{Z}^{MN} \big)
        \nonumber\\
        &\,\quad 
        - 4\mathrm{i}\alpha \big(F_{X\bullet\bullet}
        \mathcal{Z}^{\bullet\bullet}
        F_{K\bullet\bullet} \mathcal{Z}^{\bullet\bullet}\big)
        \mathcal{Z}^{X\bullet} \,F_{\bullet L\bullet} \,\mathcal{Z}^{\bullet\bullet}\, 
        \big(\omega_{\bullet} +\alpha F_{\bullet MN}\, \mathcal{Z}^{MN}\big)
        \nonumber\\
        &\, \quad 
         - 2 \mathrm{i} \alpha \, \left( F_{KLRS} - 
        F_{KL \bullet} {\cal Z}^{\bullet \bullet} F_{\bullet RS}
        - 2 F_{K \bullet R} {\cal Z}^{\bullet \bullet} F_{L \bullet S} 
         \right)
        {\cal Z}^{RT} {\cal Z}^{SU} \nonumber\\
        &\,\qquad \times \left[ \omega_{TU} - F_{TU \bullet} {\cal Z}^{\bullet \bullet}
        \left( \omega_{\bullet} + \alpha F_{\bullet X Y } {\cal Z}^{X Y} \right) \right]
        \nonumber\\
        &\,\quad 
         +\mathrm{i} \alpha^2 
        \big[ F_{KLMN\bullet\bullet}
        \mathcal{Z}^{\bullet\bullet}  
        -4\,\big(F_{KMN\bullet\bullet} \mathcal{Z}^{\bullet\bullet}
        F_{L\bullet\bullet} \mathcal{Z}^{\bullet\bullet}\big) \big]
        \,\mathcal{Z}^{MN} \nonumber\\
        &\,\quad 
        +2\mathrm{i}\alpha^2
        \big(\mathcal{Z}^{M\bullet}F_{K\bullet\bullet}
        \mathcal{Z}^{\bullet N}\big)\, \big[F_{MNPQ} -
        F_{MP\bullet}\mathcal{Z}^{\bullet\bullet} F_{\bullet NQ} \big] 
        \big(\mathcal{Z}^{P\bullet}
        F_{L\bullet\bullet}
        \mathcal{Z}^{\bullet Q}\big) \nonumber\\
        &\,\quad 
        -2\mathrm{i}\alpha^2 \big(\mathcal{Z}^{\bullet\bullet}F_{K\bullet\bullet}
        \mathcal{Z}^{\bullet\bullet}F_{X\bullet\bullet} \big)\,
        \mathcal{Z}^{XY} \,\big(F_{Y\bullet\bullet}\mathcal{Z}^{\bullet\bullet}
                F_{L\bullet\bullet} 
        \mathcal{Z}^{\bullet\bullet}\big) \nonumber\\
        &\,\quad 
        +8\mathrm{i}\alpha^2 \,F_{KMNP}\, F_{QRS}
        \,\mathcal{Z}^{MQ}\, \mathcal{Z}^{NR}\, 
        \,\big(\mathcal{Z}^{P\bullet} 
               F_{L\bullet\bullet}
               \mathcal{Z}^{\bullet S}\big) \nonumber\\
        &\,\quad 
               - 2\mathrm{i}\alpha^2
               \big(\mathcal{Z}^{M\bullet} F_{KL\bullet\bullet} 
        \mathcal{Z}^{\bullet N}
              -2 \mathcal{Z}^{M\bullet}
             F_{K\bullet\bullet} \mathcal{Z}^{\bullet\bullet}
        F_{L\bullet\bullet} \mathcal{Z}^{\bullet N} \big) \nonumber\\
        &\,\qquad\qquad \times \big[F_{MNPQ} - F_{MP\bullet}
        \mathcal{Z}^{\bullet\bullet} 
             F_{\bullet NQ} 
        \big]
        \mathcal{Z}^{PQ} \nonumber\\
        &\,\quad 
            -\tfrac43 \mathrm{i}\alpha^2
         \big(F_{KLMNP} \,F_{QRS} + F_{KMNP}
        \,F_{LQRS}\big)\, \mathcal{Z}^{MQ} \,\mathcal{Z}^{NR}\,
            \mathcal{Z}^{PS}
         \nonumber\\
        &\,\quad 
        +2\mathrm{i}\alpha^2 \,F_{KL\bullet}
        \mathcal{Z}^{\bullet X} \, 
        \mathcal{Z}^{M\bullet}\,F_{X\bullet\bullet}\,\mathcal{Z}^{\bullet N}\,
        \big(F_{MNPQ} -
        F_{MP\bullet} \,\mathcal{Z}^{\bullet\bullet} \,F_{\bullet NQ}\big)
        \,\mathcal{Z}^{PQ}       \nonumber\\
        &\,\quad 
        -\mathrm{i}\alpha^2 \,F_{KL\bullet}
        \mathcal{Z}^{\bullet\bullet}
        \big[ F_{\bullet MNPQ} -\tfrac43
        F_{MP\bullet\bullet} \,\mathcal{Z}^{\bullet\bullet}
        \,F_{\bullet NQ}\big] \,\mathcal{Z}^{MN}\,\mathcal{Z}^{PQ}\nonumber\\
        &\,\quad -2\mathrm{i}\alpha^2 \, \big( \mathcal{Z}^{\bullet
          \bullet} \, F_{\bullet K \bullet} \, \mathcal{Z}^{\bullet
          \bullet} \, F_{\bullet M \bullet} \, \mathcal{Z}^{\bullet
          \bullet} \, F_{\bullet L \bullet} \, \mathcal{Z}^{\bullet
          \bullet} \, F_{\bullet N \bullet} \big) \mathcal{Z}^{MN} 
        \Big]  +  \mathcal{O}(\alpha^3)\,.  
\end{align} 
Again we have sometimes represented indices by bullets whenever they
are contracted in an ope or closed stringlike fashion and there is no
ambiguity.

\section{Topological free energies for genus $\boldsymbol{g\leq3}$
  that satisfy the holomorphic anomaly equation }
\label{App:F(g)-s}
\setcounter{equation}{0}
In this appendix we list the topological free energies
$F^{(g)}(\mathcal{Y}, \bar{\mathcal{Y}})$ that follow from expanding
${\cal H}^{(1)}$, given in \eqref{eq:almost-harm}, order-by-order in
$\alpha$.  To order $\alpha^3$ we obtain
\begin{align}
  \mathcal{H}^{(1)}(\mathcal{Y},\bar{\mathcal{Y}};N) =&\, 4 \Big[
    F^{(1)} (\mathcal{Y},\bar{\mathcal{Y}}) + \big(F^{(2)}
    (\mathcal{Y},\bar{\mathcal{Y}}) + F^{(3)}
    (\mathcal{Y},\bar{\mathcal{Y}}) + {\rm h.c.}
    \big) \Big]\nonumber\\
  & -16\,\alpha \, \mathcal{G}_1 (\mathcal{Y},\bar{\mathcal{Y}})
  + \mathcal{O}(\alpha^4) \,, 
\end{align} 
where the function $\mathcal{G}_1$ is given in \eqref{eq:G-fcts},.
The symplectic functions $F^{(g)} (\mathcal{Y},\bar{\mathcal{Y}})$
that appear at order $\alpha^g$ are given by (for $g=1,2,3$) 
\begin{align}
   \label{eq:F1-F2-F3}
   F^{(1)} (\mathcal{Y},\bar{\mathcal{Y}}) =&\, \omega^{(1)} + \bar
   \omega^{(1)} + \alpha \ln \det N_{IJ}\;,
   \nonumber\\[2mm]
   F^{(2)}(\mathcal{Y},\bar{\mathcal{Y}}) =&\, \omega^{(2)} - N^{IJ}
   \big( \omega_I^{(1)} - \mathrm{i} \alpha \, F_{IKL} N^{KL} \big)
   \big( \omega_J^{(1)} - \mathrm{i} \alpha \, F_{JPQ} N^{PQ} \big)
   \nonumber\\
   &\, +  2 \alpha \, N^{IJ} \omega^{(1)}_{IJ} - \alpha^2 \big[\mathrm{i}
    N^{IJ} N^{KL} F_{IJKL}
     - \tfrac23  N^{IJ} F_{IKL} N^{KP} N^{LQ} F_{JPQ} \big]
   \;,\nonumber\\[2mm] 
   F^{(3)}(\mathcal{Y},\bar{\mathcal{Y}}) =&\, \omega^{(3)} - 2\,
   N^{IJ} \, \omega^{(2)}_I \, \omega^{(1)}_J + 2\,\omega^{(1)}_{IJ}
   \,N^{IK}\,
   \omega_K^{(1)}\, N^{JL} \,\omega^{(1)}_L \nonumber\\
   & \, + \tfrac{2}3 \mathrm{i} \, F_{IJK} N^{IP}\, \omega_P^{(1)}\,
   N^{JQ} \,\omega^{(1)}_Q \,N^{KL} \omega_L^{(1)}
   \nonumber\\
   &\,+ \alpha \big[ 2 \mathrm{i} \, N^{IJ} \omega^{(2)}_I F_{JKL}
   N^{KL} - 4\, N^{IJ} \omega^{(1)}_{IKL} N^{KL} \omega^{(1)}_J
   \nonumber\\ 
   &\,\qquad - 4\, \mathrm{i} N^{IJ} N^{KL} F_{I LM} N^{MN}
   \omega^{(1)}_{KN} \, \omega^{(1)}_J - 2 \mathrm{i} \, F_{IJ KL}
   N^{KL} N^{IP}
   \omega_P^{(1)} N^{JQ} \omega_Q^{(1)}  \nonumber\\
   &\, \qquad + 2\, F_{IKP} N^{KL} F_{LQJ} N^{PQ} N^{IR}
   \omega_R^{(1)} N^{JS} \omega_S^{(1)}
   \nonumber\\
   &\, \qquad -4\mathrm{i} \, \omega^{(1)}_{IJ} N^{IK}
   \omega_K^{(1)} N^{JL} F_{LPQ} N^{PQ}  \nonumber\\
   &\,\qquad +2\, F_{IJK} N^{IP} \omega_P^{(1)} N^{JQ} \omega^{(1)}_Q
   N^{KR} F_{RST} N^{ST}
   \nonumber\\
   &\, \qquad  + 2\, N^{IJ} \, \omega_{IJ}^{(2)} - 2\, N^{IJ} N^{KL}
   \, \omega^{(1)}_{IK} \,\omega^{(1)}_{JL} \big] \nonumber\\ 
   &\,+ \alpha^2 \big[
     2 \mathrm{i}\, N^{IJ} F_{IKLMN} N^{KL} N^{MN} \,\omega_J^{(1)} 
   - 4 N^{IJ} N^{KL} F_{ILM} N^{MN} F_{KNPQ} N^{PQ} \,\omega_J^{(1)}
   \nonumber\\
   &\, \qquad 
   - \tfrac83 N^{IJ} F_{IMNP} N^{MK} N^{NL} N^{PQ} F_{KLQ}\,  \omega_J^{(1)} 
   \nonumber\\
   &\, \qquad 
   - 4 \mathrm{i} \, N^{IJ} N^{KL} F_{ILM} N^{MN} F_{KRT} N^{RP}
   N^{TQ} F_{NPQ} \,  \omega_J^{(1)} 
   \nonumber\\
   &\, \qquad 
   + 4 \mathrm{i} \, N^{IJ} \omega_{IMN}^{(1)} N^{MN} F_{JKL} N^{KL}
   - 4 N^{IJ} N^{MN} F_{INP} N^{PQ} \omega^{(1)}_{MQ} F_{JKL} N^{KL} 
   \nonumber\\
   &\, \qquad 
   - 2 \,\omega^{(1)}_{IJ} N^{IK} F_{KPQ} N^{PQ} N^{JL} F_{LRS} N^{RS} 
   \nonumber\\
   &\,\qquad
   - 4 \, F_{IJMN} N^{MN} N^{IK} \omega^{(1)}_K N^{JL} F_{LRS} N^{RS} \nonumber\\
   &\, \qquad 
   - 4 \mathrm{i} \,F_{IMN} N^{MP} F_{JPQ} N^{QN}
   N^{IK} \omega^{(1)}_K N^{JL} F_{LRS} N^{RS} 
   \nonumber\\
   &\, \qquad 
   -2 \,\mathrm{i} \, F_{IJK} N^{IP} \omega_P^{(1)} N^{JQ} F_{QST}
   N^{ST} N^{KR} F_{RUV} N^{UV}     \nonumber\\
   &\, \qquad +
   2 N^{IJ} N^{KL} \,\omega^{(1)}_{IJKL} + \tfrac83 \mathrm{i} \, F_{IJK} N^{IL} N^{JP}
   N^{KQ} \, \omega^{(1)}_{LPQ}  \nonumber\\
   &\, \qquad 
     + 4 \mathrm{i} \, F_{IJRS} \, N^{IJ} N^{RK} N^{SL} \, \omega^{(1)}_{KL} 
     - 4  \, F_{IPQ} F_{JRS} N^{IK} N^{JL} N^{PR} N^{QS} \,
     \omega^{(1)}_{KL} \big] \nonumber\\ 
   &\, + \alpha^3 
   \big[
   2 \, N^{IJ} F_{IMNPQ} N^{MN} N^{PQ} F_{JKL} N^{KL} \nonumber\\
   &\, \qquad + 4 \mathrm{i} \, N^{IJ} N^{PM} F_{IMN} N^{NQ} F_{PQRS}
   N^{RS} F_{JKL} N^{KL}
   \nonumber\\
   &\, \qquad + \tfrac83 \mathrm{i} \, N^{IJ} F_{IMNP} N^{MK} N^{NL}
   N^{PQ} F_{KLQ} F_{JRS} N^{RS }
   \nonumber\\
   &\, \qquad - 4 \, N^{IJ} N^{PM} F_{IMN} N^{NQ} F_{PTU} N^{TR}
   N^{US} F_{QRS} F_{JKL} N^{KL}
   \nonumber\\
   &\, \qquad + 2 \mathrm{i} \, F_{IJMN} N^{MN} N^{IK} F_{KPQ} N^{PQ}
   N^{JL} F_{LRS} N^{RS}
   \nonumber\\
   &\, \qquad - 2 \,F_{IMN} N^{MT} F_{JTU} N^{UN} N^{IK} F_{KPQ} N^{PQ}
   N^{JL} F_{LRS} N^{RS}
   \nonumber\\
   &\, \qquad - \tfrac{2}3 \, F_{IJK} N^{IP} F_{PMN} N^{MN} N^{JQ}
   F_{QST} N^{ST} N^{KR} F_{RUV} N^{UV}
   \nonumber\\
   &\, \qquad - \tfrac23 \mathrm{i} \, F_{IJKLPQ} \, N^{IJ} N^{KL}
   N^{PQ}
   + \tfrac83 N^{IJ} \, F_{IJKLP} \, N^{KR} N^{LS} N^{PT} \, F_{RST} \nonumber\\
   &\, \qquad + 2\, N^{IJ} \, F_{IJKL} \, N^{KP} N^{LQ} \, F_{PQRS} \,
   N^{RS} \nonumber\\
   &\,\qquad 
   + \tfrac23 F_{IKPS} \, N^{IJ} N^{KL} N^{PQ} N^{SR} \, F_{JLQR} \nonumber\\
   &\,\qquad + 4\mathrm{i} \, N^{IJ} F_{IJKL}\, N^{KP} N^{LQ} \,
   F_{PST} \, N^{SU} N^{TV} \, F_{QUV} \nonumber\\ 
   &\, \qquad + 4\mathrm{i} \,  F_{IJKL}\, N^{IP} N^{JQ} N^{KR} N^{LS}
   F_{PQU} N^{UV} F_{VRS} \nonumber\\ 
   &\, \qquad - 2\, F_{IJK} F_{LPQ} F_{RST} F_{UVW} N^{IL}  N^{JP}
   N^{KR} N^{QU} N^{SV} N^{TW}\nonumber\\ 
   &\, \qquad - \tfrac43 F_{IJK} F_{LPQ} F_{RST} F_{UVW} N^{IL} N^{JR} 
   N^{KU} N^{PS} N^{QV} N^{TW} \big]\;.
\end{align}
 Here we expanded $\omega(\mathcal{Y})$ as
$\omega(\mathcal{Y}) = \sum_{n=1}^{\infty} \omega^{(n)}
(\mathcal{Y})$, where we count $\omega^{(n)} (\mathcal{Y})$ as being
of order $\alpha^n$, following \eqref{eq:series-exp-omega}. The
non-holomorphicity of $F^{(g)} (\mathcal{Y},\bar{\mathcal{Y}}) $ is
entirely contained in the quantities $N^{IJ}$.  Observe that $F^{(1)}$
is real, while the higher $F^{(g)}$ ($g\geq2$) are not.

The expressions for $F^{(g)}$ given above were obtained by explicit
construction and they satisfy the
holomorphic anomaly equations \eqref{eq:nonholo-derivative-h} of
perturbative topological string theory ($g \geq 2$),
\begin{equation}
  \label{eq:holom-top-string-g}
  \partial_{\bar I} F^{(g)} = \mathrm{i}\,  {\bar F}_{IJK} \,  N^{JM} N^{KN} \,
  \Big[ - 2 \alpha \, D_M \partial_N F^{(g-1)} + \sum_{r=1}^{g-1}
  \partial_M F^{(r)} \,
  \partial_N F^{(g-r)} \Big]  \;,
\end{equation}
where $D_M$ denotes the covariant derivative introduced in
\eqref{eq:K-connection}.  The expression for $F^{(2)}$ has been
obtained before by other methods
\cite{Bershadsky:1993cx,Aganagic:2006wq,Grimm:2007tm} based on a direct
integration of \eqref{eq:holom-top-string-g}.  Partial results for
$F^{(3)}$ have been given in \cite{Bershadsky:1993cx}.

\section{An application: the FHSV model}
\label{App:specific-model}
\setcounter{equation}{0}
In this appendix we illustrate our results in the context of the FHSV
model \cite{Ferrara:1995yx} and compare them to earlier results
obtained in \cite{Cardoso:2010gc} by means of a related but slightly
different approach. Here we restrict ourselves to second order. In
the type-II description, the FHSV model corresponds to the
compactification on the Enriques Calabi-Yau three-fold, which is
described as an orbifold
$(\mathrm{T}^2\times\mathrm{K3})/\mathbb{Z}_2$, where $\mathbb{Z}_2$
is a freely acting involution. The massless sector of the
four-dimensional theory comprises 11 vector supermultiplets, 12
hypermultiplets and the $N=2$ graviton supermultiplet. The classical
moduli space of the vector multiplet sector equals the
special-K\"ahler space,
\begin{equation}
\label{eq:4vector-special-K}
\mathcal{M}_{\mathrm{vector}}=
\frac{\mathrm{SL}(2)}{\mathrm{SO}(2)}\times
\frac{\mathrm{O}(10,2)}{\mathrm{O}(10)\times\mathrm{O}(2)}\,,
\end{equation}
which is encoded in the classical holomorphic function 
\begin{equation}
  \label{eq:F-0-FHSV}
  F^{(0)}(\mathcal{Y}) = - \frac{\mathcal{Y}^1\, 
    \mathcal{Y}^a\eta_{ab}\mathcal{Y}^b}{\mathcal{Y}^0}  \,, 
\end{equation}
where $a,b=2,\ldots,11$, and the symmetric matrix $\eta_{ab}$ is an
$\mathrm{SO}(9,1)$ invariant metric of indefinite signature. The two
factors of the special-K\"ahler space are associated with
$\mathrm{T}^2/\mathbb{Z}_2$ and the $\mathrm{K3}$ fiber, and
`special' coordinates for these two spaces are denoted by $S =
-\mathrm{i} \mathcal{Y}^1/\mathcal{Y}^0$ and
$T^a=-\mathrm{i}\mathcal{Y}^a/\mathcal{Y}^0$. This leads to the
following expression for $N \equiv \det \big[ 2 \, {\rm Im} \big[ F^{(0)}_{IJ} \big] \big] $ and $\mathcal{H}\vert_{\Omega=0}$, 
\begin{equation}
  \label{eq:N-Omega0-FHSV}
  N =c\,(S+\bar S)^{10} \big((T+\bar T)^2\big)^2 \,, \qquad
  \mathcal{H}\vert_{\Omega=0} =- (S+\bar S)(T+\bar T)^2\,
  \vert\mathcal{Y}^0\vert^2 \,,
\end{equation}
where $c$ is an irrelevant constant. Here we use the notation that $T^2\equiv
T^a\eta_{ab}T^b$ and likewise for $\vert T\vert^2$. 
Observe that $N$ is not covariant under symplectic reparametrizations, while $\mathcal{H}\vert_{\Omega=0}$
is covariant.
Note that in the
present approach we are making use of a specifc parametrization
defined by \eqref{eq:F-0-FHSV}. Therefore the covariance under
symplectic reparametrizations is not always clear, and instead we may
have to rely on the S- and T-duality invariances that we will discuss
below. This was also the strategy used in
\cite{Cardoso:2008fr,Cardoso:2010gc}.

Subsequently, we expand both $\omega(\mathcal{Y})$ and $\Omega
(\mathcal{Y},\bar{\mathcal{Y}})$ into powers of $\alpha$ as $\omega =
\omega^{(1)} + \omega^{(2)} + \mathcal{O} (\alpha^3)$ and $\Omega =
\Omega^{(1)} + \Omega^{(2)} + \mathcal{O} (\alpha^3)$, respectively.
Following the discussion in \cite{Cardoso:2010gc}, we start with the
expression for $\Omega^{(1)}$, known from threshold corrections and
from the topological string side \cite{Harvey:1996ts,Klemm:2005pd}.
In the conventions of \cite{Cardoso:2008fr}, it is given by
\begin{align}
  \label{eq:Omega-FHSV}
  \Omega^{(1)}(\mathcal{Y},\bar{\mathcal{Y}})
  = -\frac{1}{4 \pi} \Big[&\tfrac12 \ln[\eta^{24}(2S)\,\Phi(T)]
  +\tfrac12 \ln[\eta^{24}(2\bar S)\,\Phi(\bar T)]
  \nonumber\\
  & + 2 \ln [(S+\bar S)^3 (T+\bar T)^2]\Big] \,.
\end{align}
It is invariant under S-duality transformations belonging to the
$\Gamma(2)$ subgroup of $\mathrm{SL}(2;\mathbb{Z})$, and also
invariant under the T-duality group $\mathrm{O}(10,2;\mathbb{Z})$,
since $\Phi(T)$ is a holomorphic automorphic form of weight 4
\cite{Borcherds:1996}, transforming under the T-duality transformation
$T^a \rightarrow T^a\, [T^2]^{-1}$ as
\begin{equation}
  \label{eq:Phi-T}
  \Phi(T)\to [T^2]^{4} \,\Phi(T)\,.
\end{equation}
We can now recast \eqref{eq:Omega-FHSV} in the form of 
\eqref{eq:ansatz-Omega1},
\begin{equation}
  \label{eq:Omega-FHSV-1}
  \Omega^{(1)}(\mathcal{Y},\bar{\mathcal{Y}}) = \omega(\mathcal{Y})+\bar\omega(\bar{\mathcal{Y}})
  -\frac1{8\pi} \ln N + \Psi(\mathcal{Y},\bar{\mathcal{Y}})\,,   
\end{equation}
so that $\alpha = - 1/(8 \pi)$ and $\beta =1$, with
\begin{align}
  \label{eq:omega-psi-fhsv}
  \omega^{(1)} (\mathcal{Y}) =&\,  - \frac{3}{2 \pi} \ln \eta^2 (2S) -
  \frac{1}{8 \pi} \ln \Phi(T)+ \frac1{4\pi} \ln \mathcal{Y}^0  \;, 
  \nonumber\\
  \Psi(\mathcal{Y},\bar{\mathcal{Y}}) =&\,
  - \frac{1}{4 \pi} \ln \left[(S + \bar S) (T + \bar T)^2 |\mathcal{Y}^0|^2\right] \;,
\end{align}
where we note that $\Psi(\mathcal{Y},\bar{\mathcal{Y}})$ transforms as
a function because it is equal to the logarithm of $\mathcal{H}\vert_{\Omega=0}$,
the classical part of the Hesse potential.

Next, we insert these expressions into $\Omega^{(2)}$. Here we recall that in the presence
of a function $\,\Psi(\mathcal{Y},\bar{\mathcal{Y}})$, the expression for  $\Omega^{(2)}$
is not simply obtained  by the second line of \eqref{eq:ansatz-nonholo-Omega-2},
but to this we also have to add $\beta$-dependent terms, as shown in
\eqref{eq:mod-Omega-beta}. Thus, we have
\begin{align}
 \Omega^{2)}(\mathcal{Y},\bar{\mathcal{Y}}) =
 &\, \Big[  \omega^{(2)} +2\,\alpha\, N^{IJ} \omega^{(1)}_{IJ} -\alpha^2 \big[
   \mathrm{i}F_{IJKL} -\tfrac23
    F_{IKM} F_{JLN}   N^{MN}  \big]   N^{IJ}N^{KL} +\mathrm{h.c.} \Big] \nonumber\\
    &\,   +
 N^{IJ} \Big[ 2 \, \partial_I \left(  \omega^{(1)} + \alpha \ln N \right) \Psi_J  +   \Psi_I   \Psi_J  + {\rm h.c.}
    \Big] \,.
    \label{om-2-beta}
       \end{align}
Then, direct evaluation of this results in
\begin{align}
  \label{eq:Om2result}
  &\Omega^{(2)} =  \bigg\{
  \omega^{(2)}
  + \frac{1}{ (\mathcal{Y}^0)^2} \bigg[
    \frac{1}{64 \pi^2}  G_2(2S)
    \frac{ \partial \ln \Phi(T)}{\partial T^a}
    \frac{ \partial \ln \Phi(T)}{\partial T_a}
    - \frac{1}{32 \pi^2}  G_2(2S)
    \frac{ \partial^2 \ln \Phi(T)}{\partial T^a \partial T_a}
  \bigg] \nonumber\\
  &  + \frac{1}{   (\mathcal{Y}^0)^2} \bigg[ 4 \,
    \hat{G}_2 (2S, 2 \bar S)\,
    \frac{\partial \Omega^{(1)}}{\partial T^a}\,
    \frac{\partial \Omega^{(1)}}{\partial T_a}
    + \frac{1}{32\pi^2} \hat{G}_2 (2S, 2 \bar S)\,
    \Big(
    \frac{ \partial^2 \ln \Phi(T)}{\partial T^a \partial T_a} +
    \frac{ \partial \ln \Phi(T)}{\partial T^a}
    \frac{ \partial \ln \Phi(T)}{\partial T_a} \Big)
  \bigg] \nonumber\\
  &  - \frac{1}{(\mathcal{Y}^0)^2} \bigg[
    G_2 (2S)\,
    \frac{\partial \Omega^{(1)}}{\partial T^a}\,
    \frac{\partial \Omega^{(1)}}{\partial T_a}
    + \tfrac14
    \frac{\partial \ln\Phi(T)}{\partial T_a}\,
    \frac{\partial\Omega^{(1)}}{\partial T^a} \, \frac{\partial
      \Omega^{(1)}}{\partial S}
  \bigg]
  + {\rm h.c.} \bigg\}  \;,
\end{align}
where 
\begin{align}
  \label{eq:G_2}
 G_2 (2 S) =&\, \tfrac12 \partial_S \ln \eta^2 (2S)\,,\nonumber\\
 \hat{G}_2(2 S, 2 \bar S) =&\, G_2 (2S) + \frac{1}{ 2(S + \bar S)}
 \,. 
\end{align}
The first line of \eqref{eq:Om2result} contains purely holomorphic
terms, while the second line contains terms that are invariant under
S- and T-duality.  The last line contains the terms that are neither
holomorphic nor invariant under S- and T-dualities.  They were already
obtained in \cite{Cardoso:2008fr} by requiring invariance of the model
under S- and T-duality, and thus were determined up to invariant
terms. Here, the duality invariant terms are unambiguously determined and given by the second
line of \eqref{eq:Om2result}, as we just established.  The reason is that the scheme presented
in this paper ensures the validity of the holomorphic anomaly equation.
This implies that invariant terms cannot be arbitrarily included, as we discussed in section
\ref{sec:non-holom-deform}. 
Earlier results obtained in
\cite{Cardoso:2008fr,Cardoso:2010gc} are fully consistent with the
ones given above.

Next, we compute the symplectic function $F^{(2)}$, which is
constructed from $\Omega^{(2)}$ as follows.  Recalling
\eqref{om-2-beta}, we write $\Omega^{(2)}$ as $\Omega^{(2)}= \Delta +
\bar \Delta$.  Then, from \eqref{eq:F1-F2-F3} we infer the relation
$F^{(2)} = \Delta - N^{IJ} \Omega^{(1)}_I \Omega^{(1)}_J$, where
however (and differently from \eqref{eq:F1-F2-F3}) $\Omega^{(1)}$ now
also contains $\Psi(\mathcal{Y},\bar{\mathcal{Y}})$, c.f.
\eqref{eq:ansatz-Omega1}.  As we have observed in the text below
\eqref{eq:H-1-beta}, the terms depending on $\Psi$ cancel in the
higher order result for $\mathcal{H}^{(1)}$, and therefore $F^{(2)}$
will not depend on $\Psi$.  We obtain
\begin{align}
  \label{eq:F2-fhsv}
  & F^{(2)} =
  \omega^{(2)}
  + \frac{1}{ (\mathcal{Y}^0)^2} \bigg[
  \frac{1}{64 \pi^2}  G_2(2S)
  \frac{ \partial \ln \Phi(T)}{\partial T^a}
  \frac{ \partial \ln \Phi(T)}{\partial T_a}
  - \frac{1}{32 \pi^2}  G_2(2S)
  \frac{ \partial^2 \ln \Phi(T)}{\partial T^a \partial T_a}
  \bigg ]
  \nonumber\\
  & 
  + \frac{1}{32\pi^2 \, (\mathcal{Y}^0)^2 } \hat{G}_2 (2S, 2 \bar S)\,
  \Bigg(
  \frac{ \partial^2 \ln \Phi(T)}{\partial T^a \partial T_a} +
  \frac{ \partial \ln \Phi(T)}{\partial T^a}
  \frac{ \partial \ln \Phi(T)}{\partial T_a} \Bigg)
  \\
  & - \frac{3}{64 \pi^2 \, (\mathcal{Y}^0)^2} \,
  \hat{G}_2 (2S, 2 \bar S)\; \frac{\partial\log\left[\Phi(T)\,
      [(T+\bar T)^2]^4\right] }{\partial T_a} \; 
  \frac{\partial\log\left[\Phi(T)\, [(T+\bar T)^2]^4\right] }{\partial T^a}\,
     \;. \nonumber
\end{align}
The first line contains purely holomorphic terms, while the second and
third lines are given in terms of non-holomorphic combinations that
are S- and T-duality invariant.  The holomorphic contributions in the
first line should, however, be invariant as well. We can verify this
by making use of the transformation rule \eqref{eq:trans-omega-2} for
$\omega^{(2)}$. We have checked that the first line is indeed invariant
under S-duality, and we expect the same for T-duality. At this stage
we are not able to give an explicit representation of $\omega^{(2)}$
as a function of $\mathcal{Y}^0, S$ and $T^a$ that generates the
desired transformations.  The expression for $F^{(2)}$ given above is
in agreement with the finding of \cite{Grimm:2007tm}.

\end{appendix} 

\providecommand{\href}[2]{#2}
\begingroup\raggedright\endgroup
\end{document}